\documentclass[11pt]{article}
\usepackage{amsfonts,citesort}
\usepackage{amsbsy}
\usepackage{latexsym}
\usepackage{graphicx}
\usepackage{subfigure}

\usepackage{setspace}

\newcommand{\sqt}{\sqrt{(t+a)(t+b)}}

\newcommand{\oab}{\sqrt{(1-a)(1-b)}}
\newcommand{\sqx}{\sqrt{(b-x)(x-a)}}
\newcommand{\sqy}{\sqrt{(b-y)(y-a)}}

\newcommand{\rme}{{\rm e}}
\newcommand{\al}{\alpha}

\newcommand{\be}{\begin{equation}}
\newcommand{\ee}{\end{equation}}
\newcommand{\bq}{\begin{eqnarray}}
\newcommand{\eq}{\end{eqnarray}}
\newcommand{\nn}{\nonumber}
\newcommand{\ba}{\begin{array}}
\newcommand{\ea}{\end{array}}

\newcommand{\bW}{{\bf W}}

\newcommand{\co}{\textsf{Comm.}}

\newcommand{\la}{\lambda}
\newcommand{\bt}{\beta}
\newcommand{\bI}{{\bf I}}
\newcommand{\bH}{{\bf H}}
\newcommand{\bx}{{\bf x}}
\newcommand{\by}{{\bf y}}

\newcommand{\bxd}{{\bf x}^{\dag}}
\newcommand{\bHd}{{\bf H}^{\dag}}
\newcommand{\bn}{{\bf n}}

\newcommand{\Ga}{\Gamma}

\renewcommand{\v}{{\mathsf{v}}}

\newcommand{\vp}{{\mathsf{v}^{\prime}}}

\newcommand{\dn}{{\delta_n}}
\newcommand{\tD}{\tilde{D}}

\newcommand{\re}{{\rm e}}

\begin{document}



\title{\bf Perturbed Hankel Determinants: Applications to the Information Theory of MIMO Wireless Communications}

\author{Yang Chen\\
         Department of Mathematics\\
          Imperial College\\
          180 Queen's Gate\\
          London SW7 2BZ, UK\\
          ychen@ic.ac.uk\\
          \\
          Matthew McKay\\
          Electronic and Computer Engineering Department\\
          Hong Kong University of Science and Technology\\
          Clear Water Bay, Kowloon, HONG KONG\\
          eemckay.ust.hk}
\date{01-07-2010}

\maketitle

\begin{abstract}

In this paper we compute two important information-theoretic
quantities which arise in the application of multiple-input
multiple-output (MIMO) antenna wireless communication systems: the
distribution of the mutual information of multi-antenna Gaussian
channels, and the Gallager random coding upper bound on the error
probability achievable by finite-length channel codes. It turns out
that the mathematical problem underpinning both quantities is the
computation of certain Hankel determinants generated by deformed
versions of classical weight functions. For single-user MIMO
systems, it is a deformed Laguerre weight, whereas for multi-user
MIMO systems it is a deformed Jacobi weight. We apply two different
methods to characterize each of these Hankel determinants.  First,
we employ the ladder operators of the corresponding monic orthogonal
polynomials to give an exact characterization of the Hankel
determinants in terms of Painlev\'{e} differential equations. This
turns out to be a Painlev\'{e} V for the single-user MIMO scenario
and a Painlev\'{e} VI for the multi user scenario.  We then employ
Coulomb fluid methods to derive new closed-form approximations for
the Hankel determinants which, although formally valid for large
matrix dimensions, are shown to give accurate results for both the
MIMO mutual information distribution and the error exponent even
when the matrix dimensions are small. Focusing on the single-user
mutual information distribution, we then employ both the exact
Painlev\'{e} representation and the Coulomb fluid approximation to
yield deeper insights into the scaling behavior in terms of the
number of antennas and signal-to-noise ratio.  Among other things,
these results allow us to study the asymptotic Gaussianity of the
distribution as the number of antennas increase, and to explicitly
compute the correction terms to the mean, variance, and higher order
cumulants.

\end{abstract}

\vfill\eject
\vskip .4cm
\setcounter{equation}{0}

\doublespace

\section{Introduction and Preliminaries}

Over the past decade, multiple-input multiple-output (MIMO) systems
have been at the forefront of wireless communications research and
development.  Such systems, which employ multiple antennas at both
the transmitter and receiver, have revolutionized the wireless
industry and now form the basis of most emerging wireless standards,
such as next-generation wireless local area networks (WLAN) and
cellular mobile networks.  The main reason for this explosion of
interest is due to the independent discoveries of Telatar
\cite{Telatar} and Foschini \cite{Foschini}, which demonstrated that
the fundamental information-theoretic capacity of MIMO systems grows
\emph{linearly} with the number of antennas.  This is important,
since traditional methods for increasing capacity, which typically
increase valuable system resources such as the transmission power,
yield only a logarithmic capacity increase. Indeed, MIMO is now
widely recognized as a key technology for meeting the
ever-increasing demands for higher-rate data-oriented wireless
communications applications and services.

There are various metrics for characterizing the fundamental information-theoretic limits of MIMO systems, each of
which have relevance depending on the specific wireless communication configuration. The most commonly studied metric
 is the so-called \emph{ergodic capacity}, which specifies the maximum achievable average mutual information between
 the transmitter and receiver. This measure, which assumes that there is sufficient dynamics in the system such that
 a user's codeword may span a large number of ``independent channels'', has been studied extensively over the past
 decade (see e.g., \cite{Telatar,Foschini,ChianiWin03,Smith03,ShinLee03,McKayCollings05,McKaySmith08,Gao09,Grant,Lozano02} and the tutorial
 discussion \cite{Tulino}).  A closely related measure which has received far less attention is the \emph{outage capacity}.
This measure is important for characterizing the communication
limits of systems which are not highly dynamic (for example, WLANs),
or systems which have stringent delay requirements.  Compared with
the ergodic capacity, the analysis of outage capacity is much more
complicated since it requires the \emph{distribution} of the mutual
information between the transmitter and receiver, rather than simply
the average value.  Another important information-theoretic quantity
of practical interest is the error probability achievable with
block-coding schemes of a given length and rate.  Whilst an exact
characterization of this quantity is not tractable in general,
methods have been proposed by Gallager for upper bounding this
quantity \cite{Gallager}.  Such bounds have been well-established
for single-antenna systems, however much less is known for MIMO.

In this paper, we aim to present new methods for studying the outage
capacity and the error probability of MIMO systems. As we will see,
in both cases the fundamental mathematical problem of interest boils
down to characterizing the Hankel determinant \bq D_n = \det\left(
\mu_{i+j} \right)_{i,j=0}^{n-1} \eq generated from the moments of a
certain weight function $w(x)$, \bq \mu_{k} := \int_{J} x^{k} w(x)
dx \; , \hspace*{1cm} k = 0, 1, 2, \ldots \eq with $J$ denoting the
support of the weight.  The specific form of the weight depends on
the MIMO configuration under investigation.  We will consider two
important MIMO configurations pertaining to single-user and
multi-user MIMO systems respectively.  In the single-user case, the
weight function is shown to be a deformation of the classical
Laguerre weight given by \bq \label{eq:LaguerreWeight}
w(x)=x^{\al}\:\re^{-x}\:(x+t)^{\la},\quad 0\leq x<\infty,  \; \alpha
> 0, \; t> 0 \; , \eq whereas in the multi-user case the weight
function is found to be a deformation of the (shifted) classical
Jacobi weight given by \bq \label{eq:JacobiWeight} w(x) =
x^{\al_1}(1-x)^{\al_2}\left(\frac{x+t}{1-x}\right)^{\la},\;\;
x\in[0,1],\:\;\al_1>0, \:\al_2>0,\;t>0  . \eq

We employ two different methods from random matrix theory for
evaluating the corresponding Hankel determinants, both of which have
been used to great effect by the mathematical physics community.
First, we derive exact expressions for the Hankel determinants by
employing the theory of orthogonal polynomials and their
corresponding ladder operators. There exists extensive literature on
this subject; for example
\cite{bau,be-ro,bo-cl,bo-cl1,bo-lun-ne,bo-ne,chen1,chen2}.  See also
\cite{ba-ch,ch-pr,ch-fe,chen+its,ch-zh,ba-ch-eh} for recent
applications of ladder operators to Hermitian matrix ensembles.

Despite these prior contributions, this approach has not been
employed to problems in  wireless communications and information
theory.  Through the ladder operator framework, we demonstrate that
the Hankel determinant generated from the deformed Laguerre weight
(\ref{eq:LaguerreWeight}) has a simple representation involving the
Painlev\'{e} V differential equation.  We also derive a discrete
difference equation representation, which allows the Hankel
determinant to be numerically evaluated iteratively, in terms of the
Hankel matrix dimension. For the deformed Jacobi weight(
\ref{eq:JacobiWeight}), upon establishing the connection to the
multi-user MIMO system model of interest, we employ a known result
derived using the ladder operator method in \cite{zhang} to
represent the corresponding Hankel determinant in terms of the
Painlev\'{e} VI differential equation.  This presents a new
connection between the Painlev\'{e} VI differential equation and the
information theory of multi-user MIMO communication systems.

In addition to deriving exact expressions for the Hankel
determinants, we also derive approximations for these determinants
by employing the general linear statistics results from
\cite{ChenLawrence}, derived based on Dyson's Coulomb fluid
interpretation \cite{Dyson,Chen+Ismail,Chen+Manning}. These
asymptotic results are essentially the Hankel analog of Szeg\"o's
strong limit theorem on the asymptotic characterization of large
Toeplitz determinants, a component of which appeared as early as
1919 \cite{Szegohankel}, where Hankel determinants generated by
compactly supported weights were studied. We refer the reader to
\cite{bcwmsri,bcw,BasorChen} for more related contributions.

As with the ladder operator approach, Coulomb fluid techniques have
been used very successfully in the context of mathematical physics,
however, they have yet to be adopted by the wireless communications
and information theory community. (An exception is the very recent
submitted work \cite{Kasakopoulos}, which employed the Coulomb fluid
framework but not the linear statistics method to characterize the
large-antenna behavior of the single-user MIMO channel capacity.)  A
key advantage of the Coulomb fluid methodology is that the
expressions which are obtained are in closed-form, and are simpler
than those obtained via exact methods.  Moreover, although formally
valid for large matrix dimensions, we find that the approximations
are remarkably accurate for even very small matrix dimensions (e.g.,
$2 \times 2$).  By adopting this Coulomb fluid framework, our
results also establish the Gaussian behavior of the channel capacity
as the number of antennas grow large, for both the single-user and
multi-user MIMO systems. This point has been derived previously for
the single-user MIMO case, using different methods from
\cite{BaiSilverstein2004}, however we do not believe that it has
been established for the multi-user MIMO scenario.

After deriving the exact Painlev\'{e} and approximate Coulomb fluid representations, we then employ both results to gain useful insights into the mutual information distribution.  In particular, focusing on the single-user MIMO scenario (i.e., deformed Laguerre weight) with $n$ transmit and receive antennas, we demonstrate that the Coulomb fluid approximation in fact gives the \emph{exact} distribution of the mutual information, corresponding to a Gaussian, to leading order in $n$.  We also use the Painlev\'{e} V equation to compute the large-$n$ correction terms for the mean, variance, and third cumulant.  Among other things, these results demonstrate the interesting phenomenon that the distribution deviates from Gaussian as the signal-to-noise ratio (SNR) increases. The sensitivities of the mean, variance, and third moment, with respect to the SNR are also examined.

The paper is organized as follows. In the rest of this section, we
present a detailed discussion of the linear models which underpin
the single-user and multi-user MIMO wireless communication scenarios
of interest.  We also introduce the fundamental
information-theoretic measures of outage capacity and error
probability, and establish important connections with Hankel
determinants generated from deformed Laguerre and Jacobi weights.
Then, in Section \ref{sec:Ladder}, we introduce the ladder operators
of orthogonal polynomials and their associated compatibility
conditions, which provide the key ingredients for establishing an
exact finite $n$ characterization of the Hankel determinants in
terms of Painlev\'{e} differential equations. In Section
\ref{sec:Coulomb}, we introduce, for large $n$, the probability
density of a class of random variables called \emph{linear
statistics}.
These results, based on Dyson's Coulomb fluid interpretation
\cite{Dyson} and developed further in
\cite{Chen+Ismail,ChenLawrence,Chen+Manning,chemancond}, are very
general and embrace a wide class of random matrix models. By
employing these general results, we derive closed-form
approximations for the mutual information distribution for the
Hankel determinants generated by both the deformed Laguerre and
Jacobi weights.  These results permit very fast evaluation of the
error probabilities of single-user and multi-user MIMO systems,
whilst also demonstrating the Gaussian behavior of the mutual
information distribution for large $n$.  In Section
\ref{sec:Beyond}, focusing on the single-user MIMO scenario, the
Coulomb results are compared, for large $n,$ with the solutions of
the continuous $\sigma$--form of the Painlev\'{e} V. We also compute
closed-form asymptotic expressions for the recurrence coefficients
corresponding to the deformed Laguerre weight, which are basic
variables in our theory.  Our analysis will involve a number of
complicated integrals, which are derived and tabulated in the
Appendix.


\subsection{Information Theory of MIMO Wireless Systems}

Consider a MIMO communication system with $n_t$ transmit and $n_r$ receive antennas. The linear model relating the input (transmitted) signal vector $\bx_{n_t\times 1}\in \mathbb{C}^{n_t}$ and output (received) signal vector $\by_{n_r\times 1}\in \mathbb{C}^{n_r}$ takes the form
\bq \label{eq:LinearModel}
\by = \bH \bx + \bn \; .
\eq
Here, $\bn_{n_r\times 1} \in \mathbb{C}^{n_r}$ is a complex Gaussian vector with zero mean and covariance $E( \mathbf{n} \mathbf{n}^\dagger ) = \mathbf{Q}_n$. This covariance matrix can account for the effects of both receiver noise as well as multi-user interference, and as such, the selection of $\mathbf{Q}_n$ will distinguish between the single-user and multi-user MIMO models which we consider subsequently.  The matrix $\bH \in \mathbb{C}^{n_r \times n_t}$, referred to as the \emph{channel matrix}, represents the wireless fading coefficients between each transmit and receive antenna. This matrix is assumed to be known to the receiver\footnote{In practice, this information can be obtained using standard channel estimation techniques.}, but not to the transmitter. The channel is modeled stochastically, with distribution depending on the specific wireless environment. Under the realistic assumption that there are sufficient scatterers surrounding the transmit and receive terminals, the channel matrix $\bH$ is accurately modeled according to a complex Gaussian distribution with independent and identically distributed (i.i.d.) elements having zero mean and unit variance. The transmitted signal $\bx$ is designed to meet a power constraint:
\bq \label{eq:PowerConstraint}
E(\bxd\bx)\leq P  \; .
\eq

\subsubsection{Ergodic and Outage Capacity}

The Shannon capacity governs the ultimate limits of communication systems.  More specifically, this measure defines
the highest data rate that can be achieved with negligible errors by any transmission scheme.  As such, the
Shannon capacity forms a benchmark for the design of practical transmission technologies.  Mathematically, the
Shannon capacity is defined in terms of the mutual information between the input and output signals, which for the
 MIMO linear model (\ref{eq:LinearModel}) is given by:
\bq \label{eq:MIDefn}
I(\bx;\by | \bH )&=&{\cal H}(\by | \bH )-{\cal H}(\by|\bx, \bH)\nn\\
&=&{\cal H}(\by | \bH)-{\cal H}(\bn) \eq with ${\cal H}(\by | \bH)$
denoting the conditional entropy of $\by$, defined in terms of its
density $p(\by | \bH)$ as follows: \bq {\cal H}(\by | \bH)=E(-\log
p):=-\int_{\mathbb{C}^{n_r} }p(\by | \bH)\log p(\by | \bH)d \by .
\eq In general, there are two important measures of capacity---the
\emph{ergodic capacity} and the \emph{outage capacity}. The ergodic
capacity is the relevant measure for applications with highly
dynamic channels; for example,
 high-mobility wireless applications, in which case the channel matrix $\bH$ varies quickly over time, and
 therefore each transmission codeword sees a large number of ``independent'' channel realizations.
 The ergodic capacity is defined as:
\bq \label{eq:ErgodicCap} C = \max_{p(\bx)} E_{\bH} \left( I(\bx;\by
| \bH) \right) \eq where the maximum is taken over all densities
$p(\bx)$ of the input vector $\bx$, subject to the constraint
(\ref{eq:PowerConstraint}).  It has been proven in \cite{Telatar}
that the optimal input density $p^*(\bx)$ is multi-variate complex
Gaussian with zero mean. As such, the mutual information $I(\bx;\by
| \bH)$ becomes \bq I(\bx;\by | \bH) = \log \det \left( \bI_{n_r}  +
\bH {\bf Q} \bHd {\bf Q}_n^{-1} \right) \eq where ${\bf Q} = E(\bx
\bxd)$ is the input signal covariance. The capacity
(\ref{eq:ErgodicCap}) can therefore be reposed as an optimization
over ${\bf Q}$ as \bq \label{eq:ErgodicCap2a} C = \max_{{\bf Q} \geq
0} E_{\bH} \left( I(\bx;\by | \bH ) \right) \eq subject to ${\rm tr}
({\bf Q}) \leq P$.  This quantity has been studied extensively over
the past decade.  Indeed, methods have been proposed to calculate
the optimal ${\bf Q}$ for a range of MIMO channel models of interest
\cite{Jorswieck,TulinoOpt}, and the expectation has been
characterized through the use of both large- and finite-dimensional
random matrix theory
\cite{Tulino,Telatar,Foschini,ChianiWin03,Smith03,ShinLee03,McKayCollings05,Chuah,Grant}.
For the case which we consider, where $\bH$ is assumed to be complex
Gaussian with independent zero-mean unit-variance entries, the
capacity-achieving input covariance has been derived in
\cite{Telatar} and is given by \bq \mathbf{Q}^* = \frac{P}{n_t}
\bI_{n_t} \; . \eq In practical terms, this implies that the
capacity is achieved by sending independent Gaussian signals from
each of the transmit antennas with equal power.

Compared with the ergodic capacity, the outage capacity is a much
more difficult problem and has received far less attention. In
contrast to the ergodic capacity, this measure is suitable for
wireless applications with low mobility (e.g., wireless local area
networks), in which case the channel is assumed fixed during the
transmission of a codeword.  The outage capacity corresponding to an
outage probability $P_{\rm out}$ is defined as the transmission rate
which can be supported by $(1-P_{\rm out})\times 100 \%$ of the
channel realizations\footnote{By \emph{supported}, we mean that the
mutual information for a given channel realization is greater than
the transmission rate.}.  Although the optimal distribution of $\bx$
which maximizes this quantity is unknown, a sensible choice is to
employ the same distribution as that which achieves the ergodic
capacity, i.e., $p^*(\bx)$. In this case, the outage capacity
$C_{\rm out}$ satisfies: \bq \label{eq:OutageProb}
P_{\rm out} (C_{\rm out}) &=& {\rm Pr}\left(  I(\bx;\by) < C_{\rm out} \right) \; \nn \\
&=&  {\rm Pr}\left(  \log \det \left( \bI_{n_r}  + \frac{P}{n_t} \bH
\bHd {\bf Q}_n^{-1} \right) < C_{\rm out} \right) \eq with ${\bf
Q}^*$ denoting the input covariance which maximizes
(\ref{eq:ErgodicCap2a}).  Compared with (\ref{eq:ErgodicCap2a}), the
quantity on the right-hand side (r.h.s.) of (\ref{eq:OutageProb}) is
much more difficult to characterize since it involves the entire
distribution of the $\log \det( \cdot )$ random variable, rather
than simply the expected value. As such, a common approach has been
to compute the first few moments of the distribution, and then use
these to obtain a Gaussian approximation (see e.g.,
\cite{Smith03,McKayCollings05}).

The outage probability (\ref{eq:OutageProb}) can be calculated via
\bq
P_{\rm out}(C_{\rm out}) = \frac{1}{2 \pi} \int_{-\infty}^\infty {\cal M}( j \omega ) \frac{ 1 - e^{-j \omega C_{\rm out} }}{j \omega}  d \omega
\eq
where ${\cal M}(\cdot)$ denotes the moment generating function of the mutual information, taking the form
\bq
{\cal M}(\lambda) &:=& E_{\bH} \left( \exp \left( \lambda I(\bx;\by | \bH) \right)   \right) \nn \\
&=& E_{\bH} \left( \det \left( \bI_{n_r}  + \frac{P}{n_t} \bH \bHd
{\bf Q}_n^{-1} \right)^\lambda   \right) \;, \label{eq:MGFEquation}
\eq and $j:=\sqrt{-1}.$

Alternatively, the Chernoff bound may be employed, as in \cite{Grant},
to give
\bq
P_{\rm out}(C_{\rm out}) &\leq& \mathcal{M}(-s) \rme^{s C_{\rm out}}
\eq
for any $s > 0$. Let
\bq
G(s):=\log {\cal M}(-s),\nonumber
\eq
and rewrite the moment generating function in an obvious alternative form:
\bq
{\cal M}(-s)=\int \rme^{-sq(X)}\mu(dX).\nonumber
\eq
Then a straightforward computation shows that
\bq
G'(s)&=&- \frac{\int \rme^{-sq(X)}\:q(X)\mu(dX)}{\int\rme^{-sq(X)}\mu(dX)}\nonumber\\
G''(s)&=&\frac{\int\rme^{-sq(X)}[q(X)]^2\mu(dX)}{\int\rme^{-sq(X)}\mu(dX)}-
\left(\frac{\int \rme^{-sq(X)}q(X)\mu(dX)}{\int \rme^{-sq(X)}\mu(dX)}\right)^2.\nonumber
\eq
If we interpret
$$
\rme^{-sq(X)}\mu(dX)
$$
as a probability measure, then for all real $s,$
$$
G''(s)=\overline{(q-\overline{q})^2}>0,
$$
where
$$
\overline{F(X)}:=\frac{\int F(X)\rme^{-sq(X)}\mu(dX)}{\int\rme^{-sq(X)}\mu(dX)}.
$$
Since $G(s)$ is convex we can minimize the Chernoff bound by minimizing
$$
G(s)+s\:C_{\rm out}
$$
and minimum of the above is
$$
-\frac{1}{2}\:\frac{(s^*\:G''(s^*)-C_{\rm out})^2}{G''(s^*)}+\frac{1}{2}s^{*2}G''(s^*),
$$
where $G(s)$ is minimized at $s^*$ and we have assumed that $s^*>0.$

\subsubsection{Upper Bound on Error Probability}

Whilst the capacity is an important fundamental quantity, it is difficult to approach in practice since it requires infinitely long codewords
and receivers with unbounded complexity.  Thus, an important question is to
determine what data rates $R$ are achievable for practical channel coding strategies with
 fixed and finite codeword length $N$, subject to a requirement on the tolerated error probability
 $P_{\rme}$. One method for addressing this problem was proposed by Gallager \cite{Gallager},
 where a general upper bound on $P_{\rme}$ was derived, assuming that \emph{randomly-selected}
 length-$N$ block codes of rate $R$ were employed, along with maximum-likelihood receivers.
 For the MIMO transmission model (1.5), assuming that the channel
 is \emph{memoryless} (i.e., each transmission period sees an independent realization of $\mathbf{H}$),
the Gallager random coding upper bound on error probability is
expressed as \cite{Telatar} \bq \label{eq:GallagerBound} P_{\rme}
(N,R)\leq \rme^{-N E_r(R)}, \eq where $E_r(R)$ is referred to as the
\emph{error exponent}, which is independent of $N$.  This function
is given by
\bq \label{eq:Er}
E_r(R)={\rm max}_{0\leq\rho\leq 1}\;\{E_0(\rho)-\rho R\},
\eq
with
\bq \label{eq:EoDefn}
E_0(\rho)=-\log \;E_{\bH}\left[\int_{\mathbb{C}^{n_{r}}}
\left[\int_{\mathbb{C}^{n_{t}}} p(\bx)\{p( \by| \bx,\bH)\}^{1/(1+\rho)}\:d \bx \right]^{(1+\rho)} d \by\right],
\eq
where
\bq
p(\by|\bx,\bH)=\frac{{\rm exp}[-(\by-\bH\bx)^{\dag} {\bf Q}_n^{-1} (\by-\bH\bx)]}{\det(\pi {\bf Q}_n )}
\eq
and $p(\bx)$ denotes the density of the input signal $\mathbf{x}$.  To maximize the error exponent and thus minimize the error probability, $p(\bx)$ should be selected so as to maximize $E_0(\rho)$.  Evaluating this optimal input distribution is very challenging, and a sensible (and more tractable) choice is to assume that $\mathbf{x}$ takes the capacity-achieving distribution $p^*(\bx)$ presented in the previous section; i.e., it is zero-mean complex Gaussian with covariance ${\bf Q}^*$.
 In fact, it was shown in \cite{Telatar} that under the assumption
 that $\bx$ is Gaussian, $p^*(\bx)$ is optimum in terms of maximizing $E_0(\rho)$.  With this input distribution,
 it is easy to see that (\ref{eq:EoDefn}) particularizes to:
\bq \label{eq:E0Eqn}
E_0(\rho)=-\log \: E_{\bH}\left[\det\left(\bI_{n_{r}}+ \frac{P}{n_t(1+\rho)} \bH\: \bHd {\bf Q}_n^{-1} \right)^{-\rho} \right] \;  .
\eq
We remark that whilst more refined bounds on the error probability compared with
(\ref{eq:GallagerBound}) have also been derived \cite{Gallager} , these more elaborate bounds
still yield the same underlying mathematical problem as that posed in (\ref{eq:E0Eqn}).


%

\subsection{Single-User MIMO and the Deformed Laguerre Weight}

Single-user MIMO systems embrace a class of coordinated wireless
networks for which all transmissions are scheduled in an orthogonal
manner (e.g., in orthogonal time-slots or orthogonal frequency
bands), and as such, transmissions do not interfere with one
another.  In this scenario, the transmitted and received signals
conform to the linear model (\ref{eq:LinearModel}), with $\bn$
simply reflecting the receiver noise.  This noise is spatially
uncorrelated (across antennas), and without loss of generality it
has covariance \bq \mathbf{Q}_n = \bI_{n_r} \; . \label{eq:QIID} \eq
Due to the normalization of the trace of $\mathbf{Q}_n$, the
transmit power $P$ also represents the SNR.

The key quantity of interest for the outage capacity is the moment
generating function, which in this case particularizes to \bq
\label{eq:MGF_Laguerre} {\cal M}(\la) = E_{\bH} \left[ \det \left(
\bI_{n_r} + \frac{1}{t} \bH \bHd \right)^\lambda \right] \eq with
\bq t := \frac{n_t}{P} \; . \eq Similarly, the error exponent
(\ref{eq:EoDefn}) admits the same form, but with substitutions $\la
= -\rho$ and $t = (1+\rho)n_t/P$. In the following, we will focus
our discussion on the moment generating function
(\ref{eq:MGF_Laguerre}), keeping in mind that the application to the
error probability is immediate.

Let
$$
m:=\textsf{max}\{n_{r},n_{t}\},\quad  n:=\textsf{min}\{n_{r},n_{t}\} , \quad \alpha := m - n
$$
and define
\bq
\bW &:=& \left\{
\begin{array}{lr}
 \bH\bHd,  &  n_{r}<n_{t}\nn\\
\bHd\bH, &    n_{r} \geq n_{t} \nn
\end{array} .
\right.
\eq
The matrix $\bW$ is a complex Wishart random matrix with positive eigenvalues denoted by $\{x_{i}\}_{i=1}^{n}.$
It is well known that the joint probability density function of the eigenvalues read
\bq \label{eq:EigPDF_Wishart}
p(x_1,x_2,...,x_n)\propto \prod_{i=1}^{n} w_{{\rm Lag}} (x_i) \prod_{1\leq j<k\leq n}(x_j-x_k)^2 ,
\eq
where $x_i\in[0,\infty)$, and $w_{{\rm Lag}} (\cdot)$ is the classical Laguerre weight
$$
w_{{\rm Lag}} (x)=x^{\al}\:\rme^{-x} . 
$$
With these definitions, using the identity
\bq
\det(\bI+ {\bf A} {\bf B})=\det(\bI+{\bf B}{\bf A}) ,
\eq
we can evaluate the moment generating function as
\bq
\mathcal{M}(\lambda) &=& E \left[  \det \left( \bI_n + \frac{1}{t} \bW \right)^\lambda   \right] \nn\\
&=& E \left[  \prod_{k=1}^n \left( 1 + \frac{x_k}{t} \right)^\lambda   \right] \nn\\
&=&
\frac{\int_{\mathbb{R}_{+}^n}\prod_{i<j}(x_i-x_j)^2\prod_{k=1}^{n}
\:\left( 1+ \: \frac{ x_k }{ t }\right)^{\la}\: w_{{\rm Lag}} (x_k)
dx_k }{\int_{\mathbb{R}_{+}^n}
\prod_{i<j}(x_i-x_j)^2\prod_{k=1}^{n}\: w_{{\rm Lag}} (x_k) dx_k }
. \label{eq:MGFLineRatio} \eq The remaining integrals are evaluated
in determinant form via the Andreief-Heine identity: \bq
\label{eq:AnHeineIdent}
\det(\mu_{i+j})_{i,j=0}^{n-1}=D_n[w]=\frac{1}{n!}\int_{(a,b)^n}\prod_{1\leq
i<j\leq n}(x_j-x_i)^2\prod_{k=1}^{n}w(x_k)dx_k, \eq where \bq
\mu_i:=\int_a^b x^iw(x)dx,\quad i=0,1,2,... \eq are moments of the
weight $w.$ See \cite{widomheine} for a recent exposition on this
and other related matters.

Obviously the moments would depend on the parameters which may appear in the weight. With this identity, we immediately obtain
\bq \label{eq:MGF_MainResultsSec}
{\cal M}(\lambda)=t^{-n\la} \frac{D_n(t,\la)}{D_n(t,0)} \;
\eq
where
\bq
D_n(t,\la)=\det\left( \mu_{i+j}(t, \la) \right)_{i,j=0}^{n-1}
\eq
is the Hankel determinant generated from the deformed Laguerre weight
\bq \label{eq:DefLagWeight}
w_{\rm dLag}(x)=w_{\rm dLag}(x,t,\la):= (x+t)^{\la} w_{{\rm Lag}} (x) ,\;\;\; t>0 \;
\eq
with moments
\bq
\mu_{k}(t, \la) := \int_{0}^{\infty}x^{k} w_{\rm dLag}(x) dx \; , \hspace*{1cm} k = 0, 1, 2, \ldots .
\eq

\noindent {\bf Remark 1} The factor $D_n(t,0)$ is simply the Hankel determinant generated from the non-deformed Laguerre weight,
$w_{{\rm Lag}} (x)$, which can be computed exactly in terms of the  Barnes $G-$ function as
\bq \label{eq:DnLaguerre_LambdaZero}
D_n(t,0) = D_{n, \al} [w_{{\rm Lag}} ]=\frac{G(n+1)G(n+\al+1)}{G(\al+1)},\quad G(1)=1.
\eq
Similarly, for $t=0$ and fixed $\la$, it follows that
$$
D_n(0,\la) = D_{n, \al+\la} [w_{{\rm Lag}} ] \; . 
$$

\noindent
{\bf Remark 2} The moments are expressed in terms of the Kummer function of the second kind
$U(a;b;z)$ as follows:
\bq
\mu_k(t, \la)&:=&\int_{0}^{\infty}x^{\al+k}(x+t)^{\la}\:\re^{-x}dx\nn\\
&=&t^{\al+\la+k+1}\:\Gamma(\al+k+1)\:U(\al+k+1;\al+\la+k+2;t),\;\;k=0,1,2,...
\label{eq:MomentDefn} \eq  Note that this expression has been
previously reported in \cite{KangRic}, and an alternative
representation given in \cite{Wang}. Whilst this identity, combined
with (\ref{eq:DnLaguerre_LambdaZero}) and
(\ref{eq:MGF_MainResultsSec}), gives a ``closed-form'' determinantal
representation for the moment generating function, it does not
provide useful insights and it also becomes unwieldy to evaluate if
the number of antennas become large. Moreover, in this form, it is
not amenable to further analysis, such as determining
 the asymptotic scaling properties. To overcome these problems, in Section 2
  we employ the theory of orthogonal polynomials and their corresponding ladder operators to provide
  a more useful characterization, where we express the Hankel determinant generated from the deformed
   Laguerre weight in terms of the classical Painlev\'{e} V differential equation.

An alternative characterization for the moment generating function which will
also prove useful is derived as follows.  Starting with (\ref{eq:MGFLineRatio}), and  
applying the transformations $x_i \to n x_i$, $i = 1, \ldots, n$, we
obtain \bq {\cal M} (\lambda) &=& T^{- n \la}
\frac{\int_{\mathbb{R}_{+}^n} \prod_{i<j}(x_i-x_j)^2\prod_{k=1}^{n}
\:\left( T + \: x_k \right)^{\la}\:x_k^{\al} \rme^{-n x_k} dx_k
}{\int_{\mathbb{R}_{+}^n}
\prod_{i<j}(x_i-x_j)^2\prod_{k=1}^{n}\:x_k^{\al} \rme^{-n x_k} dx_k
} \eq where $T := \beta_1 / P$, with $\beta_1 = n_t/n$.
Equivalently, \bq \label{eq:MGF_Lag_Fluid} {\cal M} (\lambda) &=&
T^{-n \la} \frac{ Z_n (\la) }{ Z_n (0) } \eq where \bq
\label{eq:Zn_LagDefn} Z_n (\la) = \int_{\mathbb{R}_{+}^n} \exp
\left( - \Phi (x_1, \ldots, x_n) + \la \sum_{i=1}^n \log( T + x_i )
\right) \prod_{k=1}^n d x_k \; \eq with \bq \label{eq:PhiLag} \Phi
(x_1, \ldots, x_n) &=& - 2 \sum_{1 \leq j < k \leq n} \log | x_j -
x_k | + n \sum_{\ell =1}^n ( x_\ell - \beta \log x_\ell ) \; \eq and
we have defined $\beta := m/n - 1$.

This representation will be important for deriving an approximation
for the moment generating function based on the Coulomb fluid
approach in Section \ref{sec:Coulomb}.

\subsection{Multi-User MIMO and the Deformed Jacobi Weight} \label{sec:MUMIMO}

In this section we introduce the second communication scenario which
we will consider, corresponding to a multi-user MIMO system. Such
systems are highly relevant for cellular mobile applications, where
the transmit-receive communication channel is impaired by
interference from other users operating within the same frequency
bandwidth. In fact, the key issue of interference presents one of
the most important challenges in the successful deployment of MIMO
in practice \cite{Andrews}.  We will focus on the practical
\emph{interference-limited} scenario, where the receiver noise is
negligible compared with the multi-user interference.  We assume
that there are $K$ interferers, each equipped with $n_t$ antennas,
and transmitting (capacity achieving) independent Gaussian signals
with power $P_I / n_t$ out of each transmit antenna. We make the
common assumption (see e.g., \cite{Kang,ChianiOC}), that the
interferers have equal-power, which is valid when the interferers
are located at similar distances to the receiver.  Moreover, as
discussed in \cite{Kang}, if this assumption is not met, then the
equal power assumption leads to a lower bound on performance.  We
also assume that $n_r \leq n_t$, which is appropriate for modeling
the downlink communication (base-station to mobile transmission) of
a interference-limited cellular system, where the number of transmit
antennas deployed by the base-station may be large, but the number
of receive antennas is highly restricted due to limited space
constraints of the cellular mobile device.

Under the assumptions discussed above, the noise term $\bn$ in
(\ref{eq:LinearModel}) takes the form \bq \label{eq:MUMIMOModel} \bn
= \sum_{i=1}^K \bH_i \bx_i \eq where $\bH_i$ and $\bx_i$ denote the
channel matrix and input vector for the $i$th interferer
respectively.  These are assumed independent across $i$, and
independent of $\bH$ and $\bx$.  It is convenient to write this in
the stacked form \bq \label{eq:MUMIMOModel_Stacked} \bn= \bH_I \bx_I
\eq where \bq \bH_I = \left[ \bH_1, \cdots, \bH_K \right],
\hspace*{1cm} \bx_I = \left[ \bx_1, \cdots, \bx_K \right]^T \; . \eq
Here, $\bH_I \in \mathbb{C}^{n_r \times K n_t}$ is complex Gaussian
with independent zero-mean unit-variance entries, whilst $\bx_I \in
\mathbb{C}^{K n_t}$ is complex Gaussian with independent zero-mean
entries having variance $P / n_t$.  The noise covariance matrix,
conditioned on $\bH_I$, is therefore given by \bq \mathbf{Q}_n =
\frac{P_I}{n_t} \bH_I \bHd_I \; . \eq In this case, the moment
generating function of the capacity particularizes to \bq
\label{eq:MGF_Jacobi} {\cal M}(\la) = E_{\bH} \left[ \det \left(
\bI_{n_r} + \frac{P}{P_I} \bH \bHd (\bH_I \bHd_I )^{-1}
\right)^\lambda \right] \; . \eq Similarly, the error exponent
(\ref{eq:EoDefn}) admits the same form, but with the substitution
$\la = -\rho$ and replacing $P_I$ with $(1+\rho)P_I$. Here, $\bH
\bH^\dagger$ and $\bH_I \bH_I^\dagger$ are independent complex
central Wishart.  As for the deformed Laguerre case, in the
following we will focus our discussion on the moment generating
function (\ref{eq:MGF_Jacobi}), keeping in mind that the application
to the error probability is immediate.

For consistency with previous notation, let us define: \bq
\label{eq:ParamsJacobi}  n := n_r , \quad m_1 := n_t, \quad m_2 := K
n_t , \quad \alpha_1 := m_1 - n, \quad \alpha_2 := m_2 - n \; . \eq
Generalizing \cite[pp. 312-314]{muirhead} from real to complex
matrices, we find that the joint probability density function of the
eigenvalues $f_1,...,f_{n}$ of the random matrix $\bH \bHd (\bH_I
\bHd_I )^{-1}$ is given by
%
$$
p(f_1, f_2, \ldots, f_{n}) \propto  \prod_{k=1}^{n}\frac{f_k^{\al_1}}{(1+f_k)^{m_1 + m_2}} \prod_{1\leq i < j\leq  n}(f_i-f_j)^2 , 
$$
where $f_k\in(0,\infty),\;\;k=1,2,...,n.$  With the change of variables\footnote{We would like to
thank Iain Johnstone for pointing this out.}
$$
f_k=\frac{x_k}{1-x_k},
$$
the above density becomes
\bq \label{eq:EigPDF_Jacobi}
f(x_1, x_2, \ldots, x_{n}) \propto  \prod_{k=1}^{n} w_{\rm Jac}(x_k) \:\prod_{1\leq i<j\leq n}(x_i-x_j)^2 ,  
\eq
where $x_k \in (0,1), \; k=1,2,...,n$, and $w_{\rm Jac}(\cdot)$ denotes the shifted classical Jacobi weight
\bq \label{eq:JacobiClassical_Shifted}
w_{\rm Jac}(x) = x^{ \al_1 }(1-x)^{ \al_2 },\;\;x\in[0,1].
\eq
Note that the classical Jacobi weight has the form
\bq \label{eq:JacobiClassical}
\tilde{w}_{\rm Jac}(x) = (1-x)^{ \al_1 }(1+x)^{ \al_2 } , \quad x \in [-1, 1]
\eq
and therefore
\bq \label{eq:JacobiRelation}
w_{\rm Jac}(x) = \frac{ \tilde{w}_{\rm Jac}(1-2x) }{ 2^{\al_1+\al_2}} ,  \quad x \in [0,1] \; .
\eq
The moment generating function can be evaluated as
\bq
\mathcal{M}(\lambda) &=& E \left[  \prod_{k=1}^{n} \left( 1 + \frac{P}{P_I} f_k \right)^\lambda   \right] \nn\\
&=& E \left[  \prod_{k=1}^{n} \left( 1 + \frac{P}{P_I} \frac{x_k}{1-x_k} \right)^\lambda   \right] \nn\\
&=& t^{-n \la} E \left[  \prod_{k=1}^{n} \left( \frac{t + x_k}{1 - x_k} \right)^\lambda   \right] \nn \\
&=& t^{-n \la} \frac{ \int_{(0,1)^n}\prod_{1\leq i<j\leq n}(x_j-x_i)^2\prod_{k=1}^{n} w_{\rm Jac}(x_k)
 \left(\frac{x_k+t}{1-x_k}\right)^{\la}dx_k }{\int_{(0,1)^n}\prod_{1\leq i<j\leq n}(x_j-x_i)^2\prod_{k=1}^{n}
  w_{\rm Jac}(x_k) dx_k} , \nn
\eq
where
$$ t = \frac{P_I}{P-P_I} \; .$$

We note here that the $t$ variable has {\it no n} dependence when expressed
in terms of $P_I/(P-P_I),$ rather {\it unlike} the single user MIMO case. This is mathematically reasonable
as the Jacobi weight is compactly supported.

Applying the Andreief-Heine identity (\ref{eq:AnHeineIdent}), we
obtain \bq \label{eq:MGFJac} \mathcal{M}(\lambda) = t^{-
n\la}\frac{D_n(t,\la,m_1,m_2)}{D_n(t,0,m_1,m_2)} \eq where \bq
D_n(t,\la,m_1,m_2) = \det\left( \mu_{i+j} (t, \la, m_2, m_2)
\right)_{i,j=0}^{n-1} \eq is the Hankel determinant generated from
the deformed (shifted) Jacobi weight \bq \label{eq:DefJacWeight}
w_{\rm dJac}(x)=w_{\rm dJac}(x,t,\la):= w_{\rm Jac}(x)
\left(\frac{x+t}{1-x}\right)^{\la} \eq with moments \bq \mu_{k}(t,
\la, m_1, m_2):= \int_{0}^{1}x^{k} w_{\rm dJac}(x) dx \; ,
\hspace*{1cm} k = 0, 1, 2, \ldots \eq \noindent {\bf Remark 3} The
factor $D_n(t,0,m_1,m_2)$ is simply the Hankel determinant generated
from the (shifted) non-deformed Jacobi weight, $w_{{\rm Jac}} (x)$,
which can be computed exactly in terms of the  Barnes $G-$ function.
To this end, we apply the transformation (\ref{eq:JacobiRelation})
to give \bq
D_n(t,0,m_1,m_2) &=& D_{n, \al_1, \al_2} [w_{{\rm Jac}} ] \nn \\
&=& \frac{1}{2^{(\al_1+\al_2+n)n}}
\frac{1}{n!}\int_{(-1,1)^n}\prod_{1\leq i<j\leq n}(x_j-x_i)^2\prod_{k=1}^{n} \tilde{w}_{\rm Jac}(x_k) dx_k  \nn \\
&=& \frac{(2 \pi)^n}{4^{n(\al_1+\al_2+n)}}  \frac{ \Gamma \left( \frac{\al_1 + \al_2 + 1}{2} \right) G^2 \left( \frac{\al_1 + \al_2 + 1}{2} \right) G^2 \left( \frac{\al_1 + \al_2}{2} + 1 \right)}{ G \left( \al_1 + \al_2 + 1 \right) G \left(\al_1 + 1\right) G \left(\al_2 + 1\right) } \nn \\
&\times& \frac{ G(n+1) G(n + \al_1 + 1) G(n + \al_2 + 1) G(n + \al_1 + \al_2 + 1)}{G^2 \left( n + \frac{\al_1 + \al_2 + 1 }{2} \right) G^2 \left( n + \frac{\al_1 + \al_2}{2} + 1 \right) \Gamma \left( n + \frac{ \al_1 + \al+2 + 1 }{2} \right) }
\eq
where, to obtain the last equality, we have invoked \cite[Eq. (1.6)]{BasorChen}.

Similarly, for $t=0$ and fixed $\la$, it follows that \bq
\label{eq:D0Jacobi} D_n(0,\la,m_1,m_2) = D_{n, \al_1+\la, \al_2-\la}
[w_{{\rm Jac}} ] \; . \eq

\noindent {\bf Remark 4} The moments can be expressed in terms of
the hypergeometric function ${}_2 F_1 (\cdot)$ as follows: \bq
\mu_k(t, \la, m_1,
m_2)=t^{\la}\Ga(1+k+\al_1)\Ga(1+\al_2-\la)\:_2F_1(1+k+\al_1,-\la,2+k+\al_1+\al_2-\la,-1/t).\nn
\eq This relation was also pointed out in \cite{KangConf}. Whilst
this identity, combined with (\ref{eq:D0Jacobi}) and
(\ref{eq:MGFJac}), gives a ``closed-form'' determinantal
representation for the moment generating function, it does not
provide useful insights and it also becomes unwieldy to evaluate if
the number of antennas become large. To overcome these problems, in
Section 2   we employ the theory of orthogonal polynomials and their
corresponding ladder operators to provide a more useful
characterization, where we express the Hankel determinant generated
from the deformed Jacobi weight in terms of the classical
Painlev\'{e} VI differential equation.

 Similar to (\ref{eq:MGF_Lag_Fluid}), it will also be useful
to note the following equivalent representation for the moment
generating function, \bq \label{eq:MGFJacobi} {\cal M}(\la) = T^{-n
\la} \frac{Z_n(\la)}{Z_n(0)} \eq where $T = t$, \bq
\label{eq:Zn_Jacobi} Z_n(\la) = \int_{(0,1)^n} \exp \left( - \Phi
(x_1, \ldots, x_n) + \la \sum_{\ell=1}^n \log \left( \frac{x_\ell +
T}{1-x_\ell} \right) \right)  \prod_{k=1}^n d x_k \eq and \bq \Phi
(x_1, \ldots, x_n) = -2 \sum_{1 \leq i < j \leq n} \log | x_i - x_j
|
 - n \sum_{\ell=1}^n \left( \varphi_1 \log x_\ell + \varphi_2 \log (1 - x_\ell)  \right) \;
\eq with \bq \label{eq:varPhiDefn} \varphi_1 = \frac{m_1}{n} - 1 ,
\quad \varphi_2 = \frac{m_2}{n} - 1 \; . \eq Once again, this
representation will be critical for employing the Coulomb fluid
methodology in Section \ref{sec:Coulomb}.

\setcounter{equation}{0}

\section{Exact Characterization via the Ladder Operator Method} \label{sec:Ladder}

In the previous section, we demonstrated strong relationships between two information measures of
MIMO channels, namely, the outage capacity and the error probability, and certain Hankel determinants.
 For single-user MIMO systems, the Hankel determinant of interest was generated via the moments of a deformation
 of the Laguerre weight, whereas for multi-user MIMO systems, it was generated via the moments of a deformed Jacobi weight.
  In this section, we present an exact characterization of these
Hankel determinants by employing the theory of orthogonal
polynomials and their corresponding ``raising and lowering'' ladder
operators.  Before presenting the main results, we first introduce
some
 preliminary material which will prove useful.

\subsection{Preliminaries of Orthogonal Polynomials and their Ladder Operators}

Here we provide a brief discussion to highlight the connections
between orthogonal polynomials and the Hankel determinants of
interest, as well as provide basic properties of orthogonal
polynomials which will be needed.  From (\ref{eq:EigPDF_Wishart})
and (\ref{eq:EigPDF_Jacobi}), the joint eigenvalue distributions
arising in the single-user and multi-user MIMO scenarios admit the
generic form
$$
p(y_1, \ldots, y_n) \propto \prod_{k = 1}^n \tilde{w}(y_k) \prod_{1 \leq i < j \leq n} ( y_j - y_i )^2  ,
 \quad  y_i \in (a, b)
$$
with $\tilde{w}(\cdot)$ denoting the weight.  Moreover, in both cases, the key quantity of interest takes the generic form
$$
D_n = \frac{1}{n!} \int_{(a,b)^n}\prod_{1\leq i<j\leq n}(y_j-y_i)^2\prod_{k=1}^{n} w(y_k) dy_k
= \det \left( \int_a^b y^{i+j} w(y) dy   \right)_{i,j=0}^n \;
$$
where
$$
w(y) = \tilde{w}(y) g(y)
$$
denotes the deformed weight. As a direct consequence of the Vandermonde determinant factor,
by applying elementary operations, this Hankel determinant can be equivalently expressed as
$$
D_n = \det \left( \int_a^b P_i(y) P_j(y) w(y) dy   \right)_{i,j=0}^n \;
$$
where $P_j (\cdot)$ represents any monic polynomial of degree $j$, written as
\bq \label{eq:MonicDefn}
P_j(z)=z^{j}+\textsf{p}_1(j)\:z^{j-1}+...
\eq
If we orthogonalize the polynomial sequence $\{P_{n}(y)\}$
with respect to $w(y)$ over the interval $[a,b],$ i.e.,
\bq \label{eq:OrthRel}
\int_{a}^{b}\:P_{i}(y)P_j(y) w(y)dy=h_{i}\delta_{i,j},\;\; i,j=0,1,2,...
\eq
with $h_i$ denoting the square of the $L^2$ norm of $P_i$, then
the Hankel determinant $D_n$ evaluates to
$$
D_n = \prod_{k=0}^{n-1} h_k \; .
$$
Thus, we clearly see that the problem of computing the
Hankel determinants of interest becomes one of characterizing the
class of polynomials which are orthogonal with respect to the deformed weight function.
To attack this problem, we require some definitions and tools, as given below.

We start by noting that if
\bq
\mu_i:=\int_{a}^{b}y^i w(y)dy,\nonumber
\eq
exists for all $i=0,1,2,...$, then the theory of orthogonal polynomials states that $P_n(y)$ for $n=0,1,2,...$ satisfies the
three term recurrence relations,
\bq \label{eq:RecRel}
zP_n(z)=P_{n+1}(z)+\al_n\:P_n(z)+\bt_n\:P_{n-1}(z).
\eq
The above sequence of polynomials can be generated from the orthogonality conditions, the recurrence relations and
the initial conditions,
$$
P_0(z)=1,\;\;\bt_0P_{-1}(z)=0.
$$
For example
$$
P_1(z)=z-\al_0=z-\frac{\mu_1}{\mu_0}.
$$
Substituting (\ref{eq:MonicDefn}) into the recurrence relations, an
easy computation shows that \bq \label{eq:p1Defn}
\textsf{p}_1(n)-\textsf{p}_1(n+1)=\al_n, \eq with
$\textsf{p}_1(0):=0.$ A telescopic sum of (\ref{eq:p1Defn}) gives
\bq \al_n=-\sum_{j=0}^{n-1}\textsf{p}_1(j). \eq From the recurrence
relation (\ref{eq:RecRel}) and the orthogonality relations
(\ref{eq:OrthRel}), we find \bq\label{eq:betah}
\bt_n=\frac{h_n}{h_{n-1}}. \eq

We shall see that $\textsf{p}_1(n)$ plays an important role in
later developments. For more information on orthogonal polynomials, we give reference to Szeg\"o's treatise \cite{Sze}.

Next, we present three lemmas, which represent the ladder operators
of orthogonal polynomials, as well as some supplementary conditions.
Note that these results have been known for quite sometime; we
reproduce them here for the convenience of the reader using the
notation of \cite{chen+its}, where one can also find a list of
references to the literature. We would like to mention here that
Magnus \cite{Magnus} was perhaps the first to apply these
lemmas---albeit in a slightly different form--to random matrix
theory and the derivation of Painlev\'e equations. It should be
mentioned that Tracy and Widom also made use of the compatibility
conditions in their systematic study of finite $n$ matrix models
\cite{twdet}. See also \cite{for1,for2}.  In presenting these
lemmas, we use the following definition:
$$\v = -\log w.$$
\\
\noindent
{\bf Lemma 1} \emph{Suppose $\v=-\log w$ has a derivative in some Lipshitz class with positive exponent.
The lowering and raising operators satisfy the following:
\bq
P_n'(z)&=&-B_n(z)P_n(z)+\bt_n\:A_n(z)P_{n-1}(z)\\
P_{n-1}'(z)&=&[B_n(z)+\vp(z)]P_{n-1}(z)-A_{n-1}(z)P_n(z),
\eq
where
\bq
A_n(z)&:=&\frac{1}{h_n}\int_{a}^{b}\frac{\vp(z)-\vp(y)}{z-y}\:P_n^2(y)w(y)dy \label{eq:AnNewDefn}\\
B_n(z)&:=&\frac{1}{h_{n-1}}\int_{a}^{b}\frac{\vp(z)-\vp(y)}{z-y}P_n(y)P_{n-1}(y)w(y)dy.
\label{eq:BnNewDefn} \eq} A direct computation produces two
fundamental supplementary (compatibility) conditions valid for all
$z\in \mathbb{C}\cup\{\infty\}$ and these are stated in the next
lemma.
\\
{\bf Lemma 2} \emph{The functions $A_n(z)$ and $B_n(z)$ satisfy the conditions:
$$
B_{n+1}(z)+B_n(z)=(z-\al_n)A_n(z)-\vp(z)\eqno(S_1)\\
$$
$$
1+(z-\al_n)[B_{n+1}(z)-B_n(z)]=\bt_{n+1}A_{n+1}-\bt_nA_{n-1}(z)\eqno(S_2)
.
$$}
It turns out that there is an equation which gives better insight
into the coefficients $\al_n$ and $\bt_n$, if $(S_1)$ and $(S_2)$
are suitably combined to produce a ``sum rule'' on $A_n(z).$ We
state this in the next lemma. The sum rule, we shall see later,
gives important information about the logarithmic derivative of the
Hankel determinant.
\\
{\bf Lemma 3} \emph{The functions $A_n(z),$ $B_n(z)$, and the sum
$$
\sum_{j=0}^{n-1}A_j(z),
$$
satisfy the conditions:
$$
B_n^2(z)+\vp(z)B_n(z)+\sum_{j=0}^{n-1}A_j(z)=\bt_n\:A_n(z)\:A_{n-1}(z).\eqno(S_2')
$$}
{\bf Remark 5} If the $\vp(z)$ is a rational function in $z$ then we observe that the divided
difference
$$
\frac{\vp(z)-\vp(y)}{z-y}
$$
is also a rational function in $z$ and $y,$ from which we infer that
(\ref{eq:AnNewDefn}) and (\ref{eq:BnNewDefn}) are rational functions
of $z.$ It is now clear that the compatibility conditions in Lemmas
2 and 3 would give further insights into the recurrence coefficients
and certain auxiliary quantities that appear in the theory.

\subsection{Painlev\'e V Continuous ${\boldsymbol \sigma}$--Form and the Deformed Laguerre Weight}


%

In this subsection, we focus on the deformed Laguerre scenario, for
which the main challenge is to characterize the numerator of
(\ref{eq:MGF_MainResultsSec}). This is given by the following key
result:

{\bf Theorem 1:}  The Hankel determinant of the deformed Laguerre weight $w(x)$ in (\ref{eq:DefLagWeight})
admits the following representation:
\bq \label{eq:HankelDet_ContSigma}
D_n(t,\lambda)= t^{n\la}
\exp \left( \int_{\infty}^t \frac{H_n(x)-n\la}{x} dx \right)  \label{eq:tmp}
\eq
%
where $H_n(t)$  satisfies the Painlev\'e V continuous
Jimbo-Miwa-Okamoto $\sigma$--form: \bq \label{eq:JimboPV} (t
H_n'')^2=\left[t H_n'- H_n + H_n'(2n+\al+\la) + n \la \right]^2 -4(t
H_n'- H_n + \dn )\left[( H_n')^2+\la H_n'\right] \eq with $\dn :=
n(n+\al+\la)$.

The remainder of this subsection is devoted to the proof of Theorem $1$.

\subsubsection{Compatibility Conditions, Recurrence Coefficients and Discrete Equations}

We start by noting that since
$$
\vp(x)=-\frac{1}{w(x)}\frac{d}{dx}w(x)
$$
for the problem at hand is a rational function, the divided difference
$$
\frac{\vp(z)-\vp(y)}{z-y}
$$
will also be a rational function of $z$ and $y$. Consequently, the functions $A_n(z)$ and $B_n(z)$ are rational in $z$.
Therefore $(S_1)$ and $(S_2')$ will give insight into various $n$ and $t$ dependent auxiliary quantities, which
we shall see later.

For the purpose of applying the ladder operator method, we carry out the preliminary
computations,
\bq
\v(z)&:=&-\log w(z)=-\al \log z-\la \log(z+t)+z\nn\\
\v'(z)&=&-\frac{\al}{z}-\frac{\la}{z+t}+1\nn\\
\frac{\v'(z)-\v'(y)}{z-y}&=&\frac{\al}{zy}+\frac{\la}{(z+t)(y+t)},\nn
\eq
and we see that using the definition of our $A_n(z),$ $B_n(z)$ and applying
integration by parts,
\bq \label{eq:AnDefn}
A_n(z)&=&\frac{1-R_n(t)}{z}+\frac{R_n(t)}{z+t}\\
B_n(z)&=&-\frac{n+r_n(t)}{z}+\frac{r_n(t)}{z+t}\\
R_n(t)&:=&\frac{\la}{h_n}\int_{0}^{\infty}\frac{[P_n(y)]^2}{y+t}w(y,t)dy\\
r_n(t)&:=&\frac{\la}{h_{n-1}}\int_{0}^{\infty}\frac{P_n(y)P_{n-1}(y)}{y+t}w(y,t)dy.  \label{eq:rnDefn}
\eq

For the purpose of using $(S_1)$ and $(S_2')$, in particular $(S_2')$,
we first state the following results obtained by substituting $A_n(z)$ and $B_n(z)$ given by
(\ref{eq:AnDefn})--(\ref{eq:rnDefn}):
\bq
B_n^2(z)+\v'(z)B_n(z)+\sum_{j=0}^{n-1}A_j(z) &=&z^{-2}[(n+r_n)^2+\al(n+r_n)]\nn\\
&+&z^{-1}\Big\{n-\sum_{j=0}^{n-1}R_j+r_n[\la-\al-t-2(n+r_n)]/t+(n-\la)/t\Big\}\nn\\
&+&(z+t)^{-1}\Big\{\sum_{j=0}^{n-1}R_j+r_n[t+\al-\la+2(n+r_n)]/t-n\la/t\Big\}\nn\\
&+&(z+t)^{-2}[r_n^2-\la r_n] . \nn
\eq
Now from $(S_1)$ we find,
\bq \label{eq:S1Diff1}
-(2n+1+r_{n+1}+r_n)&=&\al-\al_n(1-R_n)\\
r_{n+1}+r_n&=&\la-R_n(t+\al_n) \label{eq:S1Diff2}
\eq
and from $(S_2')$ we find,
\bq \label{eq:S2Diff1}
(n+r_n)^2+\al(n+r_n)=\bt_n(1-R_n)(1-R_{n-1})\quad\\
n-\sum_{j=0}^{n-1}R_j+\frac{r_n}{t}[\la-\al-t-2(n+r_n)]+\frac{n(\la-t)}{t}
=\frac{\bt_n}{t}\left[(1-R_{n-1})R_n+(1-R_{n-1})R_n\right]\quad \label{eq:S2Diff2a} \\
\sum_{j=0}^{n-1}R_{j}+\frac{r_n}{t}[t+\al-\la+2(n+r_n)]-\frac{n\la}{t}=
-\frac{\bt_n}{t}\left[(1-R_n)R_{n-1}+(1-R_{n-1})R_n\right]\quad \label{eq:S2Diff2b} \\
r_n^2-\la r_n=\bt_nR_nR_{n-1}.\quad \label{eq:S2Diff3}
\eq
{\bf Remark 6} Observe that (\ref{eq:S2Diff2a}) and (\ref{eq:S2Diff2b})
 are equivalent.
We shall see later that (\ref{eq:S2Diff2a}), when combined with certain relations,
performs the sum $$\sum_{j=0}^{n-1}R_j$$
automatically in closed form.

This sum will provide an important link between the logarithmic derivative of
the Hankel determinant with respect to $t$, $\bt_n$, and $r_n$, which is an essential step in
establishing the Painlev\'e equation.

Note that although the difference relations (\ref{eq:S1Diff1})--(\ref{eq:S2Diff1}) and (\ref{eq:S2Diff3}) look rather complicated, these can be manipulated to give us insight into the recurrence coefficients $\al_n$ and $\bt_n.$

Now the sum of (\ref{eq:S1Diff1}) and (\ref{eq:S1Diff2}) gives us a simple expression for
the recurrence coefficient $\al_n$ in terms of $R_n$:
\bq \label{eq:alrn}
\al_n=2n+1+\al+\la-tR_n.
\eq
From (\ref{eq:S2Diff1}) and (\ref{eq:S2Diff3}) have
\bq
n(n+\al)+r_n(\al+\la+2n)&=&\bt_n(1-R_n-R_{n-1})\nn
\eq
or equivalently
\bq \label{eq:RnBnCond}
\bt_n(R_{n}+R_{n-1})=\bt_n-n(n+\al)-r_n(\al+\la+2n).
\eq
Now substituting (\ref{eq:S2Diff3}) and (\ref{eq:RnBnCond}) into either (\ref{eq:S2Diff2a}) or
(\ref{eq:S2Diff2b}) to eliminate $R_n$ and $R_{n-1}$
leaves us the following very simple form for $\sum_{j=0}^{n-1}R_j,$ which will play a
crucial role later,
\bq \label{eq:RSum}
t\sum_{j=0}^{n-1}R_j=n(n+\al+\la)-\bt_n-tr_n.
\eq
But in view of (\ref{eq:alrn}), we have
\bq
t\sum_{j=0}^{n-1}R_j&=&n(n+\al+\la)-\sum_{j=0}^{n-1}\al_j\nn\\
&=&n(n+\al+\la)+\textsf{p}_1(n) \label{eq:RSumb}.
\eq
Comparing (\ref{eq:RSum}) with (\ref{eq:RSumb}) gives
\bq
\textsf{p}_1(n)=-\bt_n-tr_n.\nn
\eq
Note that $\textsf{p}_1(n)$ also depends on $t$, although this is not always displayed.

We are now in a position to find an expression for $\bt_n$ in terms of $r_n$ and $R_n$. This is found by eliminating $R_{n-1}$ from (\ref{eq:RnBnCond}) and (\ref{eq:S2Diff3}) resulting in
\bq
\bt_n=\frac{1}{1-R_n}\left[r_n(2n+\al+\la)+\frac{r_n^2-\la r_n}{R_n}+n(n+\al)\right].\nn
\eq
We summarize the above in the following lemma:

{\bf Lemma 4} \emph{The recurrence coefficients $\al_n$ and $\bt_n$
are expressed in terms of the auxiliary quantities $r_n$ and $R_n$
as:} \bq \al_n&=&2n+1+\al+\la-tR_n\\ \label{eq:AlphanRelation}
\bt_n&=&\frac{1}{1-R_n}\left[r_n(2n+\al+\la)+\frac{r_n^2-\la
r_n}{R_n}+n(n+\al)\right]. \label{eq:BetanRelation} \eq
\emph{Furthermore,} \bq
\sum_{j=0}^{n-1}R_j&=&n(n+\al+\la)-\bt_n-tr_n, \label{eq:SumRjRelation} \\
\textsf{p}_1(n)&=&-\bt_n-tr_n. \label{eq:p1Relation} \eq

\subsubsection{$t$ Evolution and Painlev\'e V: Continuous ${\boldsymbol \sigma}$--Form}

In this next stage of the proof, we keep $n$ fixed and vary $t.$ The
differential relations generated here when combined with the
difference relations obtained previously will give us the desired
Painlev\'e equation.

A straightforward computation shows that
\bq
\frac{d}{dt}\log h_n=R_n.
\eq
But, from (\ref{eq:betah}), it follows that
\bq
\frac{d\bt_n}{dt}&=&\bt_n(R_n-R_{n-1})\\
&=&\bt_n R_n-\frac{r_n^2-\la r_n}{R_n}, \label{eq:BetaDiff}
\eq
where the last equality follows from (\ref{eq:S2Diff3}).

Differentiating
$$
0=\int_{0}^{\infty}x^{\al}(x+t)^{\la}\rme^{-x}P_{n}(x)P_{n-1}(x)dx
$$
with respect to $t$ produces
\bq
0&=&\la\int_{0}^{\infty}(x+t)^{\la-1}\rme^{-x}P_n(x)P_{n-1}(x)dx+
\int_{0}^{\infty}x^{\al}(x+t)^{\la}\rme^{-x}\left[\frac{d}{dt}\textsf{p}_1(n)\:x^{n-1}+...\right]P_{n-1}(x)dx\nn\\
&=&\la\int_{0}^{\infty}\frac{P_{n-1}(x)P_{n}(x)}{x+t}w(x)dx+h_{n-1}\:\frac{d}{dt}\textsf{p}_1(n),\nn
\eq finally resulting in \bq \label{eq:p1Diff}
\frac{d}{dt}\textsf{p}_1(n)&=&-r_n . \eq Upon noting
(\ref{eq:p1Defn}), this implies \bq \label{eq:AlphaDiff}
\frac{d\al_n}{dt}&=&r_{n+1}-r_{n} . \eq Now differentiating
(\ref{eq:p1Relation}) with respect to $t$ and noting
(\ref{eq:p1Diff}), we find \bq
\frac{d\textsf{p}_1(n)}{dt}&=&-\frac{d\bt_n}{dt}-\frac{d}{dt}(tr_n)\nn\\
&=&-\frac{d\bt_n}{dt}-r_n-t\frac{dr_n}{dt}\nn\\
&=&-r_n.\nn
\eq
The above result combined with (\ref{eq:BetaDiff}) gives
\bq
\frac{d\bt_n}{dt}=-t\frac{dr_n}{dt}=\bt_nR_n-\frac{r_n^2-\la r_n}{R_n}.
\eq
Because (\ref{eq:BetanRelation}) expresses $\bt_n$ as a quadratic in $r_n,$
we see that $r_n$ satisfies a Riccatti equation;
\bq \label{eq:RicEq}
t\frac{dr_n}{dt}=\frac{r_n^2-\la r_n}{R_n}-\frac{R_n}{1-R_n}\left[r_n(2n+\al+\la)+\frac{r_n^2-\la r_n}{R_n}+n(n+\al)\right].
\eq
In fact there is another Riccati equation satisfied by $R_n$, which can be found as follows.
Eliminating $r_{n+1}$ from (\ref{eq:S1Diff2}) and (\ref{eq:AlphaDiff}), and upon referring to
(\ref{eq:AlphanRelation}), we find that
\bq \label{eq:rnElim}
2r_n=t\frac{dR_n}{dt}+\la-R_n(t+2n+\al+\la-tR_n).
\eq
\\
Let us now eliminate $r_n(t)$ from (\ref{eq:RicEq}) and
(\ref{eq:rnElim}), which results in a second order ordinary
differential equation (o.d.e.) satisfied by $R_n(t)$, where $n$,
$\al$, and $\la$ appear as parameters. A further linear fractional
change of variable
$$
R_n(t)=1-\frac{1}{1-y(t)}\quad{\rm or}\quad y=1-\frac{1}{1-R_n(t)},
$$
shows that $y(t)$ satisfies a Painlev\'e V:
\bq \label{eq:PVyt}
y''=\frac{3y-1}{2y(y-1)}\:(y')^2- \frac{y'}{t}+\frac{(y-1)^2}{t^2}\:\left(\frac{\al^2}{2}y-\frac{\la^2}{2y}\right)
+\frac{(2n+1+\al+\la)\:y}{t}-\frac{y(y+1)}{2(y-1)}.
\eq
We note this is
$$
P_V\left(\frac{\al^2}{2},\;-\frac{\la^2}{2},\;2n+1+\al+\la,\;-\frac{1}{2}\right).
$$
For the continuous $\sigma$--form of this $P_V$, note that \bq
H_n&:=&t\frac{d}{dt}\log D_n\nn\\
&=&t\frac{d}{dt}\sum_{j=0}^{n-1}\log h_j\nn\\
&=&t\sum_{j=0}^{n-1}R_j\nn\\
&=&n(n+\al+\la)-\bt_n-tr_n \label{eq:Hna} \\
&=&n(n+\al+\la)+\textsf{p}_1(n) \label{eq:Hnb},
\eq
where the last two equations follow from (\ref{eq:SumRjRelation}) and (\ref{eq:p1Relation}) of Lemma 4.

From (\ref{eq:p1Diff}), (\ref{eq:Hna}), and (\ref{eq:Hnb}), we obtain expressions for $\bt_n$ and $r_n$ in terms of $H_n$ and
$H_n'$,
\bq
\bt_n&=&n(n+\al+\la)+tH_n'-H_n \label{eq:BetanHn} \\
r_n&=&-H_n',
\eq
where $'$ denotes $\frac{d}{dt}.$

All we need to do now is to eliminate $R_n$ to obtain a functional equation satisfied by $H_n,$ $H_n'$
and $H_n'',$
$$
f(H_n,H_n',H_n'',n,t)=0.
$$
For this purpose, we examine two quadratic equations satisfied by $R_n$, one of which is simply a
rearrangement of (\ref{eq:BetanRelation}) and reads
\bq \label{eq:Frac1}
\frac{r_n^2-\la r_n}{R_n}+\bt_nR_n=\bt_n-r_n(2n+\al+\la)-n(n+\al).
\eq
The other follows from a derivative of (\ref{eq:BetanHn}) with respect to $t$ and (\ref{eq:BetaDiff}),
\bq \label{eq:Frac2}
\bt_nR_n-\frac{r_n^2-\la r_n}{R_n}&=&tH_n''.
\eq
Solving for $R_n$ and $1/R_n$ from the linear system (\ref{eq:Frac1}) and (\ref{eq:Frac2}),
we find
\bq
2\bt_nR_n&=&\bt_n-r_n(2n+\al+\la)-n(n+\al)+tH_n''\nn\\
2\left(\frac{r_n^2-\la r_n}{R_n}\right)&=&\bt_n-r_n(2n+\al+\la)-n(n+\al)-tH_n'',\nn
\eq
which we rewrite as follows
\bq
2R_n&=&1+\frac{tH_n''+(t+2n+\al+\la)H_n'-H_n+n\la}{tH_n'-H_n+n(n+\al+\la)} \label{eq:Rna} \\
\frac{2}{R_n}&=&\frac{-tH_n''+(t+2n+\al+\la)H_n'-H_n+n\la}{(H_n')^2+\la\:H_n'}
\label{eq:Rnb}. \eq The product (\ref{eq:Rna}) and (\ref{eq:Rnb})
gives us the desired continuous $\sigma$--form (\ref{eq:JimboPV}).
%

It is finally worth noting that with
\bq
D_n(t,\la)=:t^{\dn}\tD_n,
\eq
then after a little computation we find that $\tD_n$ satisfies the Toda moelcule equation \cite{Toda}
\bq
\frac{d^2}{dt^2}\log\tD_n=\frac{\tD_{n+1}\tD_{n-1}}{\tD_n^2} .
\eq

\subsection{Discrete ${\boldsymbol \sigma}$--Form and the Deformed Laguerre Weight}

As an alternative to the continuous PV $\sigma$--form, the following
theorem establishes a \emph{discrete} $\sigma$--form satisfied by
the logarithmic derivative of the Hankel determinant taken with
respect $t$; which is a non-linear difference equation in $n$.

{\bf Theorem 2:}  The Hankel determinant of the deformed Laguerre
weight $w(x)$ in (\ref{eq:DefLagWeight}) admits the  representation
(\ref{eq:HankelDet_ContSigma}), with $H_n$ satisfying the discrete
$\sigma$--form \bq\label{eq:discretesigma}
&&\left[\frac{n(n+\al)t+(\delta^2H_n+t)[H_n-\delta_n]}{\delta^2H_n+2n+\al+\la+t}\right]^2
-\la\:\frac{n(n+\al)t+(\delta^2H_n+t)[H_n-\delta_n]}{\delta^2H_n+2n+\al+\la+t}\nn\\
&& \hspace*{1cm}= \left[\delta_n-H_n
+\frac{n(n+\al)t+(\delta^2H_n-t)[H_n-\delta_n]}{\delta^2H_n+2n+\al+\la+t}\right] (H_{n+1}-H_n)(H_n-H_{n-1})
\quad \eq
where
$$\delta^2H_n:=H_{n+1}-H_{n-1}$$ denotes a second order difference in $H_n$.

The initial conditions are
\bq
H_1(t) = &\frac{d}{dt}\log D_1(t,\la) \, , \; \; \; H_2(t) = \frac{d}{dt}\log D_2(t,\la) \nn
\eq
with
\bq
D_1(t,\la) = \mu_0(t) \, , \; \; \; D_2(t,\la) = \mu_0(t)\mu_2(t)-\mu_1^2(t),\nn
\eq
and the moments are defined in $(\ref{eq:MomentDefn}).$


The proof follows similar methods as in the continuous case in the previous subsection; namely,
we express the auxiliary quantities $R_n,$ $r_n$ and the recurrence coefficient $\bt_n$ in terms of
$H_n$ and $H_{n\pm 1}$, and substitute these into (\ref{eq:S2Diff3}).
To begin, note that since
$$
H_{n}=t\sum_{j=0}^{n-1}R_j,
$$
we find that
\bq
H_{n+1}-H_n&=&tR_n\nonumber\\
H_{n-1}-H_{n+1}&=&t(R_n+R_{n-1}).
\eq
Multiplying by $\bt_n$ and using (\ref{eq:RnBnCond}), we obtain the following
{\it linear} equation in $\bt_n$ and $r_n$,
\bq \label{eq:LinearEq}
(t+\delta^2H_n)\bt_n-(2n+\al+\la)t\:r_n=n(n+\al)\:t .
\eq
There is a further {\it linear} equation in $\bt_n$ and $r_n$, obtained by
rearranging (\ref{eq:RSum}), which is
\bq \label{eq:LinearRearrange}
\bt_n+t\:r_n=n(n+\al+\la)-H_n.
\eq
Solving for $\bt_n$ and $r_n$
from (\ref{eq:LinearEq}) and (\ref{eq:LinearRearrange}) leaves, \bq
\bt_n&=&n(n+\al+\la)-H_n
+\frac{n(n+\al)t+(\delta^2H_n-t)[H_n-n(n+\al+\la)]}{\delta^2H_n+2n+\al+\la+t} \label{eq:BnHnRelation} \\
t\:r_n&=&\frac{n(n+\al)t+(\delta^2H_n+t)[H_n-n(n+\al+\la)]}{\delta^2H_n+2n+\al+\la+t}.
\label{eq:rnHnRelation} \eq The discrete $\sigma$--form is found by
substituting (\ref{eq:BnHnRelation}), (\ref{eq:rnHnRelation}) and
$$
t\:R_n=H_{n+1}-H_n,
$$
into (\ref{eq:S2Diff3}), i.e.,
$$
r_n^2-\la\:r_n=\bt_n\:R_{n}\:R_{n-1} \; .
$$

\subsection{Painlev\'e IV Continuous ${\boldsymbol \sigma}$--Form and the Deformed Jacobi Weight}

We now consider the Hankel determinant generated by the deformed
Jacobi weight, \bq
x^{\al_1}(1-x)^{\al_2-\la}(x+t)^{\la},\;\;x\in(0,1), \eq discussed
in Section \ref{sec:MUMIMO}. In contrast to the deformed Laguerre
weight, existing characterizations for this case are available.
Specifically, such deformation was investigated by Magnus
\cite{Magnus}, where an auxiliary variable similar to our $R_n$ in
the last section was found to satisfy a particular Painlev\'e VI.
The continuous $\sigma$--form associated with this $P_{VI}$ was
derived recently in \cite{zhang}. Other related work dealing with
this weight can be found in \cite{Nuttal}.

To state the results of \cite{Magnus} and \cite{zhang} in our context, we must first introduce some additional notation.
Let $\{P_m(x)\}$ satisfy the orthogonality relations:
\bq
\int_{0}^{1}P_m(x)P_n(x)\:x^{\al_1}(1-x)^{\al_2-\la}(x+t)^{\la}dy=h_m(t)\delta_{m,n}
\eq
and
\bq
R_m(t)=\frac{\al_2-\la}{h_m(t)}\int_{0}^{1}[P_m(y)]^2\:y^{\al_1}\:(1-y)^{\al_2-\la-1}(y+t)^{\la}dy \; .
\eq
Then
$$
y(t)=1-\frac{(1+t)R_m(t)}{2m+\al_1+\al_2+1}
$$
satisfies the following $P_{VI}$: \bq \label{eq:JacobiyP}
y''&=&\frac{1}{2}\left(\frac{1}{y}+\frac{1}{y-1}+\frac{1}{y+t}\right)(y')^2\nonumber\\
    &-&\left(\frac{1}{t}+\frac{1}{1+t}-\frac{1}{y+t}\right)y'\nonumber\\
    &+&\frac{y(y-1)(y+t)}{t^2(1+t)^2}\left(\nu_1-\frac{\nu_2 t}{y^2}-\frac{\nu_3(1+t)}{(y-1)^2}
    +\frac{\nu_4\:t(1+t)}{(y+t)^2}\right)
\eq
where
$$
\nu_1=\frac{1}{2}(2m+\al_1+\al_2+1)^2,\quad \nu_2=-\frac{\al_1^2}{2},\quad \nu_3=\frac{(\al_2-\la)^2}{2},\quad\nu_4=\frac{1-\la^2}{2}.
$$
Furthermore, let \bq H_m(t)=-t(1+t)\frac{d \log
D_m(t)}{dt}+c_1\:t+c_2, \eq where \bq
c_1&:=&-m(m+\al_1+\al_2)-\frac{(\al_1+\al_2-\la)^2}{4}\\
c_2&:=&\frac{1}{4}[2m(m+\al_1+\al_2)+(\al_2-\la)(\al_1+\al_2-\la)-\la(\al_1-\al_2+\la)]
. \eq Then $H_m$ satisfies the following $\sigma$--form of $P_{VI}$:
\bq \label{eq:HJacobi}
-H_m'[t(1+t)H_m'']^2&-&[2H_m'(tH_m'-H_m)+(H_m')^2+r_1\:r_2\:r_3\:r_4]^2\nonumber\\
&=&(H_m'-r_1^2)(H_m'-r_2^2)(H_m'-r_3^2)(H_m'-r_4^2), \eq where
\bq
r_1=\frac{\al_1+\al_2-\la}{2},\quad
r_2=\frac{\al_2-\la+\al_1}{2},\quad
r_3=\frac{2m+\al_1+\al_2-\la}{2}, \quad
r_4=\frac{2m+\al_1+\al_2+\la}{2}, \nonumber
\eq
and $\al_1$ and
$\al_2$ are defined by (\ref{eq:ParamsJacobi}).

As for the single-user case, the Hankel determinant for the
multi-user situation can also be expressed as an integral of $H_m$:
\bq D_m(t)=D_{n,\al_1,\al_2-\la}[w_{{\rm Jac}}]\:t^{n\la}\:
\exp\left(\int_{\infty}^{s}\frac{c_1\:s+c_2-H_m(s)}{s(1+s)}ds\right).\nn
\eq

Note here that we have tacitly assumed that $t=P_I/(P-P_I)$ is
strictly positive. However, we expect the equations
(\ref{eq:JacobiyP}) and (\ref{eq:HJacobi}) to be formally valid for
all $t\in\mathbb{R}.$

\section{Characterization via the Coulomb Fluid Method} \label{sec:Coulomb}

In this section, we present an alternative characterization based on the Coulomb fluid method.
As we will see, the key benefit of this approach is that it leads to simpler expressions than
the exact results obtained via the ladder operator approach. Moreover, whilst
this method is based on large-$n$ considerations, we will show numerically that the approximations
are very accurate for very small dimensions also.  In fact, in the following section, we will demonstrate that the Coulomb fluid approach actually captures the \emph{exact} distribution of the mutual information to leading order in $n$.

Whilst the Coulomb fluid has been applied extensively in the context of statistical mechanics, it is
 relatively unfamiliar amongst the wireless communications and information theory communities.  As such,
  in the following we will first present some basic background material, based mainly on \cite{ChenLawrence},
   before deriving new results for both the single-user and multi-user MIMO systems of interest.

\subsection{Preliminaries of the Coulomb Fluid Method}

Consider a function of the form
\bq \label{eq:Ratio}
\frac{Z_n (\la)}{Z_n(0)} = \exp \left[ - ( F_n (\la) - F_n (0) ) \right]
\eq
where
\bq
Z_n (\la) := \int_{(L,U)^n} \exp \left[ - \Phi (x_1, \ldots, x_n ) - \la \sum_{i=1}^n f(x_i) \right]
\eq
with
\bq
\Phi(x_1, \ldots, x_n) := -2 \sum_{1 \leq j < k \leq n} \log | x_j - x_k | + n \sum_{j=1}^n \v (x_j) \; .
\eq
This expression embraces the moment generating function representations for both the single-user
 MIMO capacity (\ref{eq:Zn_LagDefn})--(\ref{eq:PhiLag})
 and multi-user MIMO capacity (\ref{eq:Zn_Jacobi})--(\ref{eq:varPhiDefn}),
 with appropriate selection of the functions $f(\cdot)$ and $\v(\cdot)$, and integration
 limits $L$ and $U$.

The key motivation for this representation is that it admits a
simple intuitive
 interpretation in terms of statistical physics, as observed in the seminal papers by Dyson \cite{Dyson}.
 In particular, interpreting the eigenvalues $x_1, \ldots, x_n$ as the positions of $n$ identically
 charged particles, the function $\Phi(x_1, \ldots, x_n)$ is recognized as the total energy of the
  repelling charged particles, confined by a common external potential $n\v (x)$.  The function $f(x)$
   acts as a perturbation to the system, effectively modifying the external potential. For sufficiently large $n$,
   we can approximate the particles as a continuous fluid with a certain (limiting) density, $\sigma(x)$,
 and assume that it is supported on a single interval $(a,b)$.  This density will correspond to the equilibrium
   density of the fluid, obtained via the constrained minimization
\bq
\min_{\sigma} F[\sigma]  \quad {\rm subject \; to} \; \; \int_{a}^b  \sigma (x)dx = 1 \;
\eq
with
\bq
F(\la) := \int_a^b \sigma(x) \left( n^2 \v(x) + \la n f(x) \right)dx - n^2 \int_a^b  \int_a^b  \sigma(x) \log | x - y|
\sigma(y)dxdy \; .
\eq
As a consequence of the Frostman Lemma \cite[pg. 65]{Tsuji}, the equilibrium density satisfies
the integral equation
\bq
\v(x)+ \frac{\la}{n}f(x)-2 \int_{a}^{b}\ln|x-y|\sigma(y)dy=A,\quad x\in [a,b]\nn
\eq
where $A$ is the Lagrange multiplier which fixes the constraint that the equilibrium density has total charge 1.
See \cite{Tsuji} for a detailed discussion. The above integral equation with logarithmic kernel is converted into
a singular integral equation by taking a derivative with respect to $x$ for $x\in (a,b),$
\bq
2 {\cal P} \int_{a}^{b}\frac{\sigma(y)}{x-y}dy=\v'(x)+\frac{\la}{n}f'(x),\nn
\eq
where $\mathcal{P}$ denotes Cauchy principal value.

If $\v(x)$ is convex in a set of positive measure, the solution to this problem can be found \cite{ChenLawrence},
with the optimal $\sigma(\cdot)$ taking the form
\bq
\sigma(x) = \sigma(x, \la) = \sigma_0(x) + \frac{\varrho(x,\la)}{n} ,
\eq
where
\bq \label{eq:sigma0_limiting}
\sigma_0(x) = \frac{ \sqrt{ (b-x)(x-a) }}{2 \pi^2} \int_a^b\frac{\v'(x)-\v'(y)}{(x-y) \sqrt{(b-y)(y-a)}}dy
\eq
denotes the limiting density of the original system (i.e., in the absence of any perturbation), and
\bq
\label{eq:VarRhoDefn}
\varrho(x, \la) = \la \tilde{\varrho}(x) = \frac{ \la }{ 2 \pi^2 \sqrt{ (b-x)(x-a) } }
\mathcal{P} \int_a^b \frac{ \sqrt{ (b-y)(y-a) } }{y-x} f'(y)dy
 \eq
represents the deformation of this density caused by the external perturbation. The solution theory of singular integral
equations can be found in the monographs \cite{gak}, \cite{mig}, and \cite{widominteq}. See also \cite{elas} for numerous
examples on the application singular integral equations to problems in elasticity, and \cite{widom} for the $L_p$ version of the theory.

The boundary parameters $a$ and $b$ are chosen to satisfy the supplementary conditions
\bq
\int_a^b\frac{\v'(x)}{\sqrt{(b-x)(x-a)}}dx = 0
\eq
and
\bq
\frac{1}{2\pi} \int_a^b \frac{ x \v'(x) }{ \sqrt{(b-x)(x-a)} }dx = 1 \; .
\eq
With these results, for sufficiently large $n$, the ratio
(\ref{eq:Ratio}) is then approximated by
\bq
\frac{Z (\la)}{Z(0)} = \exp
\left[ - \la^2 \frac{\mathcal{S}_1^{\co}(T)}{2} - \la \mathcal{S}_2^{\co}(T)  \right]
\eq
where
\bq \label{eq:SLaguerre}
\mathcal{S}_1^{\co}(T) = \int_a^b  f(x) \tilde{\varrho}(x)dx
, \quad \quad \mathcal{S}_2^{\co}(T) = n \int_a^b  f(x) \sigma_0 (x) dx\; .
\eq

{\bf Remark 7:} With the above results, the moment generating
functions (\ref{eq:MGF_Lag_Fluid}) and (\ref{eq:MGFJacobi}) become
\bq \label{eq:MGFCoulomb} \mathcal{M}(\la) \approx \exp \left[ -\la^2
\frac{\mathcal{S}_1^{\co}(T)}{2} - \la ( \mathcal{S}_2^{\co}(T)   +
n \log T ) \right] \; \eq which corresponds to a Gaussian
distribution with mean and variance given by \bq
\label{eq:MeanVar_Defn} \mu_{\rm Coulomb} = - \mathcal{S}_2^{\co}(T)
- n \log T , \quad \quad \sigma_{\rm Coulomb}^2 = -
\mathcal{S}_1^{\co}(T) \; . \eq Therefore, the outage probability
can be obtained via \bq P_{\rm out}(C_{\rm out}) \approx \frac{1}{2}
\left[ 1 + {\rm erf} \left( \frac{C_{\rm out} - \mu}{\sqrt{2
\sigma^2} } \right) \right] \; . \eq

{\bf Remark 8:} Based on the Coulomb fluid method, the error exponent (\ref{eq:Er}) is approximated as follows:
%
%

\bq \label{eq:E0Eqn_Coulomb}
E_r(R)={\rm max}_{0\leq\rho\leq 1}\; \left\{ \rho^2 \frac{  {\cal S}_1^{\co}(T)}{2} - \rho ( {\cal S}_2^{\co}(T) + n \log(T) ) -\rho R \right\} \; .
\eq

%

The key challenge is to
evaluate the quantities $\mathcal{S}_1^{\co}(T)$ and
$\mathcal{S}_2^{\co}(T)$ for the single-user and multi-user MIMO
scenarios. These problems are addressed in the following
subsections. As we will see, in both cases we will need to solve
numerous integrals which are quite complicated and are not readily
available. Thus, to aid the reader, we have succinctly compiled the
solutions to these integrals in the Appendix, along with some
detailed derivations.

\subsection{Coulomb Fluid and the Deformed Laguerre Weight}
\label{sec:CF_Laguerre}

In this case, we have the particularizations
\bq
f(x) = - \log(T + x) , \quad \v(x) = x - \beta \log x , \quad L = 0 , \quad U = \infty \; .
\eq
First consider the constants, $a$ and $b$.  These are determined by the equations,
\bq
\int_{a}^{b}\frac{x - \beta}{\sqrt{(b-x)(x-a)}}\frac{dx}{2\pi} = 1 ,
\quad \int_{a}^{b}\frac{ 1 - \beta/x}{\sqrt{(b-x)(x-a)}}dx = 0. \nn
\eq
With the integral identities (\ref{eq:Simple2})--(\ref{eq:Simple4}), we obtain
\bq
\frac{a+b}{4} = 1+\frac{\beta}{2} , \quad \sqrt{ab} =  \beta, \label{eq:abCondition}
\eq
which leads to
\bq
a = 2+\beta-2\sqrt{1+\beta} , \quad b = 2+\beta+2\sqrt{1+\beta} \, .
\eq
Now consider the limiting density, $\sigma_0(x)$.  In this case,
with (\ref{eq:sigma0_limiting}), (\ref{eq:abCondition}), and the integral identity (\ref{eq:Simple2}),
it can be easily verified that
\bq
\sigma_0(x) = \frac{1}{2 \pi} \frac{ \sqrt{(b-x)(x-a)}}{x} , \quad a < x < b \;
\eq
which is the celebrated Mar\^{c}enko-Pastur law \cite{Marcenko,Dyson1971}.
Substituting this distribution along with $f(x) = - \log(T + x)$ into  (\ref{eq:SLaguerre}), and
integrating using the identities (\ref{eq:Int1}), (\ref{eq:Int3}), and (\ref{eq:Int4}) gives
\bq
&&{\cal S}_2^{\rm Comm.}(T) = - \frac{n}{2} \biggl[ (a+b)\log \left( \frac{ \sqrt{T+a}+\sqrt{T+b}}{2}\right)
- \frac{ ( \sqrt{T+a} - \sqrt{T+b} )^2}{2}  \nn \\
&& \hspace*{2.5cm} - \sqrt{a b} \log \left( \frac{(\:\sqrt{a b} + \sqrt{(T+a)(T+b)}\:)^2 - T^2}
{ (\:\sqrt{a} + \sqrt{b}\:)^2 }
\right) \biggr] \; .
\label{eq:S2Laguerre}
\eq

We note that an equivalent expression can also be obtained by changing variables $x \to (\beta+1)x$
and invoking an integral result from \cite{Rapajic}.  The derivation of our result here,
based on applying the Schwinger parametrization of the $\log$ function (\ref{eq:LogParam}),
has the advantage of being much more direct, and moreover it can be used to derive other
integral expressions encountered with the Coulomb fluid approach, as shown in the Appendix. Such parametrization is ubiquitous in the
analytical computation of integrals arising in quantum field theory; see for example \cite{Schwinger}.

For $\tilde{\varrho}(x)$, substituting $f'(x) = -1/(T+x)$ into (\ref{eq:VarRhoDefn})
and using the integral identity
(\ref{eq:PInt1}), we calculate
\bq
\tilde{\varrho}(x) = \frac{1}{2 \pi \sqrt{ (b-x)(x-a)}} \left( 1 - \frac{ \sqrt{ (T+a)(T+b) } }{x+T} \right) \; .
\eq
Substituting this into  (\ref{eq:SLaguerre}), and applying the integral identities (\ref{eq:Int1}) and (\ref{eq:Int2})
we find,
\bq
\label{eq:S1Laguerre}
\mathcal{S}_1^{\rm Comm.}(T) = -2
\log \left[ \frac{1}{2} \left(\frac{ T+a }{T+b} \right)^{1/4} + \frac{1}{2} \left( \frac{ T+b }{T+a}\right)^{1/4} \right]\, .
\eq

\begin{figure}[ht]
\centering

\subfigure[$n_r = n_t = n$]{
\includegraphics[width=0.48\columnwidth]{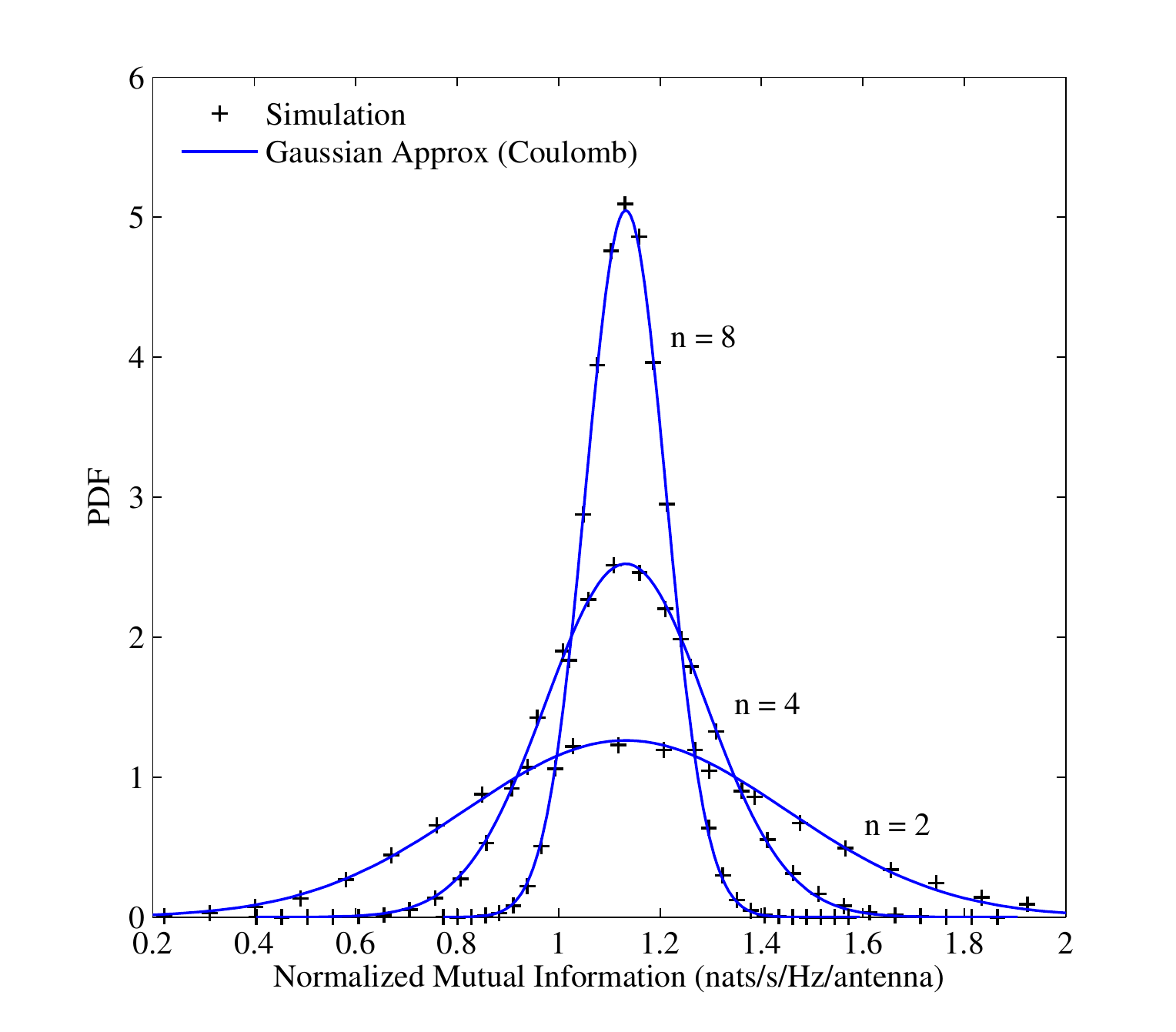}
\label{fig1:subfig1} } \subfigure[$n_r = 2 n_t \; \; (n_t = n)$]{
\includegraphics[width=0.48\columnwidth]{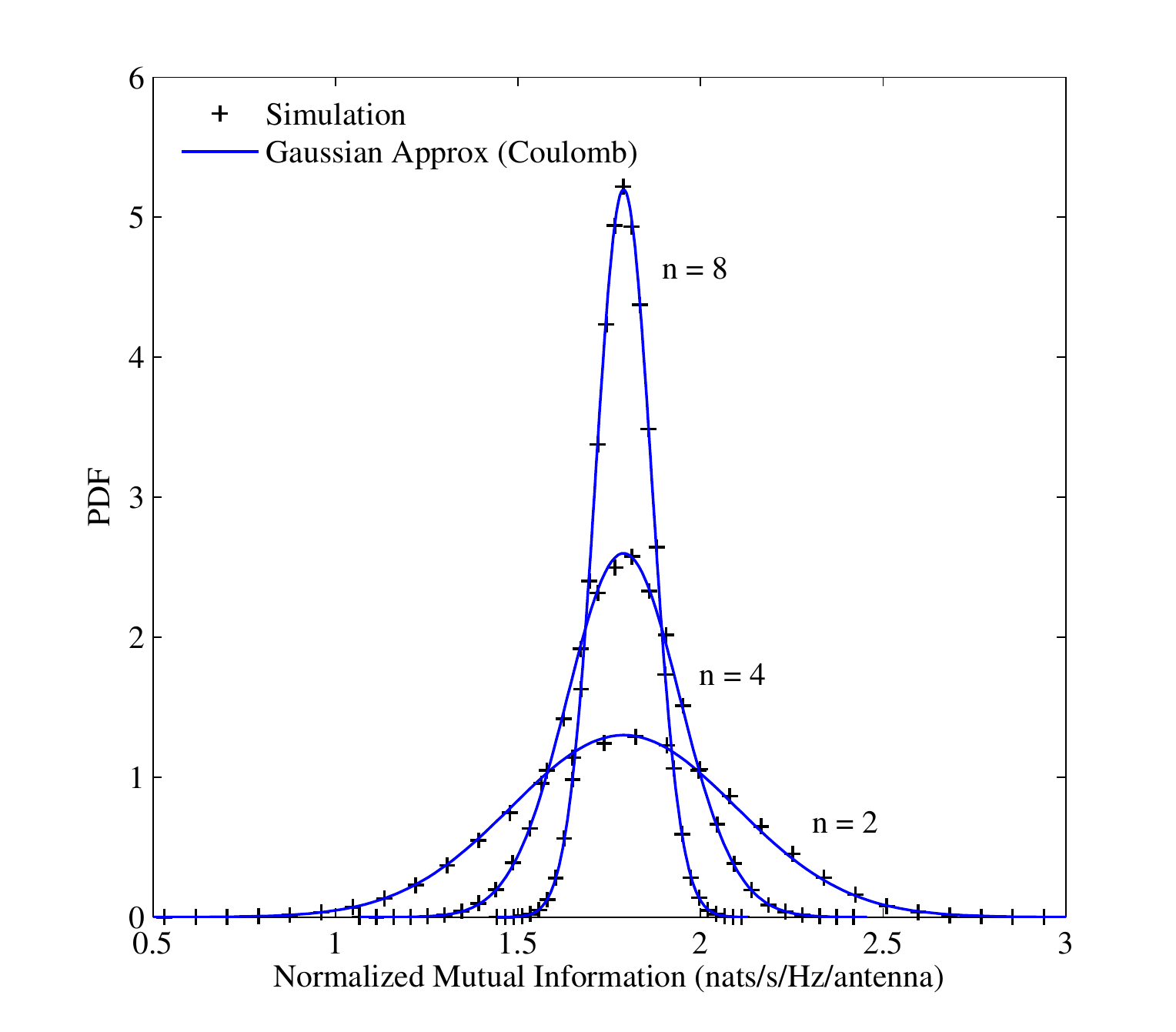}
\label{fig1:subfig2} } \label{fig1} \caption{PDF of normalized
mutual information $I(\mathbf{x}, \mathbf{y})/n$ for the single-user
MIMO scenario (deformed Laguerre case). Results shown for SNR, $P =
5$ dB, and different antenna configurations. In all cases shown, the
Coulomb fluid approximation is very accurate. }
\end{figure}

\begin{figure}[ht]
\centering

\subfigure[$n_r = 2, n_t = 2$]{
\includegraphics[width=0.48\columnwidth]{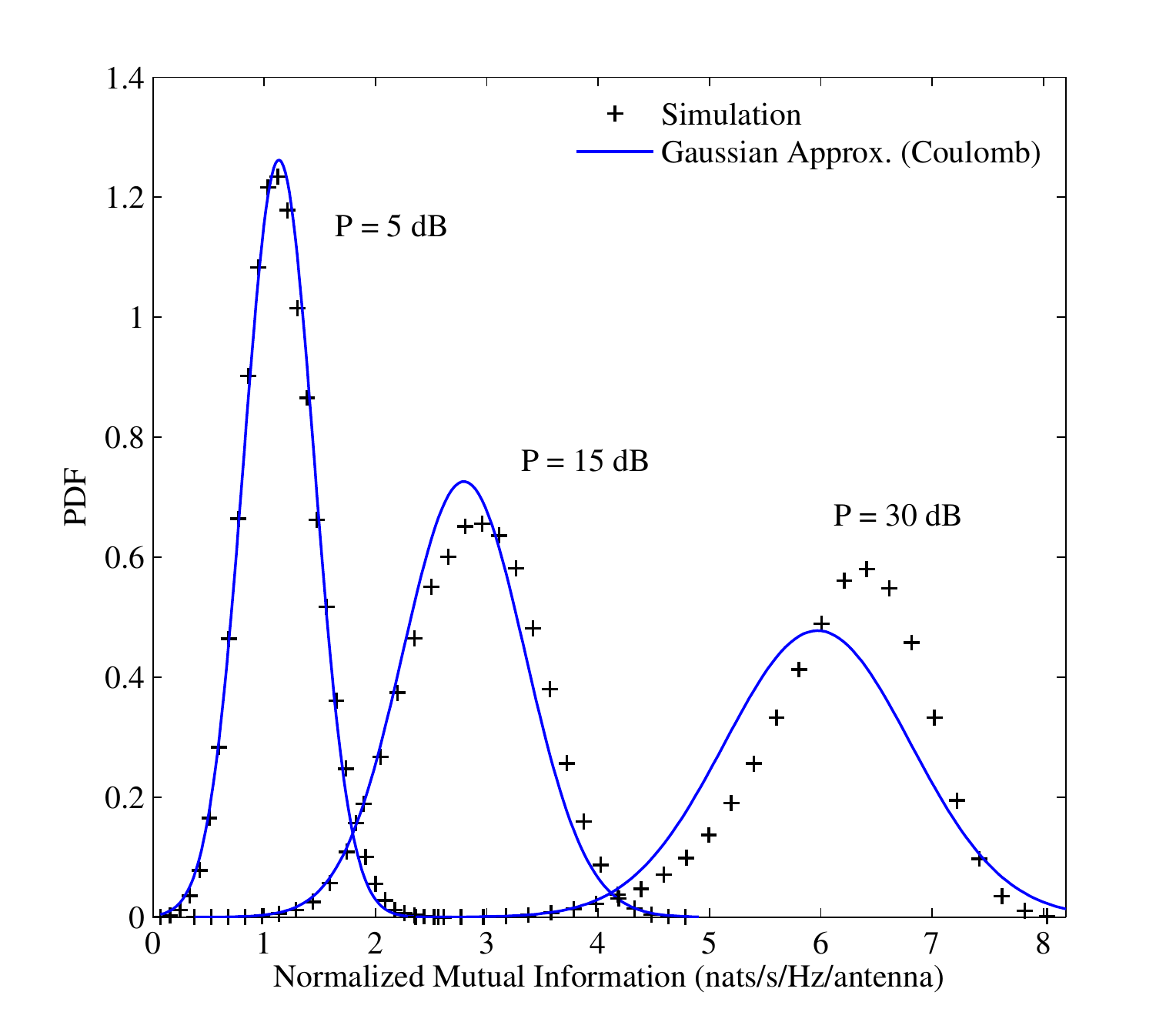}
\label{fig2:subfig1} } \subfigure[$n_r = 4, n_t = 4$]{
\includegraphics[width=0.48\columnwidth]{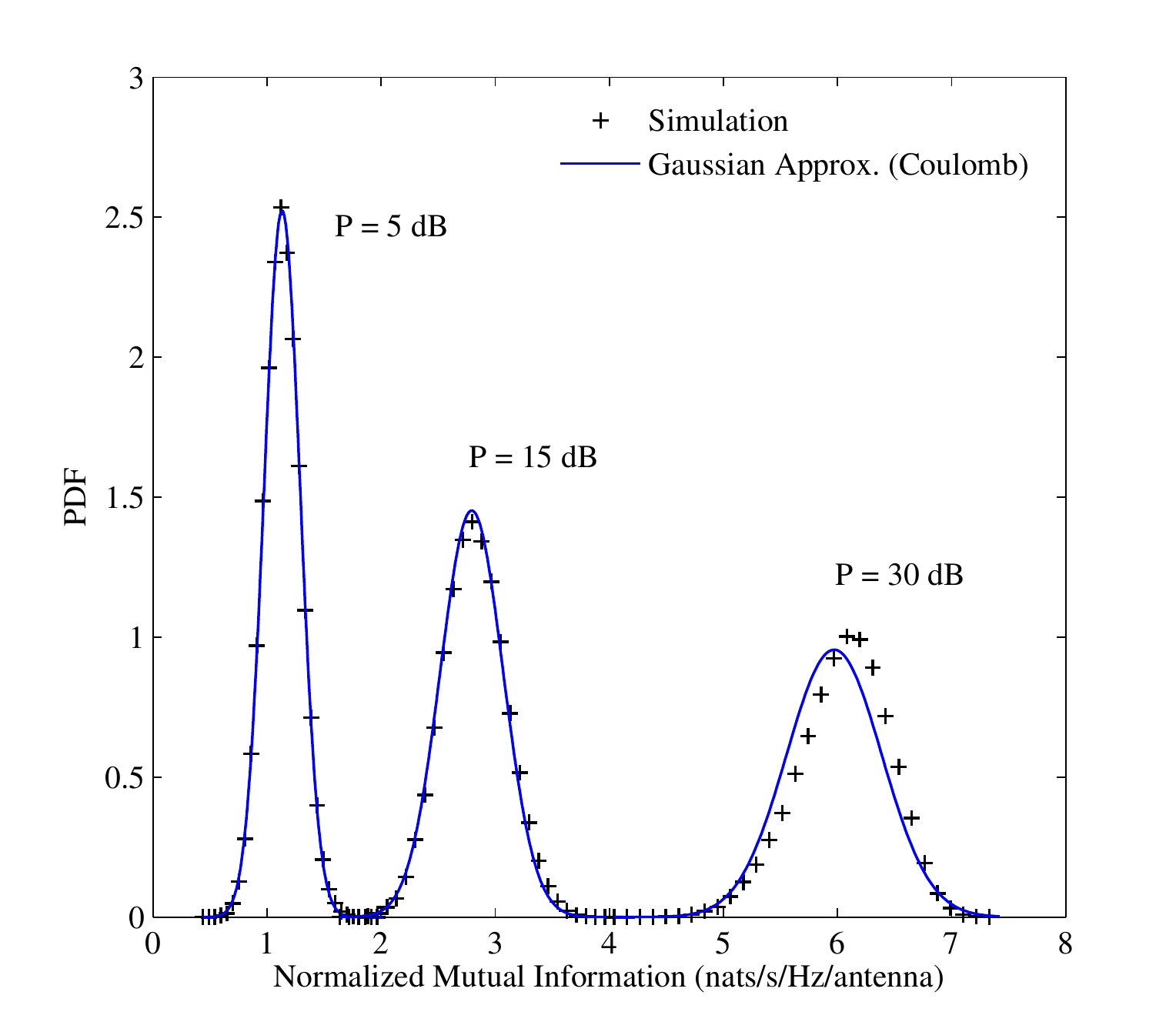}
\label{fig2:subfig2} } \label{fig2} \caption{PDF of normalized
mutual information $I(\mathbf{x}, \mathbf{y})/n$ for the single-user
MIMO scenario (deformed Laguerre case). Results shown for different
SNR values.  The Coulomb fluid approximation is very accurate when
the SNR $P$ is low, however it becomes less accurate (the mutual
information distribution deviates from Gaussian) as $P$ increases. }

\end{figure}

\begin{figure}[htp]
\centering

\subfigure[$n_r = n_t = n$, $P = 10$ dB]{
\includegraphics[width=0.45\columnwidth]{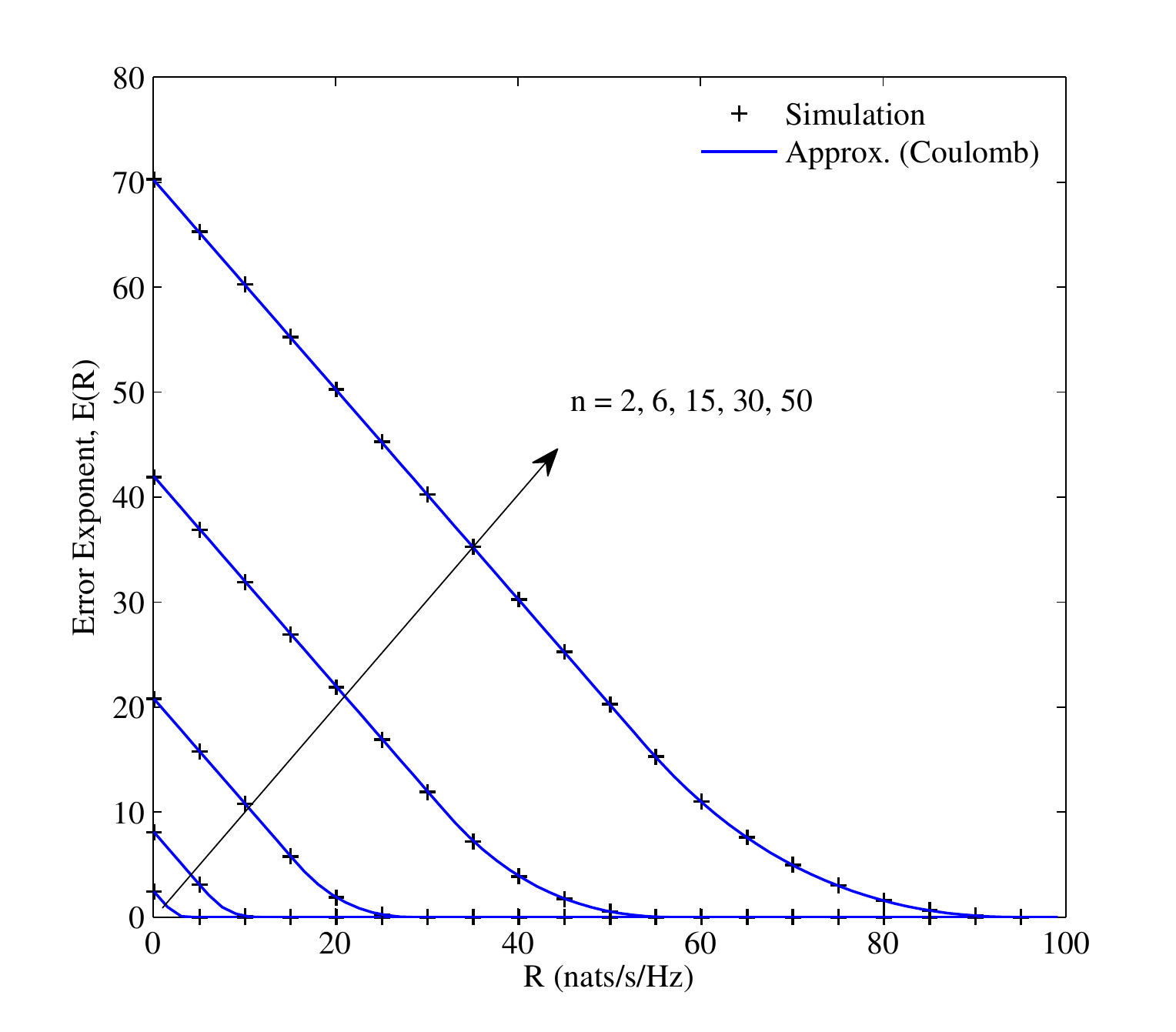}
\label{fig:subfig3} } \hspace{.3in} \subfigure[$n_r = 2 n_t \; \;
(n_t = n)$, $P = 10$ dB]{
\includegraphics[width=0.45\columnwidth]{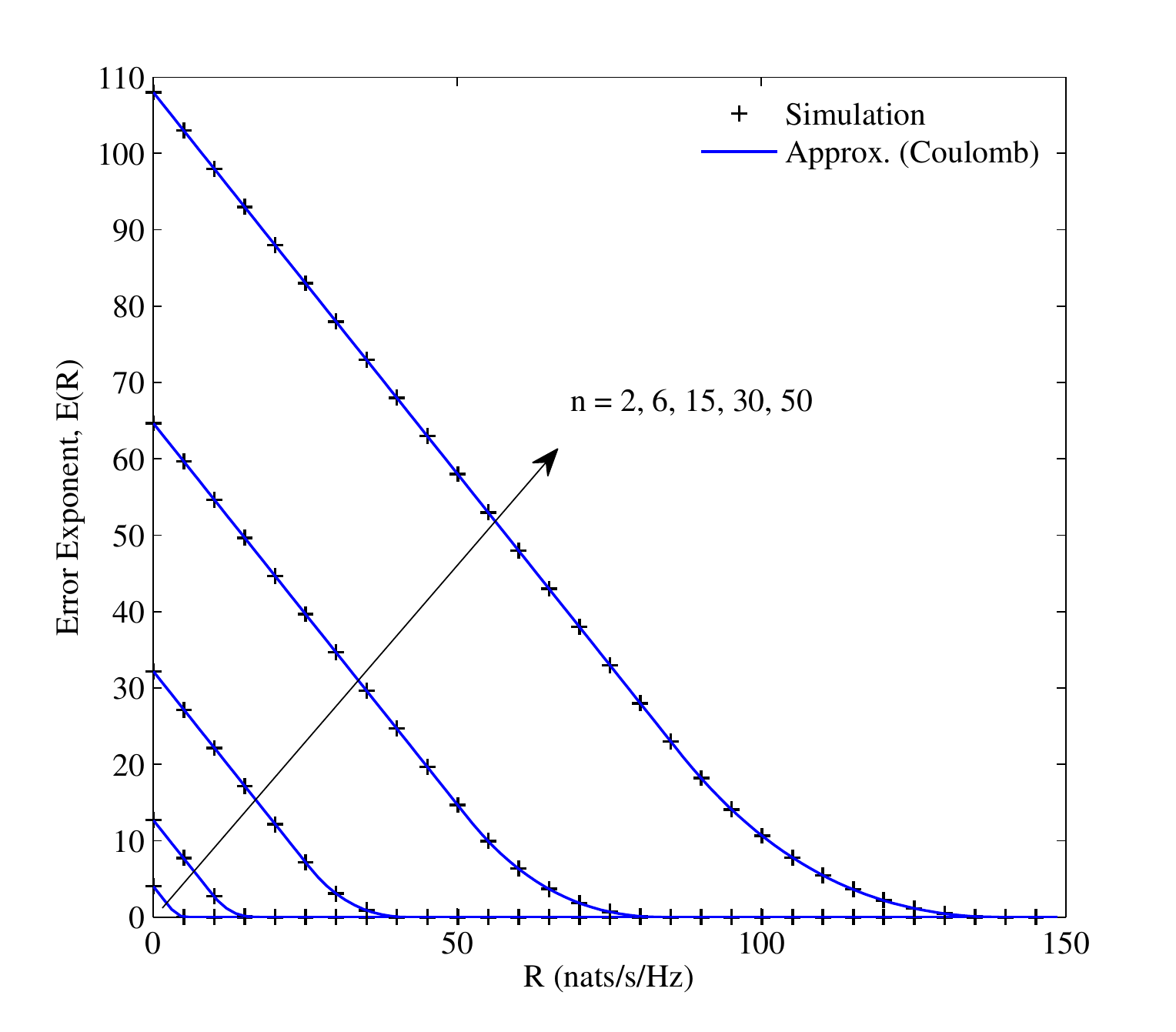}
\label{fig:subfig4} } \hspace{.3in} \subfigure[$n_r = n_t = n$, $P =
30$ dB]{
\includegraphics[width=0.45\columnwidth]{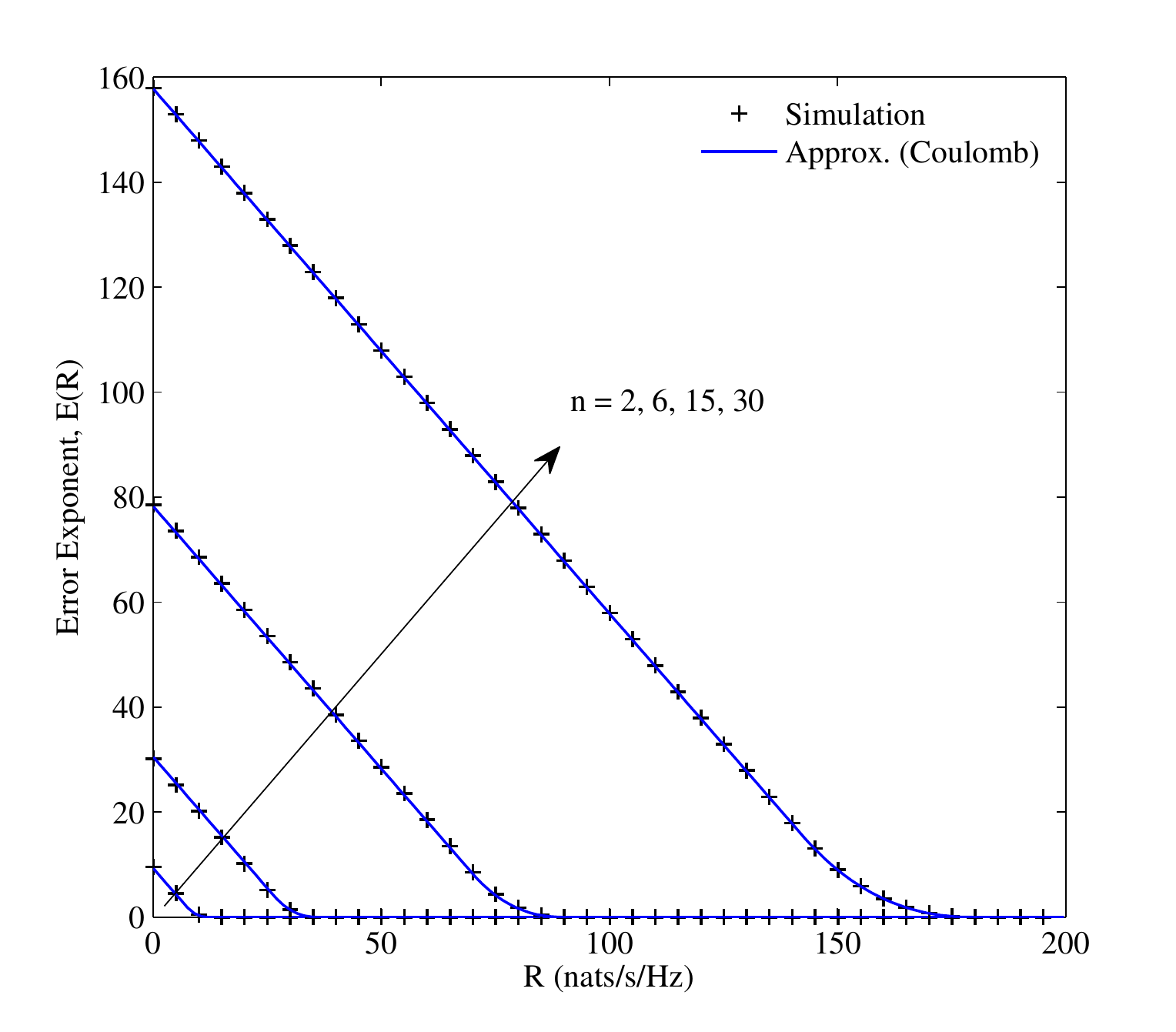}
\label{fig:subfig3} } \hspace{.3in} \subfigure[$n_r = 2 n_t \; \;
(n_t = n)$, $P = 30$ dB]{
\includegraphics[width=0.45\columnwidth]{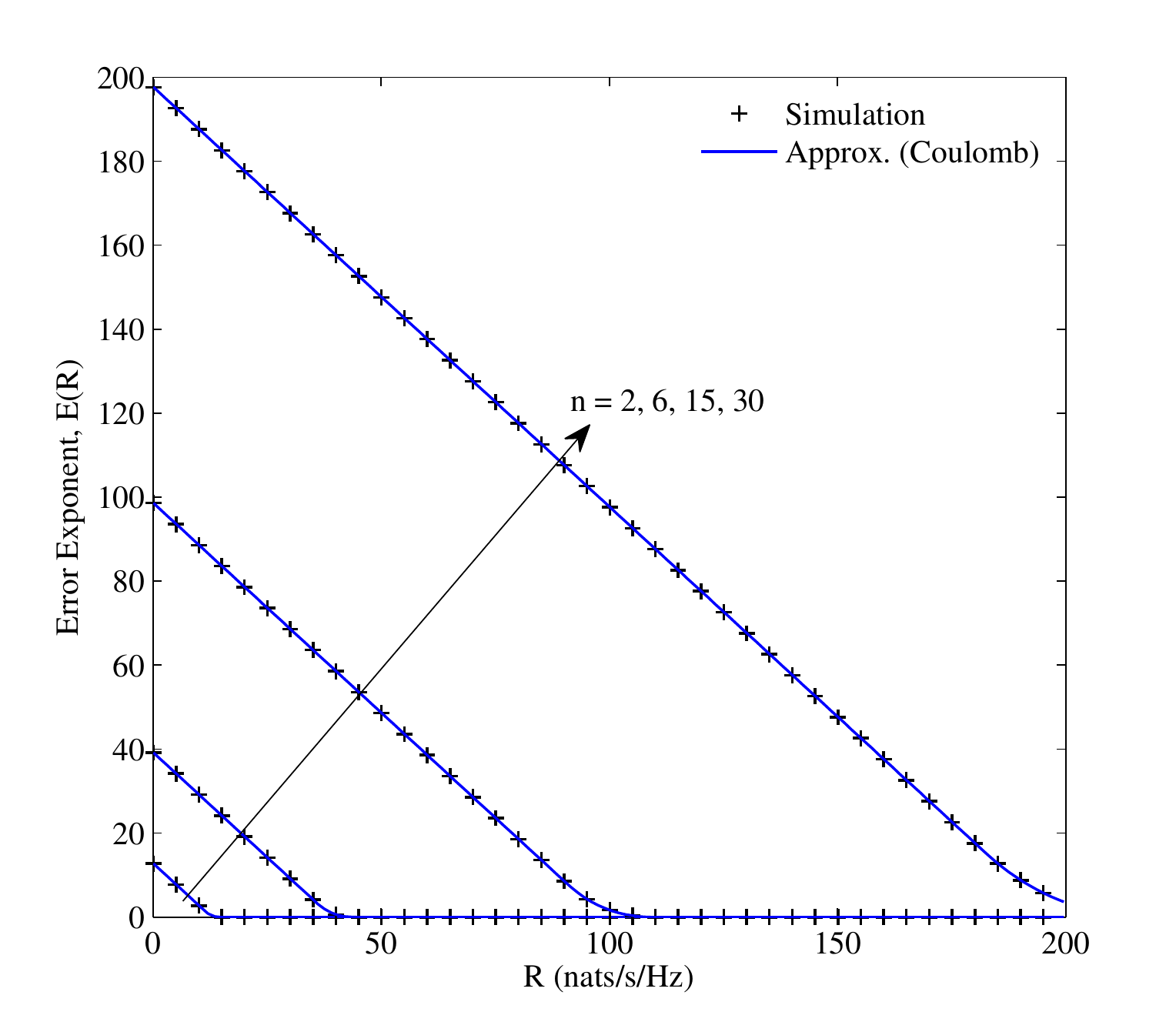}
\label{fig:subfig4} }

\label{fig:subfigureExample2} \caption{Error exponent for the
single-user MIMO scenario (deformed Laguerre case). Results shown
for SNR, $P = 10$ dB.  The Coulomb fluid approximation to the error
exponent is very accurate for both low \emph{and high} SNR ratios. }
\end{figure}


Figure 1 plots the Gaussian approximation to the distribution of the
normalized mutual information (per antenna) of single-user MIMO
systems, based on combining (\ref{eq:S1Laguerre}),
(\ref{eq:S2Laguerre}), and (\ref{eq:MeanVar_Defn}), and compares
with the true distribution generated via numerical simulations.  The
results are shown for a relatively low SNR, $P = 5$ dB, and for
various antenna configurations. In all cases the Gaussian
approximation is very accurate, even for $n$ as low as $2$.  The
situation changes, however, when $P$ is increased, as shown clearly
in Figure 2.  In particular, as $P$ increases, it is evident that
the distribution starts to deviate from Gaussian, and that this
deviation is most significant for small $n$.  This behavior is in
line with the results of \cite{Zheng}, which shows that the tail of
the distribution behaves like an exponential random variable, rather
than a Gaussian, when $P \to \infty$ (and $n$ fixed). We investigate
this phenomenon in more detail in Section \ref{sec:Beyond} (see also
\cite{Kasakopoulos}).


Figure 3 compares the Coulomb fluid approximation for the error exponent, based on combining (\ref{eq:S1Laguerre}), (\ref{eq:S2Laguerre}) and (\ref{eq:E0Eqn_Coulomb}), with the true error exponent computed via numerical simulation of (\ref{eq:E0Eqn}) and (\ref{eq:Er}).  Interestingly, we see that in all cases, including both low \emph{and high} $P$, the Coulomb fluid approximation is extremely accurate.

\subsection{Coulomb Fluid and the Deformed Jacobi Weight}  \label{sec:CF_Jacobi}

In this case, we have the particularizations:
\bq
f(x) = - \log \left( \frac{x + T}{1-x} \right) ,
\quad \v(x) =  - \varphi_1 \log x - \varphi_2 \log(1-x) , \quad L = 0, \quad U = 1 \; .
\eq
First consider the limiting density $\sigma_0(x)$.  This can be obtained by direct evaluation of (\ref{eq:sigma0_limiting}).
Alternatively, we may apply a simple transformation of the limiting density
$\rho(\cdot)$ corresponding to the classical Jacobi weight (\ref{eq:JacobiClassical}), given by \cite{BasorChen}
\bq
\rho(y)&=&\frac{n}{\pi}\:
\frac{1+(\varphi_1+\varphi_2)/2}{1-y^2}\sqrt{(B_n-y)(y-A_n)},\quad y\in(A_n,B_n) \\
A_n&:=&\frac{1}{(2n+\al_1+\al_2+2)^2}\left[\al_2^2-\al_1^2-4\sqrt{n(n+\al_1)(n+\al_2)(n+\al_1+\al_2)}\right] \\
B_n&:=&\frac{1}{(2n+\al_1+\al_2+2)^2}\left[\al_2^2-\al_1^2+4\sqrt{n(n+\al_1)(n+\al_2)(n+\al_1+\al_2)}\right] \\
A_n\to A&:=&\frac{1}{(\varphi_1+\varphi_2+2)^2}\left[\varphi_2^2-\varphi_1^2-4\sqrt{(1+\varphi_1)(1+\varphi_2)
(1+\varphi_1+\varphi_2)}\right] \\
B_n\to B&:=&\frac{1}{(\varphi_1+\varphi_2+2)^2}\left[\varphi_2^2-\varphi_1^2+4\sqrt{(1+\varphi_1)(1+\varphi_2)
(1+\varphi_1+\varphi_2)}\right].
\eq
Here, the quantities $A$ and $B$ were obtained by recalling that $\al_1 = n \varphi_1$ and $\al_2 = n \varphi_2$, and
taking the limit $n \to \infty$. To relate this density to $\sigma_0(x)$, we first note that
\bq
\v(x) = - \frac{\log w_{\rm Jac}(x)}{n}
\eq
where $w_{\rm Jac}(x)$ is the classical Jacobi weight
(\ref{eq:JacobiClassical_Shifted}).  Now, from the relation (\ref{eq:JacobiRelation}) we have
\bq
\v'(x) &=& - \frac{w_{\rm Jac}'(x)}{n w_{\rm Jac}(x)}  \nn \\
&=&  - \frac{\tilde{w}_{\rm Jac}'(1-2x)}{n \tilde{w}_{\rm Jac}(1-2x)}
\eq
which after substituting into (\ref{eq:sigma0_limiting}) gives the desired result
\bq
\sigma_0(x) &=& \frac{2}{n} \rho ( 1 - 2 x )  \nn \\
&=&  \left[1 + (\varphi_1+\varphi_2)/2\right]\:\frac{\sqrt{ (b-x)(x-a) } }{\pi\:x(1-x)},\:\quad   a < x < b
\eq
with
\bq
a := \frac{ 1-B}{2} , \quad \quad b := \frac{1-A}{2} \; .
\eq
Substituting this distribution along with $f(x) = - \log[(T + x)/(1-x)]$ into  (\ref{eq:SLaguerre}),
applying the partial fraction decomposition
\bq
\frac{\sqrt{(b-x)(x-a)}}{x(1-x)}
= \frac{1}{\sqrt{(b-x)(x-a)}} \left( 1 - \frac{a b}{x} + \frac{ (1-a)(1-b) }{ x-1 } \right)
\eq
and integrating using the identities (\ref{eq:Int1}), (\ref{eq:Int3})--(\ref{eq:Int7}), we find
\bq \label{eq:S2Jacobi}
\frac{{\cal S}_2^{\rm Comm.}(T)}{n[1 + (\varphi_1 + \varphi_2)/2]} &=&
2 \log\left( \frac{ \sqrt{1-a} + \sqrt{1-b} }{ \sqrt{T+a} + \sqrt{T+b} } \right)  -
\sqrt{ab}\log \left( \frac{1 - (\:\sqrt{a b} - \sqrt{ (1-a)(1-b)}\:)^2 }
{(\:\sqrt{a b} + \sqrt{ (T+a) (T+b)}\:)^2 - T^2 }  \right) \nn \\
&+& \sqrt{ (1-a)(1-b) } \log \left( \frac{ (T+1)^2 -
(\:\sqrt{(T+a)(T+b)} - \sqrt{(1-a)(1-b)}\:)^2 }{4(1-a)(1-b)}\right)
. \nn \eq We note that an alternative solution was also computed in
\cite{Lozano02}, requiring the numerical evaluation of a certain
fixed-point equation.

\begin{figure}[ht]
\centering

\subfigure[$P/P_I = 10$ dB]{
\includegraphics[width=0.48\columnwidth]{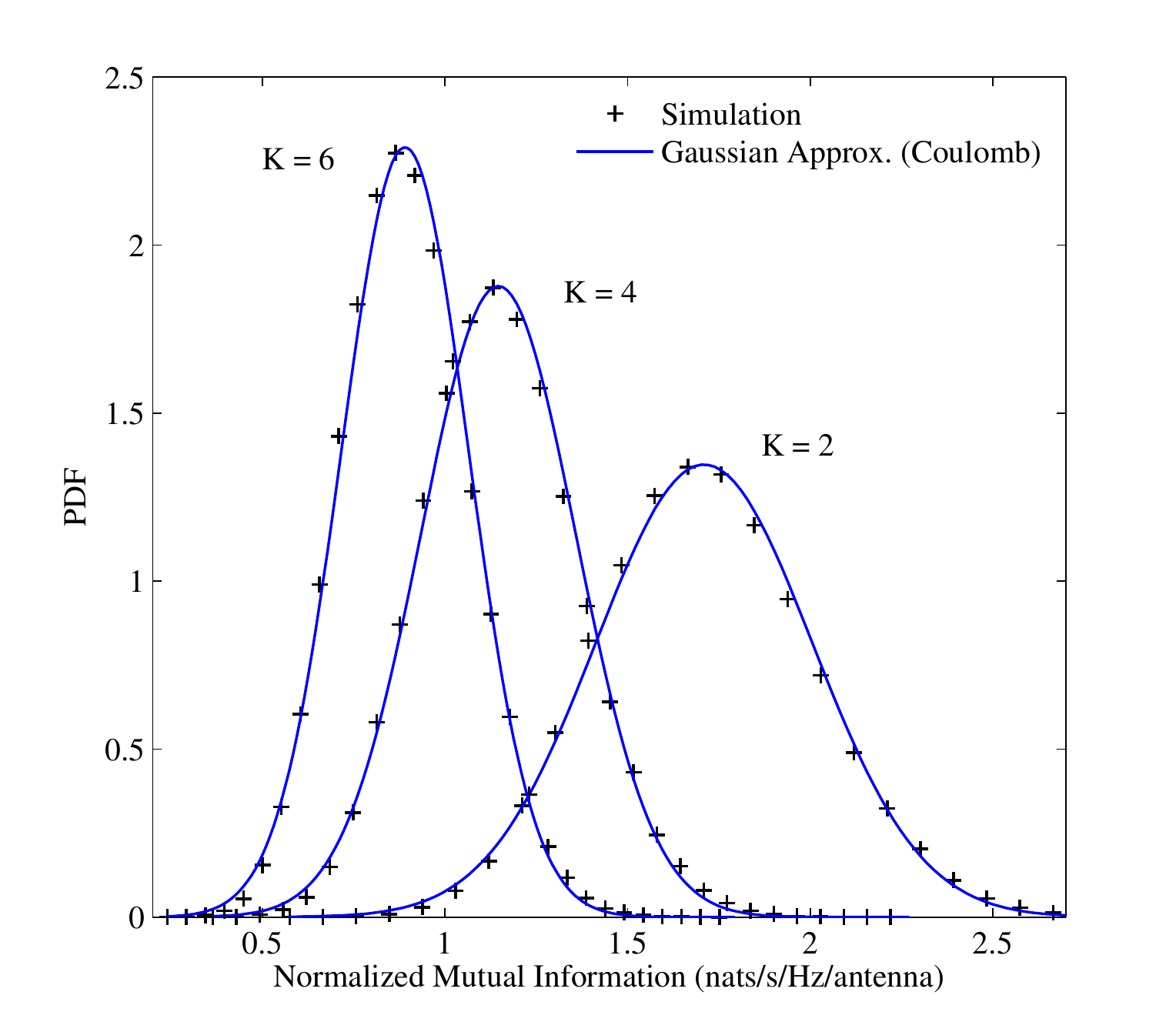}
\label{fig:JacCapacitya}
}
\subfigure[$P/P_I = 30$ dB]{
\includegraphics[width=0.48\columnwidth]{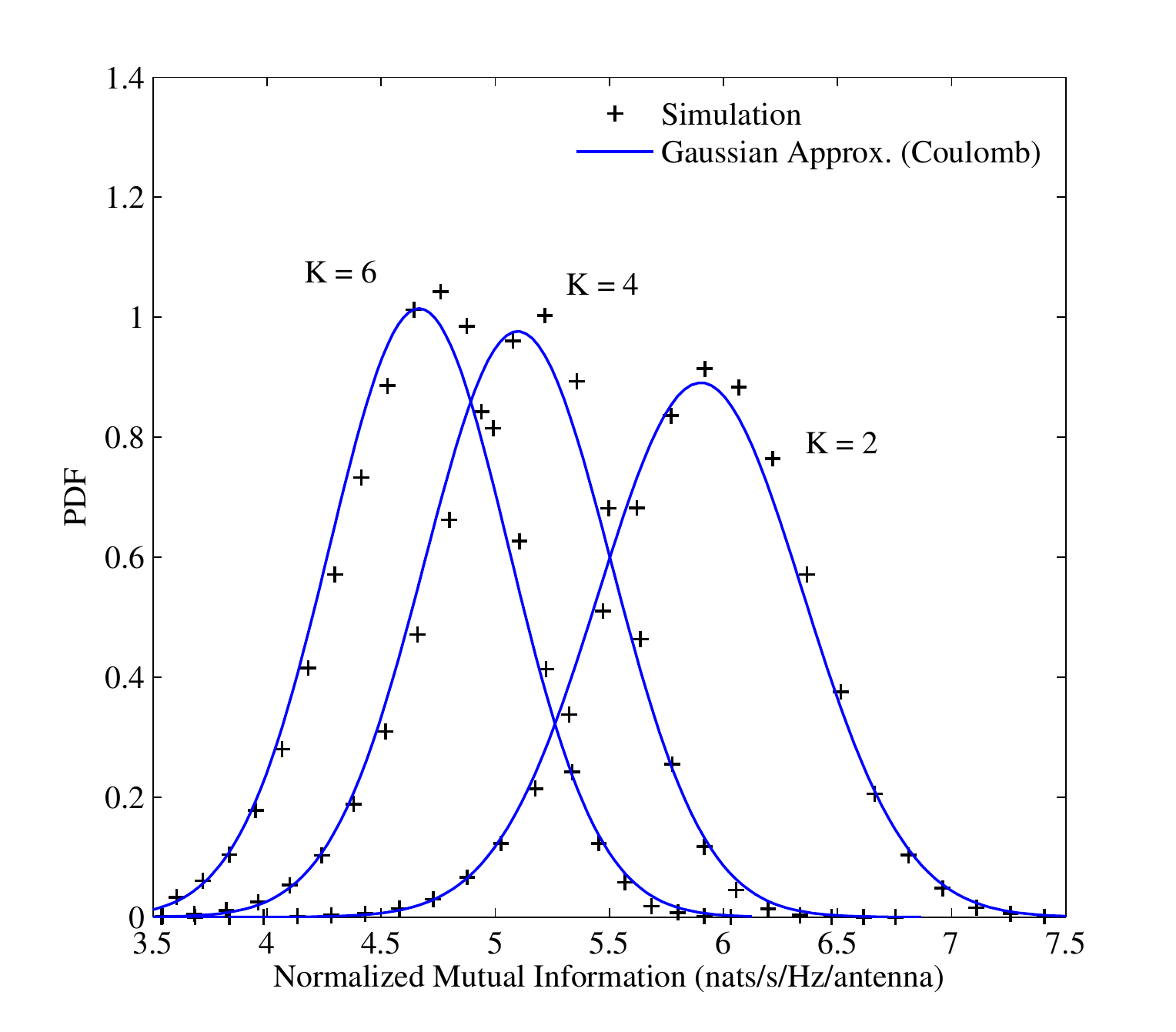}
\label{fig:JacCapacityb} } \label{fig:JacCapacity} \caption{PDF of
normalized mutual information $I(\mathbf{x}, \mathbf{y})/n$ for the
multi-user MIMO scenario (deformed Jacobi case). Results shown for
$n_t = 4$ and $n_r = 3$.  The Coulomb fluid approximation is very
accurate when the signal-to-interference ratio $P/P_I$ is low,
however it becomes less accurate (the mutual information
distribution deviates from Gaussian) as $P/P_I$ increases.}

\end{figure}

\begin{figure}[ht]
\centering

\subfigure[$P/P_I = 10$ dB]{
\includegraphics[width=0.48\columnwidth]{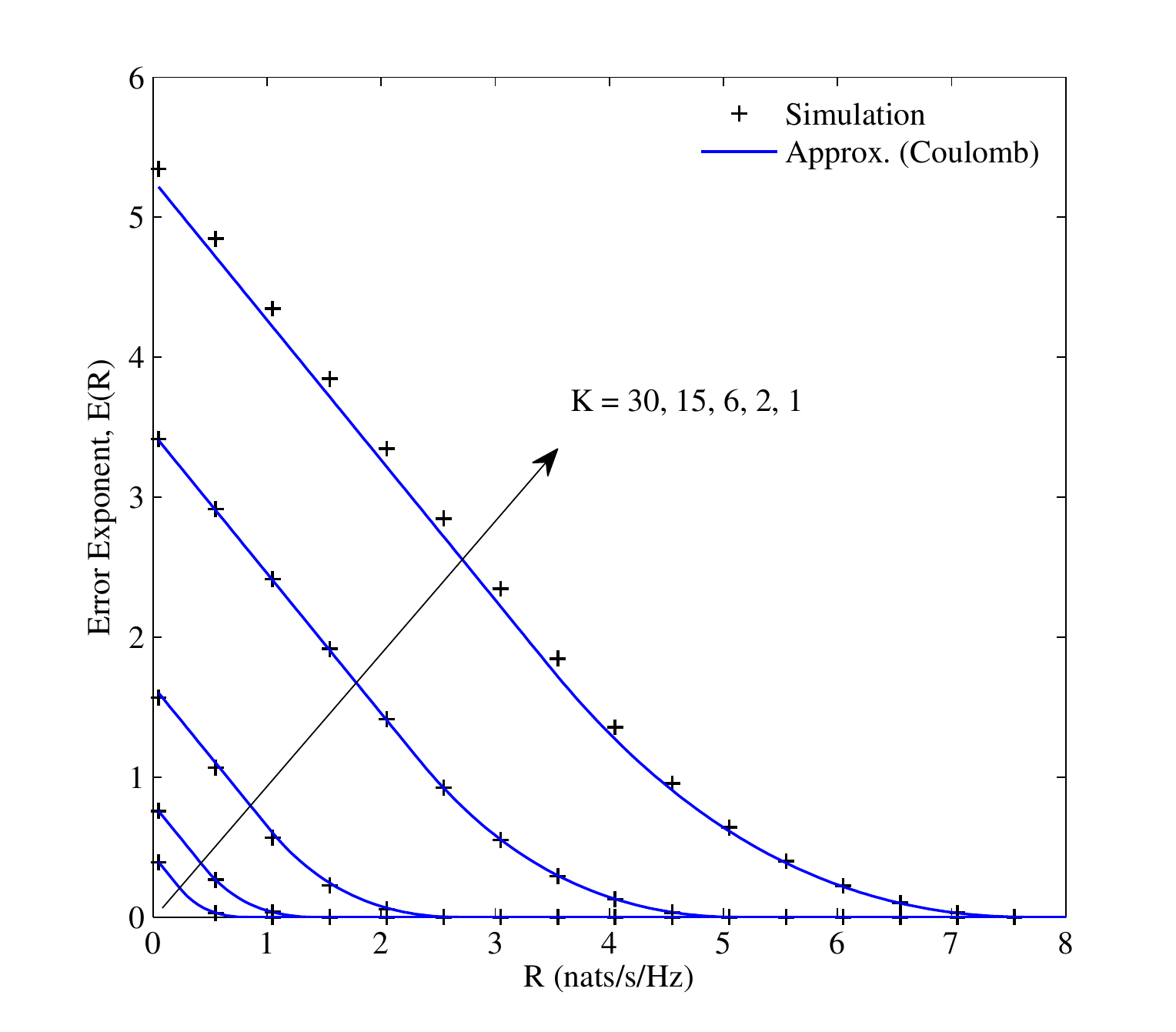}
\label{fig:JacErrorExpa}
}
\subfigure[$P/P_I = 30$ dB]{
\includegraphics[width=0.48\columnwidth]{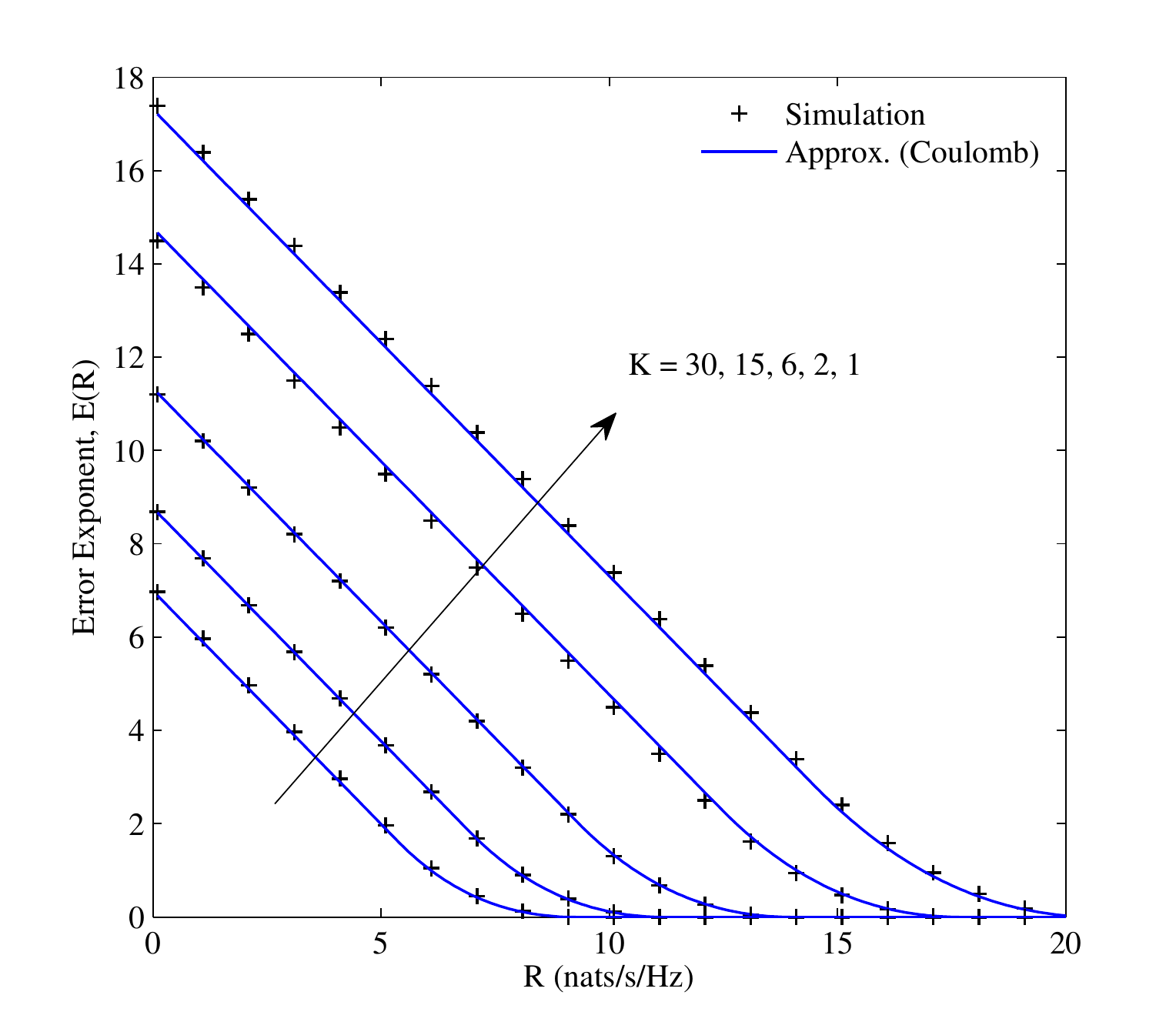}
\label{fig:JacErrorExpb}
}
\label{fig:JacErrorExp}
\caption{Error exponent for the multi-user MIMO scenario (deformed Jacobi case).
Results shown for $n_t = 4$ and $n_r = 3$. Similar
to the single-user MIMO case, the Coulomb fluid approximation to the error exponent of multi-user MIMO channels
is very accurate for both low and high signal-to-interference ratios.}

\end{figure}

For $\tilde{\varrho}(x)$, we substitute $f'(x) = -1/(T+x) - 1/(1-x)$ into (\ref{eq:VarRhoDefn}) and apply the integral identity (\ref{eq:PInt1}) along with (\ref{eq:PInt2}) to yield
\bq
\tilde{\varrho}(T) = \frac{1}{2 \pi \sqrt{ (b-x)(x-a) } } \left[ \frac{ \sqrt{(1-a)(1-b)}}{1-x} - \frac{ \sqrt{(T+a)(T+b)}}{x+T}  \right] \; .
\eq
Substituting this into
(\ref{eq:SLaguerre}), and applying the integral identities
(\ref{eq:Int2}),  (\ref{eq:Int4}), (\ref{eq:Int7}), (\ref{eq:Int8}),
we obtain
\bq \label{eq:S1Jacobi}
{\cal S}_1^{\rm Comm.}(T) =
\log \left[\:\frac{4 \sqrt{(T+a) (T+b) (1-a) (1-b)}}{(T+1)^2 - ( \sqrt{(1-a)(1-b)} - \sqrt{(T+a)(T+b)}\:)^2}\right]  .
\eq

Figure 4 plots the Gaussian approximation to the distribution of the normalized mutual information (per antenna) of multi-user MIMO systems, based on combining (\ref{eq:S1Jacobi}), (\ref{eq:S2Jacobi}), and (\ref{eq:MeanVar_Defn}), and compares with the true distribution generated via numerical simulations.  As for the single-user MIMO scenario, the Gaussian approximation is accurate regardless of the number of interferers $K$, especially when the signal-to-interference ratio $P/P_I$ is not high.  As $P/P_I$ is increased, once again the distribution starts to deviate from Gaussian; however, this deviation is seemingly less significant than that seen previously for the single-user case.


Figure 5 compares the Coulomb fluid approximation for the error exponent, based on combining (\ref{eq:S1Jacobi}), (\ref{eq:S2Jacobi}) and (\ref{eq:E0Eqn_Coulomb}), with the true error exponent computed via numerical simulation of (\ref{eq:E0Eqn}) and (\ref{eq:Er}).  Again, we see that in all cases, the Coulomb fluid approximation is extremely accurate.

\section{Beyond the Coulomb Fluid Approximation} \label{sec:Beyond}

In this section, we take a closer look at the Painlev\'e and Coulomb
fluid representations of the mutual information distribution.  We
will focus on the Laguerre scenario (i.e., the single-user MIMO
case), although the analysis can be extended to the Jacobi scenario
also.  Our main objective is to establish relationships between the
exact characterization of the mutual information distribution via
the Painlev\'e equation, and the simpler Coulomb fluid
approximation.  As a key result, we will show that for both the mean
and the variance, the Coulomb fluid gives an \emph{exact}
representation to leading order in $n$.  We also employ the
Painlev\'e equation to establish the first-order correction terms to
the mean, variance, and third cumulant, which helps to characterize
the deviation of the mutual information distribution from Gaussian.


\subsection{Initial Study using Power Series Expansion}

To get an intuitive feel for the problem, we start by looking for a $1/t$ expansion in the Painlev\'e equation as $t \to \infty$. This is essentially the case where the SNR, $P$, is small.  For simplicity, we consider the case $m = n$; although the analysis can be extended to the case $m \neq n$.

\subsubsection{Evaluating Cumulants from the Painlev\'e}

To proceed, let
\bq
H_n(t) =: n \la + G_n (t) \;
\eq
such that
\bq  \label{eq:MGF_G}
{\cal M}(\la) = \exp \left( \int_\infty^t \frac{G_n(x)}{x} d x   \right) \; .
\eq
Note also that
\bq
t \frac{d}{dt} \log {\cal M}(\la) = G_n(t) .
\eq
From the Painlev\'e equation (\ref{eq:JimboPV}), $G_n(t)$ satisfies
\bq \label{eq:PEquation}
(t G_n'')^2 = \left( G_n' (t + 2 n + \la ) - G_n  \right)^2 - 4 ( t G_n' - G_n + n^2 ) (G_n'^2 + \la G_n' )
\eq
where the derivatives are with respect to $t$.  Suppose that $G_n(t)$ has a formal power series expansion in $1/t$:
\bq  \label{eq:GnSeriesinA}
\frac{G_n(t)}{n^2\la} = \sum_{k=1}^\infty \frac{b_k}{t^k} \;
\eq
where the coefficient $b_k$ depends on $n$ and $\la$.
As a first step, we substitute the power series into (\ref{eq:PEquation}) and find the first few $b_k$'s as follows:
\bq
b_1&=&-1\nn\\
b_2&=&2n-\la\nn\\
b_3&=&-(1+5n^2 -6n\la+\la^2)\nn\\
b_4&=&10n+14n^3-(5+29n^2)\la + 12n\la^2-\la^3\nn\\
b_5&=&-(8+70n^2+42n^4-(80n+130n^3)\la+(15 +95n^2)\la^2 -20n\la^3+\la^4)\nn\\
b_6&=&168n+420n^3+132n^5-(84+806n^2+562n^4)\la+(350n+624n^3)\la^2 \nn \\
&& -(35+235n^2)\la^3+30n\la^4-\la^5\nn\\
b_7&=&-\{180+2121n^2+2310n^4+429n^6-(2366n+6510n^3+2380n^5)\la\nn\\
&+&(469+4795n^2+3682n^4)\la^2-(1120n+2128n^3)\la^3
+(70+490n^2)\la^4-42n\la^5+\la^6\}.\nn
\eq
From these examples, it is clear that $b_k$ takes the form
\bq \label{eq:bk_GeneralForm}
b_k = \sum_{\ell = 0}^{k-1} \la^\ell \left[ B_{k, \ell} n^{k-\ell-1} + C_{k, \ell} n^{k-\ell-3} + {\rm O} ( n^{k-\ell-5} ) \right]  ,
\eq
where the coefficients $B_{k,\ell}$ and $C_{k,\ell}$ are \emph{independent} of $n$ and $\la$, and are computed via
\bq
B_{k, \ell} = \frac{1}{(k-\ell-1)! \ell!} \frac{d^{k-\ell-1}}{d n^{k-\ell-1} }  \, \frac{d^\ell}{d \la^\ell} b_k \biggr|_{\la = 0, n = 0}  , \quad \quad k = 1, 2, \ldots, \infty, \quad \ell = 0, \ldots, k-1
\eq
and
\bq
C_{k, \ell} = \frac{1}{(k-\ell-3)! \ell!} \frac{d^{k-\ell-3}}{d n^{k-\ell-3} }  \, \frac{d^\ell}{d \la^\ell} b_k \biggr|_{\la = 0, n = 0}  , \quad \quad k = 3, 4, \ldots, \infty, \quad \ell = 0, \ldots, k-3
\eq
respectively. Note that $C_{k, \ell} = 0$ for $k = 1$, $k = 2$, or $\ell \geq k-2$.  The $B_{k,\ell}$ coefficients represent the leading order terms in $n$, whereas the $C_{k,\ell}$ coefficients represent the next (non-zero) lower order correction terms; for example,
\bq
B_{1,0} = -1, \quad B_{2,0} = 2, \quad B_{3,0} = -5, \quad B_{2,1} = -1, \quad B_{3,1} = 6, \quad B_{3,2} = -1 \nonumber
\eq
and
\bq
C_{3,0} = -1, \quad C_{4,0} = 10, \quad C_{5,0} = -70, \quad C_{3,1} = 0, \quad C_{4,1} = -5, \quad C_{5,1} = 80 \; . \nonumber
\eq

We aim to investigate the cumulants of the mutual information distribution, which in turn requires an expansion of the form
\bq \label{eq:GnSeries}
G_n(t) = \la g_1 (t) + \la^2 g_2(t) + \cdots
\eq
Together with (\ref{eq:GnSeriesinA}) and (\ref{eq:bk_GeneralForm}), we then get
\bq \label{eq:gkDefn}
g_k(t) = g_k (n/P) =  n^{2-k} \sum_{\ell=k}^\infty \left( B_{\ell, k-1} + \frac{C_{\ell, k-1}}{n^2} + {\rm O} \left( \frac{1}{n^4} \right) \right) P^\ell , \quad \quad k = 1, 2, \ldots
\eq
Plugging this into (\ref{eq:MGF_G}) and integrating, the logarithm of the moment generating function becomes
\bq  \label{eq:MGFExpansion}
\log {\cal M}(\la) =  \sum_{k=1}^\infty \kappa_k \frac{\la^k}{k!}  \;
\eq
where $\kappa_k$ denotes the $k$th cumulant given by
\bq
\kappa_k &=& - k! \int_0^P \frac{ g_k (n/y) }{y} d y \nonumber \\
&=& - k! \, n^{2-k} \sum_{\ell=k}^\infty  \frac{B_{\ell, k-1} + \frac{C_{\ell, k-1}}{n^2} + {\rm O} \left( \frac{1}{n^4} \right)}{\ell} P^\ell \; .
\label{eq:KappaDefn}
\eq
The mean is
\bq  \label{eq:MeanSum}
\kappa_1 &=& n \left( P - P^2 + \frac{5}{3} P^3 - \frac{7}{2} P^4 + \cdots \right) + \frac{1}{n} \left( \frac{1}{3} P^3 - \frac{5}{2} P^4 + \frac{70}{5} P^ 5 + \cdots \right) + {\rm O} \left( \frac{1}{n^3} \right) \nonumber
\eq
the variance is
\bq
\kappa_2 &=& P^2 - 4 P^3 + \frac{29}{2}P^4 - \frac{260}{5} P^5 + \cdots  + \frac{1}{n^2} \left(\frac{5}{2}P^4 - 32 P^5 + \frac{806}{3}P^6 + \cdots \right) + {\rm O} \left( \frac{1}{n^4}  \right)\; \nonumber
\eq
the third cumulant is
\bq
\kappa_3
&=& \frac{1}{n} \left[ 2 P^3 - 18 P^4 + 114 P^5 + \cdots + \frac{1}{n^2} \left( 18 P^5 - 350 P^6 + \cdots  \right) \right] + {\rm O} \left( \frac{1}{n^5} \right) \nonumber
\eq
and so on.

Importantly, this result demonstrates that as $n$ grows large, the $k$th cumulant scales as
\bq
\kappa_k = {\rm O} (n^{2-k}) , \quad \quad k = 1, 2, \ldots
\eq
Thus, as expected, we see that the mean of the mutual information grows linearly with $n$, the variance converges to a constant, and all other cumulants disappear as $n \to \infty$. This reaffirms that the distribution becomes Gaussian for asymptotically large $n$.

Whilst in the analysis above we have substituted the numerical values of the constants $B_{\ell, k}$, explicit formulae can also be derived by directly using the Painlev\'e differential equation.  We demonstrate the procedure by considering the coefficients in the mean summation (\ref{eq:MeanSum}), namely $\{ B_{\ell, 0} \}$.  The same technique can be used to derive the formulae for the coefficients of the higher moments also.

Plugging the series (\ref{eq:GnSeriesinA}) into the differential equation (\ref{eq:PEquation}), keeping only the lowest order terms in $\la$ and taking $n$ large, the r.h.s. of (\ref{eq:PEquation}) becomes
\bq
 (P n \la)^2 \left[  (4 B_{1,0}^2 + 4 B_{1,0} ) + \left( 12 B_{1,0} B_{2,0} + 8 B_{1,0}^2 + 8 B_{2,0} \right) P  + \sum_{ i = 2 }^\infty \chi_{i+2} P^i \right]
\eq
with
\bq
\chi_{i+2} = - 4 B_{i+1,0} - 8 i B_{i,0} + \sum_{j=1}^{i-1} ( 2 + j ) B_{j+1,0} \left( (i-j+2) B_{i-j+1,0} + 4 (i - j) B_{i-j,0} \right)  \; .
\eq
Now consider the l.h.s. of (\ref{eq:PEquation}).  We have
\bq
\frac{ (t G_n'')^2 }{ (n \la )^2 }   \sim   \left( \sum_{i=1}^\infty \frac{ i (i+1) B_{i,0} P^{i+1} }{n }       \right)^2 \to 0
\eq
as $n \to \infty$.  Thus, dividing both the l.h.s. and r.h.s. by  $(P n \la)^2$ and taking $n$ large we get
\bq
(4 B_{1,0}^2 + 4 B_{1,0} ) + \left( 12 B_{1,0} B_{2,0} + 8 B_{1,0}^2 + 8 B_{2,0} \right) P  + \sum_{ i = 2 }^\infty \chi_{i+2} P^i  = 0 \; .
\eq
We calculate the $B_{i,0}$'s recursively, since every coefficient of $P$ must equate to zero.  Trivially, considering the constant and linear terms in $P$, we get
\bq \label{eq:aiInitial}
B_{1,0} = -1, \quad \quad B_{2,0} = 2 \; .
\eq
For higher order coefficients, we have the recurrence relation
\bq \label{eq:aiHigher}
 B_{i+1,0} =  - 2 i B_{i,0} + \frac{1}{4} \sum_{j=1}^{i-1} ( 2 + j ) B_{j+1,0} \left( (i-j+2) B_{i-j+1,0} + 4 (i - j) B_{i-j,0} \right)  \;
\eq
for $i \geq 2$.  Based on this, the next few coefficients are evaluated as
\bq
B_{3,0} = -5, \quad \quad B_{4,0} = 14, \quad \quad B_{5,0} = -42 \; .
\eq
It can be verified that this difference equation also admits the explicit solution
\bq
B_{\ell, 0} =  \frac{ (-1)^\ell (2\ell)!}{ \ell! (\ell+1)! } \; , \quad \quad \ell = 1, 2, \ldots
\eq
and therefore the mean of the mutual information takes the explicit form (to leading order of $n$)
\bq
\kappa_1 = - n \sum_{\ell = 1}^\infty \frac{ (-1)^\ell (2\ell)!}{ \ell! (\ell+1)! \ell }  P^\ell  + {\rm O} \left( \frac{1}{n} \right) \; .
\eq

\subsubsection{Comparison with the Coulomb Fluid}

Now, consider the corresponding quantity derived based on the Coulomb fluid method:
\bq
\tilde{G}_n(1/P) &=& t \frac{d}{dt} \log {\cal M}(\la) \nonumber \\
&=& \la \tilde{g}_1(1/P) + \la^2 \tilde{g}_2(1/P)
\eq
where
\bq
\tilde{g}_1(1/P) &=& P \frac{d}{d P} \left( {\cal S}_2^{\rm Comm.}(1/P) + n \log P \right) \nonumber \\
\tilde{g}_2(1/P) &=& P \frac{d}{d P} \frac{{\cal S}_1^{\rm Comm.}(1/P)}{2} \; . \label{eq:g2TDefin}
\eq
Here we have substituted the expression for ${\cal M}(\la)$ given in (\ref{eq:MGFCoulomb}), and used the fact that $t = n/P$.


From (\ref{eq:S2Laguerre}), we compute
\bq \label{eq:gT1}
\tilde{g}_1(1/P) = - n \left[ 1 + \frac{1}{2 P} \left( 1 - \sqrt{1 + 4 P} \right) \right]
\eq
which, after applying a Taylor expansion of $\sqrt{ 1 + 4 P }$ around zero, gives
\bq
\tilde{g}_1(1/P) = n \sum_{k=1}^\infty  B_{k,0} P^k  \; .
\eq
To leading order in $n$, this agrees precisely with $g_1(t)$ in (\ref{eq:gkDefn}), thereby establishing that the Coulomb fluid method gives the \emph{exact} value of the mean mutual information for large $n$.

From (\ref{eq:S1Laguerre}), we compute
\bq
{\cal S}_1^{\rm Comm.} (1/P) =  - P^2 + 4 P^3 - \frac{29}{3} P^4 + 52 P^5 - \frac{562}{3} P^6 + 680 P^7 - \cdots
\eq
giving
\bq
\tilde{g}_2 (1/P) =  - P^2 + 6 P^3 - 29 P^4 + 130 P^5 - 562 P^6 + \cdots
\eq
This agrees precisely with $g_2 (t)$ in (\ref{eq:gkDefn}) to leading order in $n$, thereby confirming that the Coulomb fluid gives the correct  asymptotic variance, in addition to the correct asymptotic mean.

In summary, we have
\bq
\log {\cal M}(\la) &=&  \la \left( \mu_{\rm Coulomb} + \frac{1}{n} \left( \frac{1}{3} P^3 - \frac{5}{2} P^4 + \frac{70}{5} P^ 5 + \cdots \right) + {\rm O} \left( \frac{1}{n^3} \right) \right) \nonumber \\
&+& \frac{\la^2}{2!} \left( \sigma^2_{\rm Coulomb} + \frac{1}{n^2} \left(\frac{5}{2}P^4 - 32 P^5 + \frac{806}{3}P^6 + \cdots \right) + {\rm O} \left( \frac{1}{n^4}  \right)  \right)  \nonumber \\
&+& \frac{\la^3}{3!} \, \left( \frac{1}{n} \left( 2 P^3 - 18 P^4 + 114 P^5 + \cdots \right) + \frac{1}{n^3} \left( 18 P^5 - 350 P^6 + \cdots  \right) + {\rm O} \left( \frac{1}{n^5} \right) \right)  \nonumber \\
&+& \sum_{k=4}^\infty \frac{\la^k}{k!}  {\rm O} \left( \frac{1}{ n^{k-2} } \right) \; .
\eq
Here, $\mu_{\rm Coulomb}$ and $\sigma^2_{\rm Coulomb}$ represent the mean and variance respectively, calculated based on the Coulomb fluid method.  All other terms represent correction terms, which essentially account for the deviation of the mutual information distribution from Gaussian for finite values of $n$.




\subsection{Refined Analysis for All $P$}

We now present a more refined analysis, which does not require a power series representation of $1/t$. This analysis is based on evaluating a non-perturbative summation of the perturbation series in $P$, obtained from the terms of $G_n(t)$ which are linear in $\la$, and to leading order in $n$. For this purpose, we substitute the series representation (\ref{eq:GnSeries}) into (\ref{eq:PEquation}).

\subsubsection{Analysis of the Mean}
We start by considering the first cumulant (i.e., the mean). By comparing the coefficients of $\la^k$ on the l.h.s. and r.h.s., we find that
the coefficient of $\la$ is identically equal to $0$, and the coefficient of $\la^2$ satisfies:
\bq
(g_1)^2-4n^2 g_1'-2[t+ 2 n]g_1g_1'+[t^2 + 4 n t ](g_1')^2-t^2(g_1'')^2=0 . \nn
\eq
Note that
$$
t=\frac{n}{P}=:nT.
$$
We have introduced $T$ so that the differential equation does not get too complicated.
After this change of variable in $t$ to $nT$, without introducing further notation in place of
$g_1$, the differential equation becomes
\bq (g_1)^2- 4 n g_1'-2(T+2) g_1g_1'+[T^2+4T
](g_1')^2-\frac{T^2(g_1'')^2}{n^2}=0 \nn \eq where now $'$ denotes
the derivative with respect to $T$.  Letting
\bq \label{eq:g1Expand}
g_1 (n T) =n Y(T)
\eq
we find that
$$
Y^2-4 Y'-2(T+2 )YY'+[T^2+4T ]Y'^2-\frac{T^2(Y'')^2}{n^2}=0.\eqno(Y)
$$
Taking $n\to\infty,$ $Y$ is seen to satisfy
$$
Y^2-4 Y'-2(T+2 )YY'+[T^2+4T ]Y'^2=0.\eqno(Y0)
$$
Now consider $\tilde{g}_1(T)$, the term analogous to $g_1(nT)$ but derived based on the Coulomb fluid.  From (\ref{eq:gT1}),
\bq
\tilde{g}_1(T) =  n Y_0 (T) 
\eq
where
\bq \label{eq:Y0Expand}
Y_0(T) 
= -\frac{4+T-\sqrt{T(4+T)}}{4+T+\sqrt{T(4+T)}} \; .
\eq
This expression is found to satisfy Eq.\ (Y0) identically, thus confirming that the Coulomb fluid approach gives the exact value for the mean of the mutual information to leading order in $n$ for \emph{all values of P}.  Note that this result is stronger than that derived in the previous section, since it applies even for values of $P$ (or values of $1/t$) for which a formal convergent power series in $1/t$ does not exist.

To compute the $1/n$ correction to the previously obtained $Y$,
we substitute 
\bq \label{eq:YDefn}
Y(T) = Y_0(T) + \frac{1}{n^2} Y_1(T) + {\rm O}
\left( \frac{1}{n^4} \right)
\eq
into Eq.\ (Y), and then obtain
$Y_1$ by setting the coefficient of $1/n^2$ equal to $0$. This gives
$$
2Y_0Y_1-2(2+T)Y_1Y_0'-4Y_1'-2(2+T)Y_0Y_1'+2T(4+T)Y_0'Y_1'-T^2\:Y_0''=0
$$
which is a first order linear equation in $Y_1$. However, the coefficient of $Y_1'$ vanishes identically
when we make use of $Y_0$ from (\ref{eq:Y0Expand}). The solution of the algebraic equation reads:
$$
Y_1(T)=-\frac{1}{\sqrt{T}(4+T)^{5/2}} \; .
$$

With these results, we can compute the asymptotic mean of the mutual information, including the first-order corrections, as
\bq
\kappa_1 &=& - \int_0^P \frac{g_1 (n/z) }{z} d z \nonumber \\
&=& \mu_{\rm Coulomb} + \frac{1}{n} \mu_{\rm Correction} + {\rm O}\left(\frac{1}{n^3} \right)
\label{eq:MuColomb}
\eq
where $\mu_{\rm Coulomb}$ is the mean value computed via the Coulomb fluid, which from (\ref{eq:S2Laguerre}) and (\ref{eq:MeanVar_Defn}) with $\beta = 0$ is given by
\bq \label{eq:CoulombMean}
\mu_{\rm Coulomb} = n \left[ 2  \log \left( \frac{ 1 + \sqrt{ 1 + 4 P } }{2} \right)
- \frac{ \left( 1 - \sqrt{1 +  4 P } \right)^2 }{ 4 P } \right] \;
\eq
and $\mu_{\rm Correction}$ is the first order correction term given by
\bq \label{eq:CorrectionMean}
\mu_{\rm Correction} = - \int_0^P \frac{Y_1(1/z)}{z} d z = \frac{1}{12} \left[  \frac{ 1 + 6 P + 6 P^2 }{ (4 P + 1)^{3/2}} -  1 \right] \; .
\eq
Note that if we expand this expression for $\kappa_1$ around  $P = 0$, the series matches precisely with (\ref{eq:KappaDefn}) as expected.


\subsubsection{Analysis of the Variance}

Now we consider the second cumulant.   To this end, after substituting (\ref{eq:GnSeries}) into (\ref{eq:PEquation}) and setting the coefficient of $\la^3$ equal to $0$, we get
\bq
&&2g_1'(1+g_1')(g_1-tg_1')-2n^2(1+2g_1')g_2'\nn\\
&& \hspace*{1cm} + \{-g_1+(2n+t)g_1'\}\{-g_2+g_1'+(2n+t)g_2'\}-t^2g_1''\;g_2''=0.
\eq
Once again, applying the change of variable $t=nT$, but without introducing new notation for $g_1$ and $g_2$,  we find
\bq \label{eq:Diffg2}
&&\frac{2}{n}g_1'\left(1+\frac{g_1'}{n}\right)\left(g_1-Tg_1'\right)-2n\left(1+2\frac{g_1'}{n}\right) g_2' \nn\\
&& \hspace*{1cm} + \{-g_1+(T+2)g_1'\}\{-g_2+g_1'/n+(2+T)g_2'\}-\frac{1}{n^2}T^2\:g_1''g_2''=0
\eq
with $'$ denoting $d/dT$.

Let
\bq \label{eq:g2Expand}
g_2 (n T) = Z_0 (T) + \frac{1}{n^2} Z_1 (T) + {\rm O} \left(
\frac{1}{n^4} \right) \; .
\eq
We first compute $Z_0 (T)$, the leading order term in $n$.  To this end, substituting (\ref{eq:g2Expand}) along with
(\ref{eq:g1Expand}) and (\ref{eq:YDefn}) into (\ref{eq:Diffg2}), and then keeping only the leading order terms in $n$ (the terms which
are linear in $n$), we obtain
%
\bq \label{eq:g2Eqn}
&& 2Y_0'(1+Y_0')(Y_0-T\:Y_0')-2(1+2Y_0') Z_0' \nn \\
&& \hspace*{1cm} +\{-Y_0+(T+2)Y_0'\}\{-Z_0+(T+2)Z_0'+Y_0'\}=0 .
\eq
Interestingly, if we plug in the expression for $Y_0(T)$ given in (\ref{eq:Y0Expand}), we find that the coefficient of $Z_0'$ is identically equal to $0$.
Thus, (\ref{eq:g2Eqn}) reduces to a simple algebraic equation, whose solution is:
$$
 Z_0(T) = -\frac{1}{2}+\frac{1}{2}\sqrt{\frac{T}{4+T}}+\frac{1}{4+T} \, .
$$
After making the substitution $T = 1/P$, it can be verified that this expression matches precisely with $\tilde{g}_2(T)$ in (\ref{eq:g2TDefin}), derived
based on the Coulomb fluid method. This result confirms that the Coulomb fluid approach gives the exact value for the variance of the mutual
information to leading order in $n$ for \emph{all values of P}. Again, this result is stronger than that
derived in the previous section, since it applies for values of $P$ (or values of $1/t$) for which a formal
convergent power series in $1/t$ does not exist.

Now consider the correction term, $Z_1(T)$, in (\ref{eq:g2Expand}).  Again we substitute (\ref{eq:g2Expand}) along with
(\ref{eq:g1Expand}) and (\ref{eq:YDefn}) into (\ref{eq:Diffg2}). In this case, however, we extract only the terms of order $1/n$, which gives
a rather large first order equation in the unknown $Z_1(T)$.
%
%
Fortunately, we find that by plugging in the previously determined equations for $Y_0(T),$ $Y_1(T)$, and $Z_0(T)$, the coefficient of $Z_1'(T)$ vanishes
identically, and so we are left with a {\it linear equation} in $Z_1(T)$. This is easily solved and we find
\bq
Z_1(T)=-\frac{8+16T+20T^2+5T^3+6\sqrt{T(4+T)}+10T^{3/2}\sqrt{4+T}+5T^{5/2}\sqrt{4+T}}
{T(4+T)^4\left(2+4T+T^2+2\sqrt{T(4+T)}+T^{3/2}\sqrt{4+T}\right)} \; .
\eq

With these results, we can compute the asymptotic variance of the mutual information, including the first-order corrections, as
\bq
\kappa_2 &=& - 2 \int_0^P \frac{g_2 (n/z) }{z} d z \nonumber \\
&=& \sigma^2_{\rm Coulomb} + \frac{1}{n^2} \sigma^2_{\rm Correction} + {\rm O}\left(\frac{1}{n^4} \right)
\label{eq:kappa2}
\eq
where $\sigma^2_{\rm Coulomb}$ is the variance computed via the Coulomb fluid, which from (\ref{eq:S1Laguerre}) and (\ref{eq:MeanVar_Defn}) with $\beta = 0$ is given by
\bq
\sigma^2_{\rm Coulomb} = 2 \log \left[ \frac{ (4 P + 1)^{1/4} + (4 P + 1)^{-1/4} }{2} \right] .
\eq
and $\sigma^2_{\rm Correction}$ is the first order correction term given by
\bq  \label{eq:CorrectionVar}
\sigma^2_{\rm Correction} = - 2 \int_0^P \frac{Z_1(1/z)}{z} d z =  \frac{1}{12} \left[ 1 - \frac{ 8 P^3 ( 1 - 3 P )}{(4 P + 1 )^3 }  -
\frac{ 12 P^3 + 30 P^2 + 10 P + 1 }{(4 P + 1 )^{5/2} } \right] .
\eq
Note that if we expand this expression for $\kappa_2$ around  $P = 0$ we get a series which matches precisely with (\ref{eq:KappaDefn}), as expected.

\subsubsection{Analysis of the Third Cumulant}

Now consider the third cumulant.  After substituting (\ref{eq:GnSeries}) into (\ref{eq:PEquation}), setting the coefficient of $\la^4$ equal to $0$, and then
going through the same procedure as before (i.e., applying $t=nT$) we get
\bq
&&4\left(1+2\frac{g_1'}{n}\right)(g_1-Tg_1')\frac{g_2'}{n} + \frac{4}{n}g_1'\left(1+\frac{g_1'}{n}\right)(g_2-Tg_2')\nn\\
&& \hspace*{0.5cm} + \left(-g_2+\frac{g_1'}{n}+(2+T)g_2'\right)^2 +
2(-g_1+(T+2)g_1')\:\left(-g_3+\frac{g_2'}{n}+(2+T)g_3'\right)\nn\\
&& \hspace*{1cm} -4n^2\left[\left(\frac{g_2'}{n}\right)^2+\left(1+2\frac{g_1'}{n}\right)\frac{g_3'}{n}\right]-
\frac{T^2}{n^2}(g_2''+2g_1''g_3'')=0
\eq
where the derivatives are taken with respect to $T$.  Now substitute (\ref{eq:g2Expand}) along with
(\ref{eq:g1Expand}) and (\ref{eq:YDefn}) and
%
\bq
g_3(T) = \frac{1}{n} X_0(T) + \frac{1}{n^3} X_1(T) + {\rm O}\left( \frac{1}{n^5} \right),
\eq
where $X_0(T)$ and $X_1(T)$ are $n$ independent. We find that the highest order term in $n$ is in fact $n$ independent,
and using the previously determined equations for $Y_0(T)$, $Y_1(T)$, $Z_0(T)$, and $Z_1(T)$, once again we find that the coefficient of $X_0'(T)$ vanishes
identically. The resulting linear equation in $X_0(T)$ has solution
\bq
X_0(T)= \frac{1}{2}
\left( \frac{2 \sqrt{T} }{ (4 + T)^{5/2} } + \frac{T}{(4+T)^{2}} - \frac{\sqrt{T} }{(4+T)^{3/2}}\right) .
\eq
Proceeding in a similar manner, we find
\bq X_1(T)= && - \frac{16} {T^{3/2}(4+T)^{11/2}} - \frac{30}{ T^{1/2}(4+T)^{11/2} } - \frac{4 \sqrt{T}}{ (4+T)^{11/2}} + \frac{9 T^{3/2}}{ (4+T)^{11/2}} - \frac{T^{5/2}}{ 2 (4+T)^{11/2}} \nonumber \\
&& + \frac{8 }{ T (4+T)^{5}} + \frac{10 }{ (4+T)^{5}} - \frac{10 T }{ (4+T)^{5}} + \frac{ T^2 }{ 2(4+T)^{5}} \; .
\eq
With these results, we can compute the asymptotic third cumulant of the mutual information as
\bq
\kappa_3 &=& - 3! \int_0^P \frac{g_3 (n/z) }{z} d z \nonumber \\
&=& \frac{1}{n} \kappa_{3, {\rm Correction \, A}} + \frac{1}{n^3}
\kappa_{3, {\rm Correction \, B}} + {\rm O}\left(\frac{1}{n^5}
\right) \label{eq:3Cum} \eq
where
\bq \label{eq:Kappa3A}
\kappa_{3, {\rm Correction \, A}} &=& - 6  \int_0^P \frac{X_0(1/z)}{z} d z \nonumber \\
&=& 1 - \frac{1}{\sqrt{1 + 4 P}} - \frac{ 3 P }{ 1 + 4P } + \frac{ P }{ (1 + 4P)^{3/2} }
\eq
and
\bq
\kappa_{3, {\rm Correction \, B}} &=& - 6  \int_0^P \frac{X_1(1/z)}{z} d z \nonumber \\
&=&  \frac{ 16 P^6}{ (4 P + 1)^4} - \frac{48 P^6}{ (4 P + 1)^5 } - \frac{380 P^5}{3 (4 P + 1)^5 } - \frac{610 P^4}{3 (4 P + 1)^5 } - \frac{323 P^3}{3 (4 P + 1)^5 } \nonumber \\
&&- \frac{80 P^2}{3 (4 P + 1)^5 } - \frac{10 P}{3 (4 P + 1)^5 } - \frac{1}{6 (4 P + 1)^5 }
+ \frac{42 P^5 }{(4 P + 1)^{9/2} } + \frac{93 P^4 }{ (4 P + 1)^{9/2} } \nonumber \\
&&  + \frac{71 P^3 }{ (4 P + 1)^{9/2} }  + \frac{21 P^2 }{ (4 P +
1)^{9/2} }  + \frac{3 P }{ (4 P + 1)^{9/2} }  + \frac{1}{6 (4 P +
1)^{9/2} } \; . \eq
%
%
%
%
Note that if we expand this around  $P = 0$ we get a series which matches precisely with (\ref{eq:KappaDefn}).

The correction terms to the mean, variance, and third cumulant are
plotted in Figure 6.  In particular, the ``Mean'' curves represent
$$n (E [ I(\mathbf{x}, \mathbf{y}) ] - \mu_{\rm Correction})$$
with $\mu_{\rm Correction}$ given by (\ref{eq:CorrectionMean}); the
``Variance'' curves represent
$$n^2 ({\rm Var} [ I(\mathbf{x},
\mathbf{y}) ] - \sigma^2_{\rm Correction})$$ with $\sigma^2_{\rm
Correction}$ given by (\ref{eq:CorrectionVar}); and the
``$\kappa_3$'' curves represent
$$n (\kappa_3 [ I(\mathbf{x},
\mathbf{y}) ] - \kappa_{3, {\rm Correction \, A}})$$ with
$\kappa_{3, {\rm Correction \, A}}$ given by (\ref{eq:Kappa3A}).
Here, the mean $E [I(\mathbf{x}, \mathbf{y})]$, variance ${\rm Var}
[ I(\mathbf{x}, \mathbf{y}) ]$, and third cumulant $\kappa_3 [
I(\mathbf{x}, \mathbf{y}) ]$ of the mutual information were
calculated using numerical integration procedures in Maple.  From
the figure, we can make some interesting observations.  First, it is
clearly evident that for low SNR $P$, all three correction terms
converge very quickly to zero, confirming the near-Gaussian behavior
of the distribution even for small $n$, as seen previously in Figure
2. However, as $P$ increases, all three correction terms become much
more significant. This, in turn, leads to a larger deviation from
Gaussian, which again is in line with the numerics presented in
Figure 2. It is also particularly interesting to note that the
correction terms for the higher-order moments tend to \emph{deviate
quicker} than the lower order moments. To understand this
phenomenon, it is useful to look closer at the correction terms as
$P$ grows large.  This is the focus of the next subsection.


\begin{figure}[htp]
\centering

\subfigure[$P = 0$ dB]{
\includegraphics[width=0.46\columnwidth]{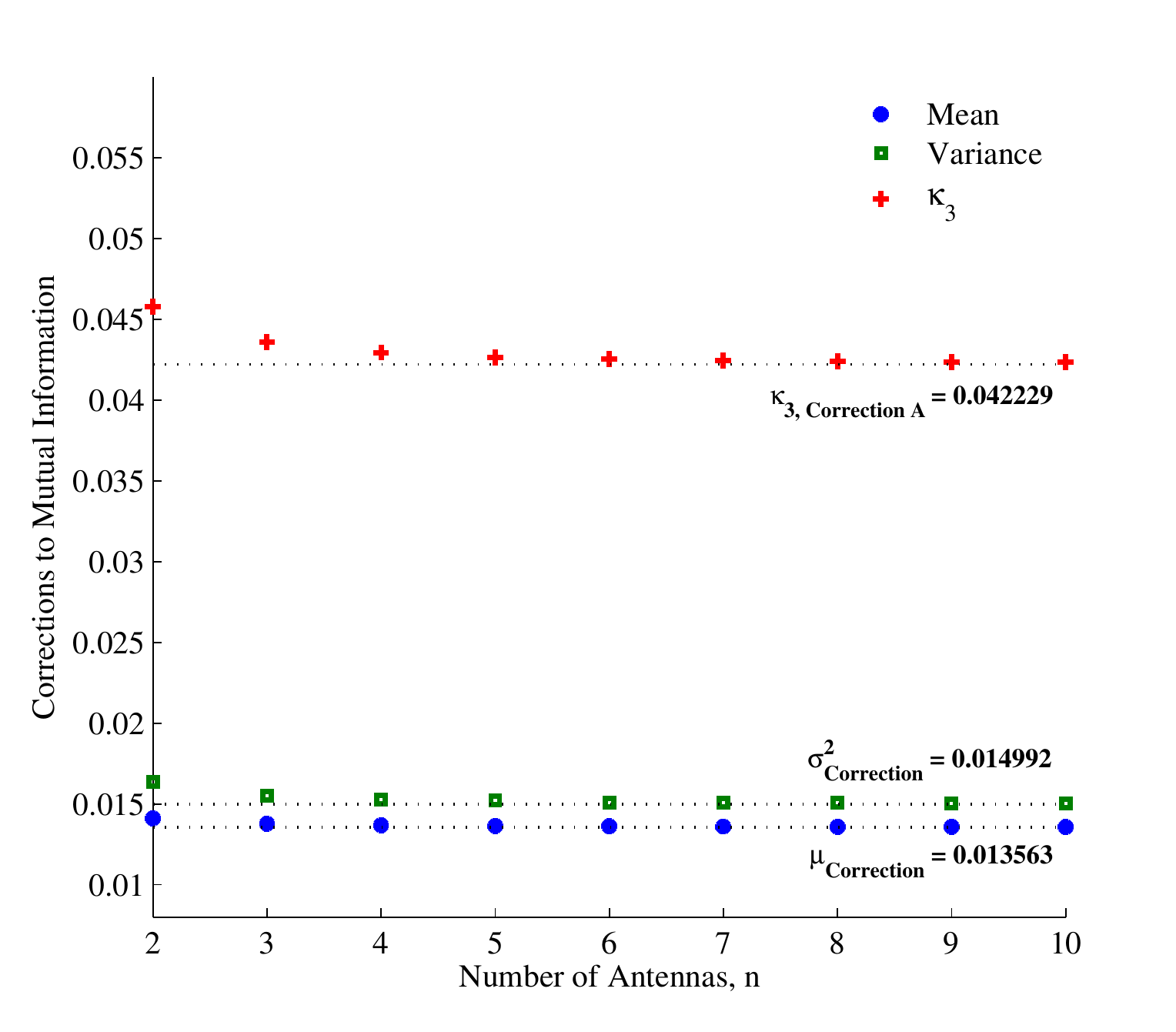}
\label{fig:Corr1} } \hspace{.1in} \subfigure[$P = 5$ dB]{
\includegraphics[width=0.46\columnwidth]{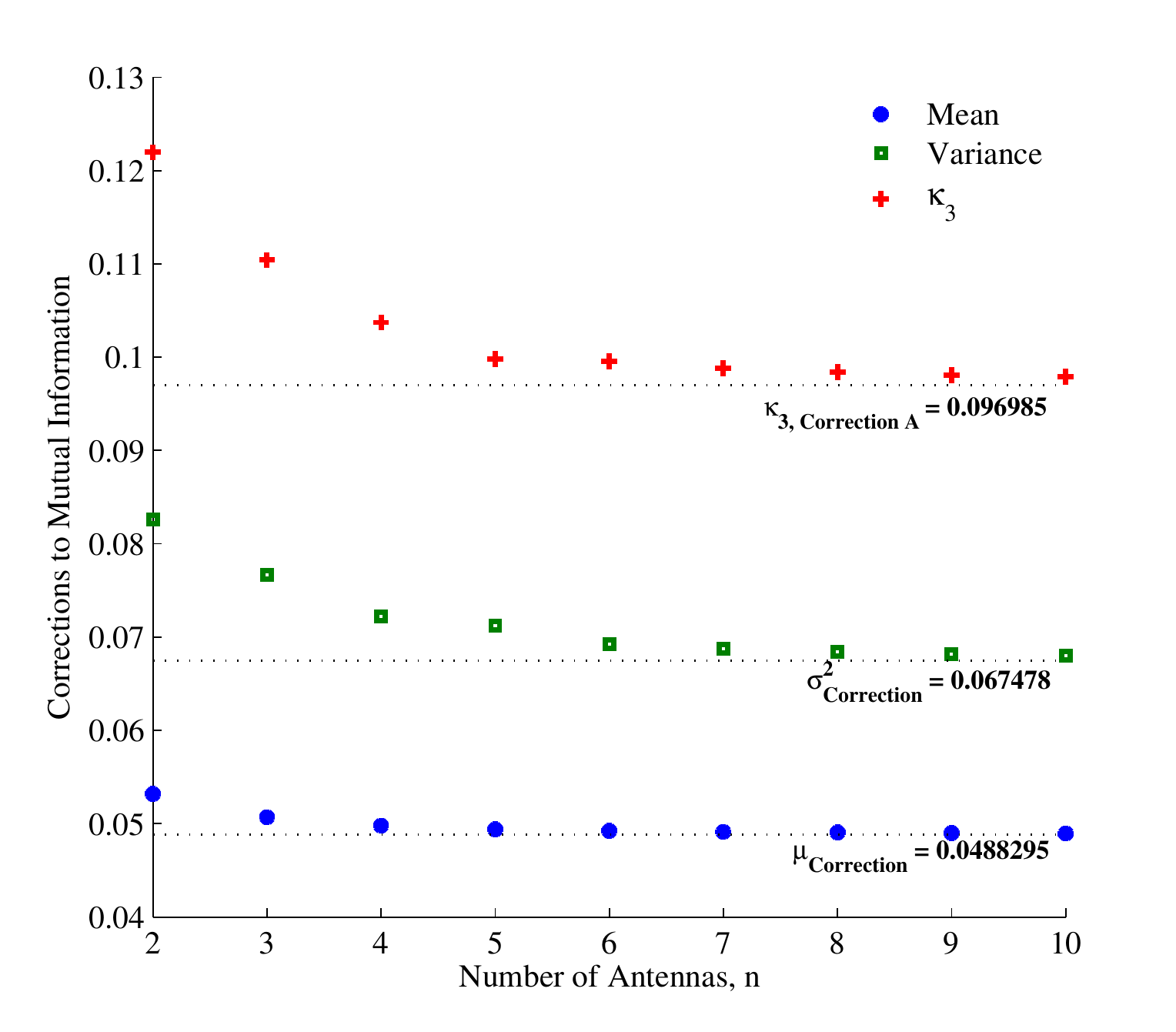}
\label{fig:Corr2} } \hspace{.1in} \subfigure[$P = 10$ dB]{
\includegraphics[width=0.46\columnwidth]{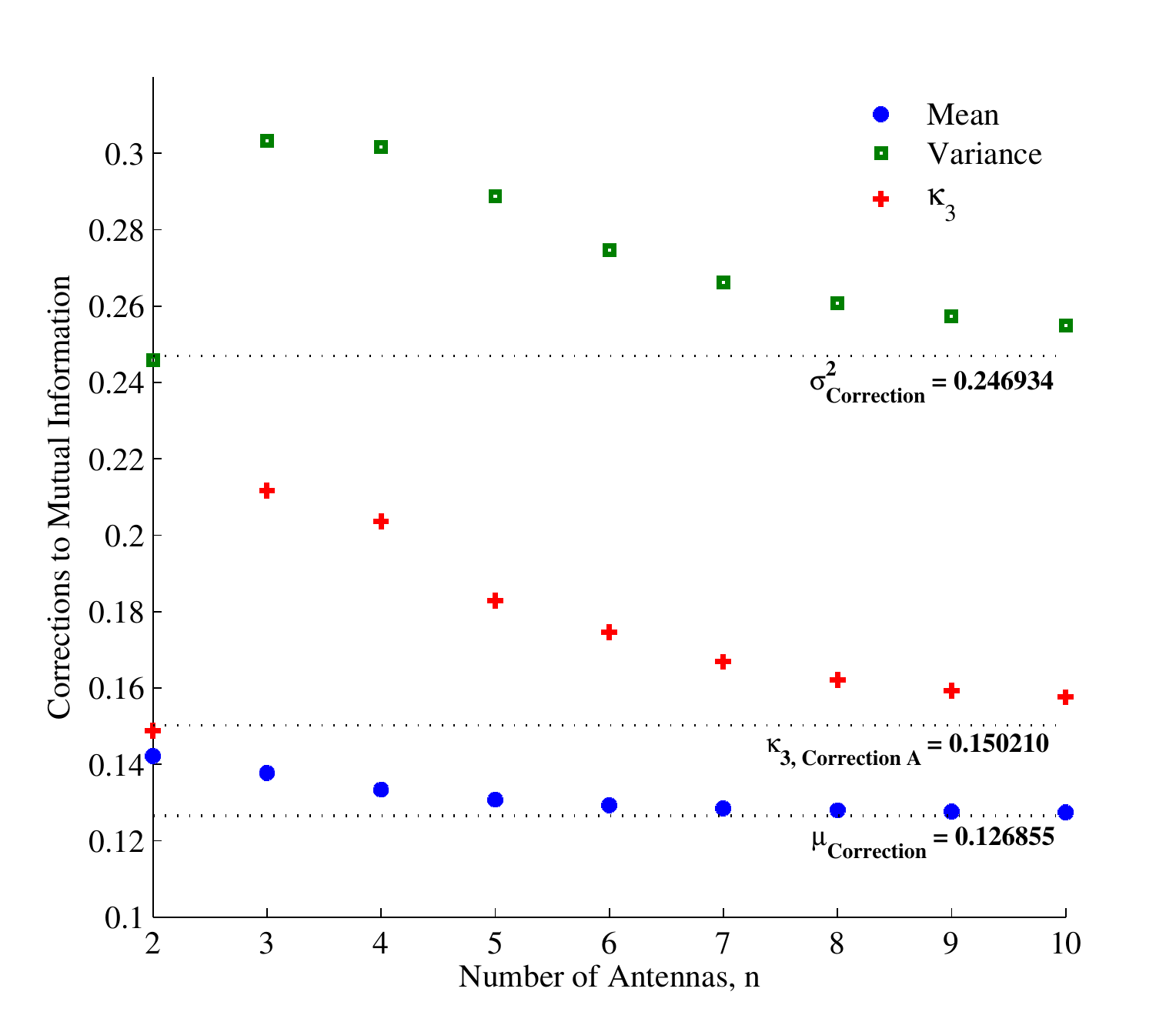}
\label{fig:Corr3} } \hspace{.1in} \subfigure[$P = 15$ dB]{
\includegraphics[width=0.46\columnwidth]{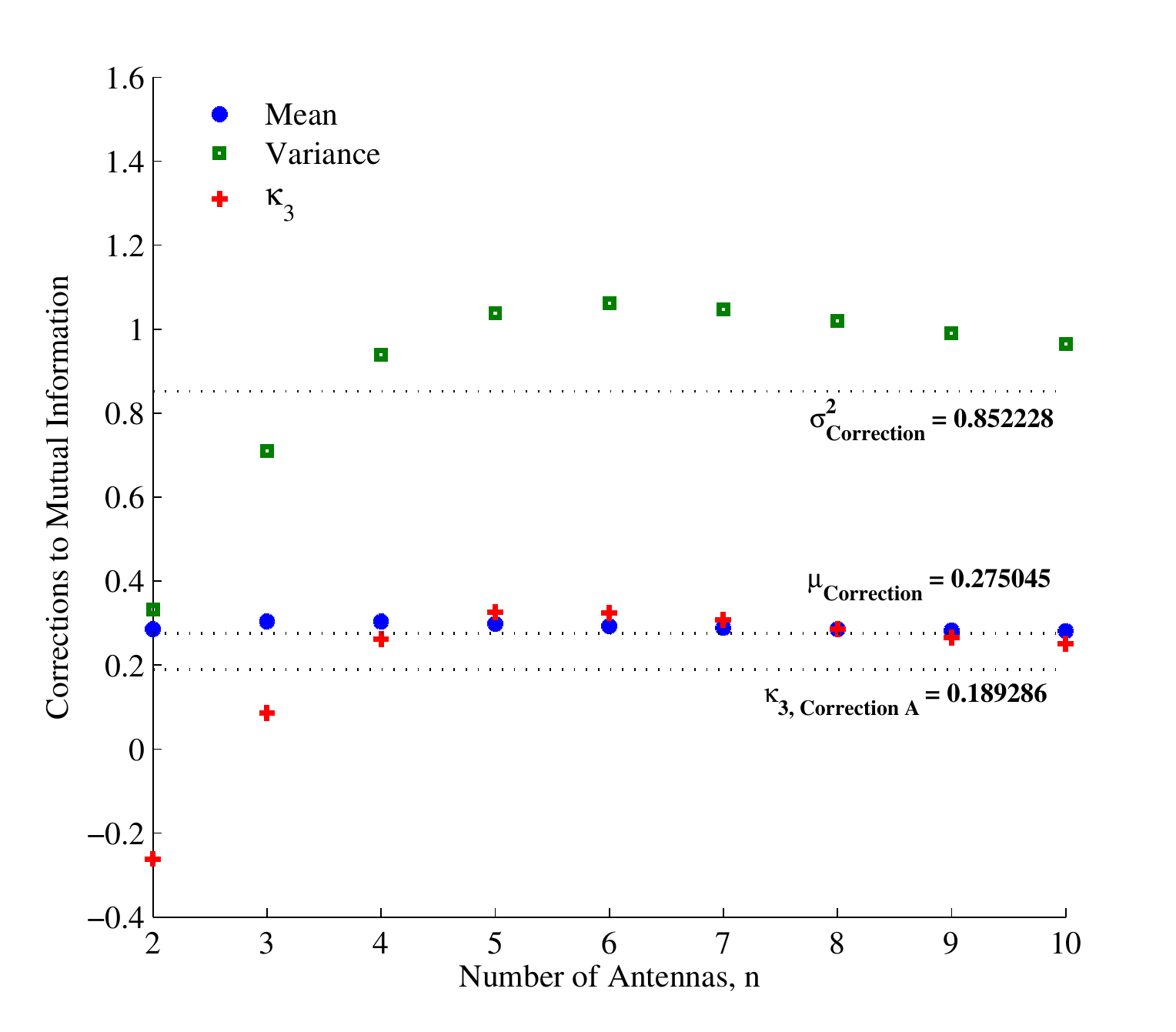}
\label{fig:Corr1} }

\label{fig:MeanVarCorrections} \caption{Correction terms to the mean
and variance of the mutual information for the single-user MIMO
scenario (deformed Laguerre case). Results are shown for $n_r = n_t
= n$.}
\end{figure}

\subsection{Analysis at Large $P$}

For large $P$, the asymptotic mean (\ref{eq:MuColomb}), variance (\ref{eq:kappa2}), and third cumulant (\ref{eq:3Cum}) become
%
\bq
&&\mu \sim  n \log P + \frac{\sqrt{P}}{n} \left( \frac{1}{16}
\right) \nonumber \\
&&\sigma^2 \sim   \frac{ \log (2 P) }{2} + \frac{P}{n^2} \left(
\frac{3 }{8} \right) \nonumber \\
&&\kappa_3 \sim \frac{1}{n} \left( \frac{1}{4} \right) + \frac{P^2}{n^3} \left( \frac{1}{16} \right) \; .
 \eq
From these expressions, we notice that in all three cases, the
correction terms (i.e., the second) are increasing in $P$.  Since
these terms give corrections to the Coulomb fluid Gaussian
approximation, this implies that as $P$ increases, the Coulomb fluid
approximation loses accuracy, and the distribution deviates away
from Gaussian, as our previous numerical results have indicated. We
can also gain insights into the \emph{relative} effect of $P$ on
each of the correction terms. Specifically, for the mean, we see
that for the correction term to dominate the leading term, it must
be at least ${\rm O}(n^4)$.  The variance, on the other hand, must
only be ${\rm O}(n^2)$, whist the third cumulant is even more
sensitive to $P$, and only requires $P$ to be $O(n)$. This confirms
that the higher moments are more sensitive to the variations in $P$,
and moreover, it allows insights into the Gaussianity of the mutual
information distribution in terms of \emph{both} $n$ and $P$.

We would like to mention, however, that some caution should be
exercised in interpreting these results. In particular, since our
analytical results which are asymptotic expansions in $n$ are based
on the Painlev\'{e} $V$ representation of the moment generating
function, they are formally valid for large $n$ but \emph{fixed} $T$
or $P$. Therefore, it is to be expected that when $P$ increases to
the point where the corrections are comparable with and/or overwhelm
the main terms, the Coulomb fluid approximation will break down.
Providing a rigorous investigation of the situation where both $n$
and $P$ increase simultaneously is the subject of on-going work.

\subsection{Asymptotic Recurrence Coefficients}

In addition to deriving the asymptotic moments, having determined the large $n$ expansion of $G_n(nT)$, we can also compute the recurrence coefficients,
$\al_n$ and $\bt_n$ for $\al=0$ and large $n$ with $t=nT$. For this purpose, note that $\al_n$
is easily related to $R_n$ (see Eq. (\ref{eq:S2Diff2a})) and $R_n$ is in turn expressed in terms of
$H_n$ (or equivalently $G_n$) and its derivatives (see Eq. (\ref{eq:PVyt})), while $\bt_n$ related to $H_n$
and its derivative through Eq. (\ref{eq:AlphaDiff}).

From our previous analysis, we have learned that for $t=nT$ and $n$ large, $G_n(nT)$ has the expansion
\bq
G_n(nT)&=&\la\left(nY_0(T)+\frac{Y_1(T)}{n}+...\right)
+\la^2\left(Z_0(T)+\frac{Z_1(T)}{n^2}+...\right)\nn\\
&+&\la^3\left(\frac{X_0(T)}{n}+\frac{X_1(T)}{n^3}+...\right) + ... \nn
\eq
From this series, after a straightforward if lengthy computation we find that
\bq
\al_n(nT)
&=&\left(2-\frac{T}{2}\right)\:n
+\la\left(1-\sqrt{\frac{T}{4+T}}\right)\nn\\
&+&\la\frac{\sqrt{T}}{(4+T)^2}\left[\sqrt{4+T}+\la(\sqrt{4+T}-\sqrt{T})\right]
\:\frac{1}{n}+{\rm O}\left(\frac{1}{n^2}\right),
\eq
and
\bq
\bt_n(nT)&=&n^2+\frac{4\la}{4+T+\sqrt{T(4+T)}}\:n\nn\\
&-&\frac{\left[2T+(2+T)(4+T)(-1+\sqrt{T}/\sqrt{4+T})\right]\la^2}{2(4+T)^2}\nn\\
&+&\la\frac{6+4T+\la^2(2T^2+T^3-T^{5/2}\sqrt{4+T}+2T)}{\sqrt{T}(4+T)^{7/2}}
\:\frac{1}{n}+{\rm O}\left(\frac{1}{n^2}\right).
\eq

\section{Concluding Remarks}

This paper has focused on the computation of Hankel determinants
which arise in the information-theoretic study of MIMO communication
systems.  We considered two practical scenarios; namely, single-user
systems, in which case the determinants of interest are generated
from a certain deformed Laguerre weight, and multi-user systems, in
which case they are generated from a deformed Jacobi weight.  In
both cases, we derived new exact characterizations of the Hankel
determinants in terms of classical Painlev\'{e} differential
equations, as well as closed-form expressions which are formally
valid for large matrix dimensions, but give accurate results for
small dimensions also.

We also demonstrated that, in contrast to most prior work dealing with MIMO information theory, our exact and approximate results can be used together to derive explicit formula for the mean, variance, and higher order cumulants of the mutual information, as well as their corresponding correction terms. This is important, since it allows one to gain insights into the Gaussianity of the mutual information distribution in terms of key system parameters, such as the number of antennas and the signal-to-noise ratio.  For example, by focusing on the single-user MIMO scenario, we showed that the mutual information distribution may deviate strongly from Gaussian when the signal-to-noise ratio is relatively high, and we explicitly captured this effect through the cumulant formulas.

\noindent
\setcounter{equation}{0}

\section{Appendix: Some Relevant Integral Identities}

For the Coulomb fluid derivations, we will require the following integrals:

\bq
\label{eq:Int1}
\int_{a}^{b}\frac{\log(x+t)}{\sqx}dx &=& 2 \pi\log \left( \frac{ \sqrt{t+a}+\sqrt{t+b}}{2} \right) \\
\label{eq:Int2}
\int_{a}^{b}\frac{\log(x+t)}{\sqx(x+t)}dx
&=& -\frac{2 \pi}{\sqt} \log \left( \frac{1}{2 \sqrt{t+a}} + \frac{1}{2 \sqrt{t+b}} \right)  \\
\label{eq:Int3}
\int_{a}^{b}\frac{\log(x+t)}{\sqx\:x}dx  &=& \frac{\pi}{\sqrt{ab}}
 \log \left( \frac{ (\:\sqrt{a b} + \sqrt{(t+a)(t+b)}\:)^2 - t^2  }{ ( \sqrt{a} + \sqrt{b})^2} \right) \\
\label{eq:Int4a}
\int_{a}^{b}\frac{x \log(x+t)}{\sqx }dx
&=& \pi  \frac{(\sqrt{a+t}-\sqrt{b+t})^2}{2} \nn \\
&& + \pi \frac{a+b}{2} \log \left( \frac{ (\:\sqrt{(a+t)(b+t)}+t\:)^2- a b}{ 4 t}
  \right)
 \label{eq:Int4a}  \hspace*{1cm}  \\
\label{eq:Int4}
\int_{a}^{b}\frac{\log(x+t)}{\sqx\:(x-1)}  dx &=& \pi
\frac{\log \left(\frac{ (\:\sqrt{1-a} + \sqrt{1-b}\:)^2  }
{ (t+1)^2 - (\:\sqrt{(t+a)(t+b)}-\sqrt{(1-a)(1-b)}\:)^2 } \right) }{\sqrt{(1-a)(1-b)}} \\
\label{eq:Int5}
\int_{a}^{b}\frac{\log(1-x)}{\sqx}dx &=& 2 \pi \log \left(\frac{\sqrt{1-a} + \sqrt{1-b}}{2} \right) \\
\label{eq:Int6}
\int_{a}^{b}\frac{\log(1-x)}{\sqx\:x}dx &=& \frac{\pi}{\sqrt{ab}}
\log \left(\frac{ 1 - (\:\sqrt{a b} - \sqrt{ (1-a)(1-b)}\:)^2 }{(\:\sqrt{a} + \sqrt{b}\:)^2} \right) \\
\label{eq:Int7}
\int_{a}^{b}\frac{\log(1-x)}{\sqx\:(x-1)}dx &=&  \frac{2 \pi}{\sqrt{(1-a)(1-b)}}
\log \left( \frac{1}{ 2 \sqrt{1-a}} + \frac{1}{ 2 \sqrt{1-b}} \right) \\
\label{eq:Int8}
\int_{a}^{b}\frac{\log(1-x)}{\sqx\:(x+t)}  dx &=& \pi  \frac{\log
\left( \frac{ (t+1)^2 - ( \sqrt{(t+a)(t+b)}-\sqrt{(1-a)(1-b)} )^2 }{ (\:\sqrt{t+a} + \sqrt{t+b}\:)^2  }  \right) }
{\sqrt{(t+a)(t+b)}}
\eq

\bq
 \label{eq:PInt1}
{\cal P} \int_{a}^{b}\frac{\sqy}{(y-x)(y+t)}dy= \pi\left(\frac{\sqt}{x+t}-1\right)
\eq

\bq \label{eq:PInt2}
{\cal P} \int_{a}^{b}\frac{\sqy}{(y-x)(y-1)}dy=\pi\left(\frac{\oab}{1-x}-1\right)
\eq

\noindent
Before proving these results, we state for reference the following identities:
\bq
\int\frac{dx}{x\sqrt{ax^2+bx+c}}&=&-\frac{1}{\sqrt{c}}\:\log\frac{2\sqrt{c}\sqrt{ax^2+bx+c}+bx+2c}{x},\quad c>0, \label{eq:Simple0} \\
\int\frac{dx}{\sqrt{(t+ax)(t+bx)}}&=&\frac{1}{\sqrt{ab}}\log \left[2\sqrt{ab}\sqt+2abx+(a+b)t\right] \label{eq:Simple1} \\
\int \frac{ dx}{x^2 \sqrt{(x+a)(x+b)}}&=&  \frac{a+b}{2 (a b)^{3/2}} \log \left( \frac{ (\sqrt{(x+a)(x+b)}+\sqrt{a b})^2-x^2}{  x} \right)  \nn \\
&& \hspace*{2cm} - \frac{ \sqrt{(x+a)(x+b)}}{a b x}    \label{eq:Simple5} \\
\log(A+B)&=&\log A+ \int_{0}^{1} \frac{Bd\la}{A+\la B} \label{eq:LogParam}\\
\int_{a}^{b}\frac{dx}{(x+t)\sqx}&=&\frac{\pi}{\sqt} \label{eq:Simple2} \\
\int_{a}^{b}\frac{dx}{ \sqx }&=& \pi  \label{eq:Simple3} \\
\int_{a}^{b}\frac{x dx}{\sqx}&=& \pi \frac{a+b}{2} \label{eq:Simple4} \\
{\cal P} \int_{a}^{b}\frac{dy}{(x-y)\sqrt{(b-y)(y-a)}}&=&0  \label{eq:PIntSimple}
\eq

We start with some brief remarks concerning the identities (\ref{eq:Simple0})--(\ref{eq:PIntSimple}), before focusing on the proofs of the main results (\ref{eq:Int1})--(\ref{eq:PInt2}).
The integral (\ref{eq:Simple0}) is \cite[Eq. (2.266)]{grad}, (\ref{eq:Simple1}) is a minor variation of \cite[Eq. (2.261)]{grad},
and (\ref{eq:Simple5}) follows from \cite[Eq. (2.269.2)]{grad}. Note also that the r.h.s. of equation (\ref{eq:LogParam})
is
$$
\log A+ \int_{0}^{B}\frac{dt}{A+t}
$$
which obviously equals the l.h.s. To give an indication how (\ref{eq:Simple2})--(\ref{eq:PIntSimple}) may be proved, first consider the
analytic function
$$
R(z)=\sqrt{(z-a)(z-b)}
$$
defined in the complex plane slit along $[a,b].$ Here we assume that
$$
0<a<b<1
$$
without loss of generality. The branch of $R(z)$ is chosen in such a way that
$$
R(z)\rightarrow z,\quad as \quad\Re\:z\rightarrow \infty.
$$
Let $F(x)$ be defined for $x\in \mathbb{R}$ and extended to $F(z),$
a meromorphic function with poles. Let $\Lambda$ be a ``dog bone''
contour traversed {\it clockwise} above and below the segment
$[a,b],$ where the point of $\infty$ is contained in the {\it
interior} of $\Lambda.$ Keeping in mind that
$$
R_{\pm}(x)=\pm\;j\sqrt{(b-x)(x-a)},\quad x\in(a,b),
$$
where $R_{\pm}(x)$ is defined to be the analytic continuation of $R(z)$ to above and below the segment $(a,b)$,  we have that
$$
\int_{a}^{b}\frac{F(x)dx}{\sqrt{(b-x)(x-a)}}= \frac{j}{2}\int_{\Lambda}\frac{F(z)dz}{\sqrt{(z-a)(z-b)}},
$$
and the r.h.s. of the above equation can evaluated using residue
calculus. The equation (\ref{eq:Simple2}) follows immediately by
computing the residue at $t$, while in computing (\ref{eq:Simple3})
and (\ref{eq:Simple4}) we should keep in mind the contributions from
the residues at $\infty$. To compute the principal value integral
(\ref{eq:PIntSimple}), we first define
$$
g(t):=\int_{a}^{b}\frac{dy}{(y+t)\sqrt{(b-y)(y-a)}}=\frac{\pi}{\sqrt{(t+a)(t+b)}} .
$$
The principal value integral is then evaluated as
$$
-\frac{1}{2}\lim_{\epsilon\to 0}\left[g(-x+j\epsilon)+g(-x-j\epsilon)\right], \quad x\in(a,b).
$$
An easy computation gives (\ref{eq:PIntSimple}).

We now come to the main integrals, (\ref{eq:Int1})--(\ref{eq:PInt2}).  Start by considering (\ref{eq:Int1})--(\ref{eq:Int4}).
Of these, we will explicitly derive (\ref{eq:Int2}); the other integrals are evaluated in a similar way
with the help of the properties (\ref{eq:Simple0})--(\ref{eq:Simple5}). Using (\ref{eq:LogParam})
 along with (\ref{eq:Simple2}), we obtain
\bq
&&\int_{a}^{b}\frac{\log(x+t)dx}{(x+t)\sqx} \nn \\
&& \hspace*{1cm} = \frac{\pi\log t}{\sqt}
+\pi\int_{0}^{1}\frac{1}{\la-1}\left(\frac{1}{\sqt}-\frac{1}{\sqrt{(t+a\la)(t+b\la)}}\right)d\la\nn\\
&& \hspace*{1cm} =\frac{\pi\log t}{\sqt} + \pi\int_{0}^{1}\frac{1}{x}
\left(\frac{1}{\sqrt{(-ax+t+a)(-bx+t+b)}}-\frac{1}{\sqt}\right)dx\nn\\
&& \hspace*{1cm} =\frac{\pi\log t}{\sqt} + \pi
\lim_{\epsilon\to 0}\int_{\epsilon}^{1}\frac{1}{x}\left(\frac{1}{\sqrt{(-ax+t+a)(-bx+t+b)}}-\frac{1}{\sqt}\right)dx, \nn
\eq
where we have made the substitution $x = 1-\la$ and have replaced
$\int_{0}^{1}...$
by $
\lim_{\epsilon\to 0}\int_{\epsilon}^{1}...,$
so that we may invoke (\ref{eq:Simple0}). The integration is now completed as
\bq
&&\int_{a}^{b}\frac{\log(x+t)}{(x+t)\sqx}dx=\frac{\pi\log t}{\sqt}\nn\\
&& \hspace*{1cm} +\lim_{\epsilon\to 0}
\frac{\pi}{\sqt}\left(-\log\frac{\epsilon}{4(t+a)(t+b)}-\log t[2\sqt+2t+a+b]-\log(1/\epsilon)\right)\nn\\
&& \hspace*{1cm} = \frac{\pi}{\sqt}\log\left(\frac{4(t+a)(t+b)}{2\sqt+2t+a+b}\right)\\
&& \hspace*{1cm} \sim \pi\frac{\log t}{t},\quad t\to\infty. \label{eq:larget}
\eq
Note that the correct large $t$ behavior is reproduced in (\ref{eq:larget}). Some trivial algebra yields (\ref{eq:Int2}).

Now consider (\ref{eq:Int5})--(\ref{eq:Int8}). We will explicitly derive (\ref{eq:Int8});
the integral (\ref{eq:Int7}) is then obtained by the analytical continuation of (\ref{eq:Int8}) to $t=-1$,
whereas the integrals (\ref{eq:Int5}) and (\ref{eq:Int6}) are obtained by taking $t\to\infty$ and $t\to 0$
in (\ref{eq:Int8}) respectively. With the Schwinger parametrization (\ref{eq:LogParam}) and the partial
fraction decomposition
$$
\frac{x}{(x-1/\la)(x+t)}=\frac{1/\la}{(x-\la)(t+1/\la)}+\frac{t}{(t+1/\la)(x+t)},
$$
the integral becomes
\bq
\int_{a}^{b}\frac{\log(1-x)}{(x+t)\sqx}dx&=&
\int_{0}^{1}\frac{d\la}{1+\la t}\int_{a}^{b}\left(\frac{1/\la}{x-1/\la}+\frac{t}{x+t}\right)\frac{dx}{\sqx}\nn\\
&=&\pi\int_{0}^{1}\left(-\frac{1}{\sqrt{(1-\la a)(1-\la b)}}+\frac{t}{\sqt}\right)\frac{d\la}{1+\la\:t}.\nn
\eq
The last equation was obtained by invoking (\ref{eq:Simple2}) and taking the analytic continuation of $t$ to $-1/\la$, together with
the implicit assumption that $\Re\:t>0$ in (\ref{eq:Simple2}).  From a further change of variable $1+\la\;t=x,$ we have
\bq
&&\int_{a}^{b}\frac{\log(1-x)}{(x+t)\sqx}dx=\pi
\int_{1}^{1+t}\left(\frac{1}{\sqt}-\frac{1}{\sqrt{(t+a-ax)(t+b-bx)}}\right)\frac{dx}{x}\nn\\
&&  \hspace*{0.5cm} = \frac{\pi}{\sqt}\Bigg[\log(1+t)\nn\\
&& \hspace*{0.5cm} + \log\frac{2\sqt\sqrt{(t+a-ax)(t+b-bx)}-[t(a+b)+2ab]x+2(t+a)(t+b)}{x}\Bigg|_{x=1}^{x=1+t}\Bigg]\nn\\
&& \hspace*{0.5cm} = \frac{\pi}{\sqt}\log\frac{2\sqt\sqrt{(1-a)(1-b)}+[2-a-b]t+a+b-2ab}{2\sqt+2t+a+b}\nn\\
&& \hspace*{0.5cm} \sim\frac{\pi}{t}\log\frac{2\oab+2-a-b}{4},\quad t\to\infty.\nn
\eq
Simple algebra yields (\ref{eq:Int8}).


Finally, consider the principal value integrals (\ref{eq:PInt1}) and (\ref{eq:PInt2}).  These results are obtained by taking the square root to the denominator of the integrand,
followed by performing a partial fraction decomposition and invoking (\ref{eq:Simple4}) and (\ref{eq:PIntSimple}).

\end{document}